\documentclass[useAMS,usenatbib]{mnras}

\usepackage[T1]{fontenc}
\usepackage{graphicx}	% Including Fig.\ files
\usepackage{amsmath}	% Advanced maths commands
\usepackage{amssymb}	% Extra maths symbols
\usepackage{enumitem}
\usepackage{tabularx}

\usepackage[]{natbib}
\usepackage[utf8]{inputenc}
\usepackage{times}
\usepackage{booktabs}
\let\textbf\relax
\usepackage{newtxtext,newtxmath}

\newcommand{\aanda}{A\&A}

\newcommand{\Msol}{M_{\odot}}
\newcommand{\eps}{\epsilon}
\newcommand{\yeeq}{Y_e^{\mathrm{eq}}}
\newcommand{\yeeqem}{Y_e^{\mathrm{eq,em}}}
\newcommand{\yeeqabs}{Y_e^{\mathrm{eq,abs}}}
\newcommand{\yeeqmu}{Y_e^{\mathrm{eq},\mu_\nu=0}}
\newcommand{\tauem}{\tau_{\mathrm{em}}}
\newcommand{\tauabs}{\tau_{\mathrm{abs}}}

\newcommand{\tauexp}{\tau_{\mathrm{exp}}}
\newcommand{\tauopt}{\tau_{\mathrm{opt}}}
\newcommand{\dd}{\mathrm{d}}
\newcommand{\gcm}{\,g\,cm$^{-3}$}
\newcommand{\alpvis}{\alpha_{\mathrm{vis}}}
\newcommand{\MBH}{M_{\mathrm{BH}}}
\newcommand{\ABH}{A_{\mathrm{BH}}}
\newcommand{\XLA}{X_{\mathrm{LA}}}
\newcommand{\Xlan}{X_{\mathrm{lan}}}
\newcommand{\Xact}{X_{\mathrm{act}}}
\newcommand{\qheat}{q}
\newcommand{\lturb}{l_{\mathrm{t}}}
\newcommand{\yeavg}{\langle Y_e\rangle}
\newcommand{\yeeqavg}{\langle \yeeq \rangle}
\newcommand{\yeeqemavg}{\langle \yeeqem \rangle}

\newcommand{\yeeqmuavg}{\langle \yeeqmu \rangle}

\newcommand{\yeeqbar}{\bar{Y}_e^{\mathrm{eq}}}
\newcommand{\yeeqembar}{\bar{Y}_e^{\mathrm{eq,em}}}

\newcommand{\mtor}{m_{\mathrm{tor}}}

\newcommand{\yemin}{\yeavg_{\mathrm{tor}}^{\mathrm{min}}}
\newcommand{\yeej}{Y_{e,\mathrm{ej}}}

\def\ga{\,\,\raise0.14em\hbox{$>$}\kern-0.76em\lower0.28em\hbox{$\sim$}\,\,}
\def\la{\,\,\raise0.14em\hbox{$<$}\kern-0.76em\lower0.28em\hbox{$\sim$}\,\,}

\title[Neutrino absorption in black-hole disks]
      {Neutrino absorption and other physics dependencies in neutrino-cooled black-hole accretion disks}
\author[Just et al.]{O.~Just$^{1,2}$\thanks{E-mail: o.just@gsi.de}, S.~Goriely$^{3}$, H.-Th.~Janka$^{4}$,
   S.~Nagataki$^{2,5}$, \& A.~Bauswein$^{1,6}$ \\
  $^1$GSI Helmholtzzentrum f\"ur Schwerionenforschung, Planckstraße 1, 64291 Darmstadt, Germany \\
  $^2$Astrophysical Big Bang Laboratory, RIKEN Cluster for Pioneering Research, 2-1 Hirosawa, Wako, Saitama 351-0198, Japan \\
  $^3$Institut d'Astronomie et d'Astrophysique, CP-226, Universit\'e Libre de Bruxelles, 
  1050 Brussels, Belgium \\
  $^4$Max-Planck-Institut f\"ur Astrophysik, Postfach 1317, 85741 Garching, Germany \\
  $^5$RIKEN Interdisciplinary Theoretical \& Mathematical Science Program (iTHEMS), 2-1 Hirosawa, Wako, Saitama, Japan 351-0198 \\
  $^6$Helmholtz Research Academy Hesse for FAIR (HFHF), GSI Helmholtz Center for Heavy Ion Research, \\
  Campus Darmstadt, Planckstra{\ss}e 1, 64291 Darmstadt, Germany  
}

\begin{document}
%% \label{firstpage}
%% \pagerange{\pageref{firstpage}--\pageref{lastpage}}
\maketitle

\begin{abstract}
  Black-hole (BH) accretion disks formed in compact-object mergers or collapsars may be major sites of the rapid-neutron-capture (r-)process, but the conditions determining the electron fraction ($Y_e$) remain uncertain given the complexity of neutrino transfer and angular-momentum transport. After discussing relevant weak-interaction regimes, we study the role of neutrino absorption for shaping $Y_e$ using an extensive set of simulations performed with two-moment neutrino transport and again without neutrino absorption. We vary the torus mass, BH mass and spin, and examine the impact of rest-mass and weak-magnetism corrections in the neutrino rates. We also test the dependence on the angular-momentum transport treatment by comparing axisymmetric models using the standard $\alpha$-viscosity with viscous models assuming constant viscous length scales ($\lturb$) and three-dimensional magnetohydrodynamic (MHD) simulations. Finally, we discuss the nucleosynthesis yields and basic kilonova properties. We find that absorption pushes $Y_e$ towards $\sim$0.5 outside the torus, while inside increasing the equilibrium value $\yeeq$ by $\sim$0.05--0.2. Correspondingly, a substantial ejecta fraction is pushed above $Y_e=0.25$, leading to a reduced lanthanide fraction and a brighter, earlier, and bluer kilonova than without absorption. More compact tori with higher neutrino optical depth, $\tau$, tend to have lower $\yeeq$ up to $\tau\sim$~1--10, above which absorption becomes strong enough to reverse this trend. Disk ejecta are less (more) neutron-rich when employing an $\lturb$=const. viscosity (MHD treatment). The solar-like abundance pattern found for our MHD model marginally supports collapsar disks as major r-process sites, although a strong r-process may be limited to phases of high mass-infall rates, $\dot{M}\ga 2\times 10^{-2}$~$\Msol$\,s$^{-1}$.
\end{abstract}


%% \newpage

%%%%%%%%%%%%%%%%%%%%%%%%%%%%%%%%%%%%%%%%%%%%%%%%%%%%%%%%%%%%%%%%%%%%%%%%%%%%%%%%%%
\section{Introduction}\label{sec:introduction}

The recent discovery of a binary neutron-star (NS) merger via gravitational waves and electromagnetic counterparts, GW170817/AT2017gfo/GRB170817 \citep[e.g.][]{Abbott2017a,Abbott2017b,Chornock2017a,Villar2017a,Kasen2017a,Metzger2019a,Tanvir2017a,Waxman2018a,Perego2017a,Gottlieb2018c,Kawaguchi2018a,Mooley2018b}, provided long-sought observational support for the idea that NS mergers are prolific sites of the rapid neutron capture (r-) process \citep[e.g.][]{Lattimer1977, Eichler1989, Freiburghaus1999, Goriely2005, Goriely2011, Korobkin2012, Wanajo2014a, Perego2014a, Just2015a}, can be observed as a kilo- or macronova in optical frequency bands \citep[e.g.][]{Metzger2010c, Roberts2011, Tanaka2013, Grossman2014,  Kasen2015}, can produce short gamma-ray burst (GRB) jets \citep[e.g.][]{Eichler1989, Ruffert1998a, Rosswog2003a, Nakar2007, Lee2007, Rezzolla2011, Paschalidis2015, Just2016}, and may serve as unique laboratories for exploring the high-density regime of matter \citep[e.g.][]{Bauswein2017b,Margalit2017a,Radice2018b,Abbott2018a,Rezzolla2018a}. One of the main ejecta components encountered in NS mergers (and also in mergers of NSs with black holes, BH) is thought to originate during the first few seconds post merger after the central object has formed a BH and the surrounding disk disintegrates as a result of turbulent angular momentum transport and neutrino cooling \citep{Popham1999, Kohri2002, Beloborodov2003, Setiawan2004, Shibata2007, Metzger2008c,Fernandez2013b, Just2015a, Siegel2018c, Hossein-Nouri2018a, Siegel2019b, Janiuk2019a, Miller2020s,Fujibayashi2020a}.

Apart from NS mergers, neutrino-cooled BH-accretion disks may also form during the collapse of a strongly rotating massive star \citep[e.g.][]{MacFadyen1999}. Once the innermost core has collapsed to a BH the surrounding layers of the star with sufficient angular momentum settle on circular orbits. These systems, so-called collapsars, are not only potential candidates for powering long GRBs \citep{Woosley1993}, but have also been considered as nucleosynthesis sites \citep[e.g.][]{Pruet2004, Surman2005,Nagataki2006a,Nakamura2015b}. While early studies found a strong r-process only for very specific conditions, more recent investigations \citep{Siegel2019b} based on time-dependent numerical simulations instead report massive outflows with generically favorable conditions for the r-process and even predict that the galactic enrichment of r-process material could be mainly due to these collapsar outflows.

While theoretical modeling of neutrino-cooled BH-accretion disks has seen tremendous progress in the recent years -- from stationary-state semi-analytic spherically symmetric (1D) models \citep[e.g.][]{Popham1999} to time-dependent three-dimensional (3D) general relativistic (GR) magnetohydrodynamic (MHD) models \citep[e.g.][]{Siegel2018c} -- significant uncertainties are still connected to the composition of the disk outflows. One major challenge is the treatment of neutrinos, of which the total release rates control the disk dynamics, whereas the relative emission and absorption rates of electron-neutrinos ($\nu_e$) to electron-antineutrinos ($\bar\nu_e$) characterize the lepton number transport and regulate the electron fraction, $Y_e$. In general, a consistent description of neutrinos requires solving the radiative transfer (i.e. Boltzmann) equation with a total of six (three spatial plus three momentum) degrees of freedom, which calls for enormous computational efforts \citep[e.g.][]{Mihalas1984} even without accounting for the possibility of neutrino flavor oscillations (e.g. \citealp{Malkus2012, Wu2017a, Deaton2018a, Richers2019a}). While the first simulations evolving the Boltzmann equations have recently become available \citep{Miller2019a}, their computational demands yet pose strict limits on the evolution times and prohibit extensive parameter explorations.

The typically rather low masses of neutrino-cooled disks and correspondingly low optical depths to neutrinos, compared to neutron stars formed in core-collapse supernovae \citep[e.g.][]{Janka2017c} or as remnants of NS mergers \citep[e.g.][]{Perego2014a}, spur the notion that the impact of neutrino absorption might be minor or possibly even negligible. From the modeling point of view this situation would be advantageous, because the smaller the impact of neutrino absorption is, the more accurate and credible are results obtained with schemes incorporating neutrino absorption only approximately or not at all, such as purely local trapping schemes \citep{Shibata2007}, leakage schemes \citep[e.g.][]{Fernandez2020a}, leakage plus post-processing schemes \citep[e.g.][]{Siegel2019b}, or M1 schemes (i.e. two-moment transport schemes with a local closure, e.g. \citealp{Just2015b}). Indirect evidence for a relatively minor relevance of neutrino absorption might come from the fact that simulations including neutrino absorption, if only approximately, only found very small amounts of neutrino-driven compared to viscously driven ejecta \citep{Just2015a, Fujibayashi2020a}. Moreover, the recent simulations by \citet{Fujibayashi2020a} of viscous disks, performed in GR using a combination of energy-independent M1 and leakage schemes, seem to suggest that neutrino absorption is even irrelevant for disks more massive than $0.1\,M_\odot$. These results are, however, in stark contrast to the findings of \citet{Miller2020s}, whose GRMHD simulations with Boltzmann neutrino transport advocate a substantial sensitivity of $Y_e$ to absorption-related effects even for a $0.02\,M_\odot$ disk. Thus, the role of neutrino absorption and its sensitivity to other modeling ingredients still remains unclear and detailed investigations are overdue.

The difficulties connected to the neutrino treatment are aggravated by the existence of another, comparably challenging modeling ingredient, namely angular momentum transport, i.e. the mechanism that is mainly responsible for accretion, heating, and ejection of disk material. Being a consequence of MHD turbulence driven by the magneto-rotational instability (MRI, e.g. \citealp{Balbus1991}), angular momentum transport in MHD disks requires, in order for it to be modeled properly, that the simulation is performed in three dimensions\footnote{As pointed out by the anti-dynamo theorem \citep{Moffatt1978}, axisymmetric models suffer from the inability to efficiently create poloidal magnetic fields from toroidal fields, a mechanism that is needed to keep the MRI alive and therefore to sustain angular momentum transport.} and with sufficiently high spatial resolution in order to resolve the wavelengths of MRI growth and the relevant scales of MHD turbulence. Hence, 3D MHD models, of which the first have recently become available \citep{Siegel2018c, Hossein-Nouri2018a, Fernandez2019b, Christie2019a, Miller2019a} are computationally quite expensive even without neutrino transport, and many of their properties still remain unexplored or poorly understood, particularly concerning the $Y_e$ evolution and its sensitivity to details of the neutrino interactions.

Both of the aforementioned requirements for MHD models can be relaxed by reverting to an approximate mean-field description of turbulent angular momentum transport, such as embodied by the $\alpha$-viscosity approach \citep{Shakura1973} that has been employed in numerous 1D and 2D studies \citep[e.g.][]{Popham1999, DiMatteo2002, Chen2007, Metzger2009b, Fernandez2013b, Just2015a, Fujibayashi2020a}. Given their computational efficiency, viscous disk models have been studied already for a much broader range of conditions and input parameters than MHD models. However, although taken into account by various published results with different degrees of sophistication (e.g. \citealp{Just2015a, Fujibayashi2020a, Miller2019a}), we still lack a comprehensive understanding of the importance of neutrino absorption. Moreover, relatively little attention has been drawn so far to the sensitivity of the ejecta properties with respect to other components of modeling, such as using a non-standard prescription for the dynamic viscosity \citep{Fujibayashi2020a}, neglecting rest-mass terms in the $\beta$-reaction rates or including weak magnetism corrections \citep{Horowitz2002a}, or using different initial $Y_e$ values in the torus.

Whether occurring in the course of a NS merger or of a collapsar, the possibility that neutrino-cooled disks may be major sites of r-process elements calls for a profound understanding of all processes and modeling assumptions that have a leverage on the ejecta $Y_e$, first and foremost the interplay between neutrino emission, neutrino absorption, and angular momentum transport. In this study we therefore systematically investigate, on the basis of two- and three-dimensional viscous and MHD simulations including M1 neutrino transport, the impact of neutrino absorption and the sensitivity to the treatment of angular momentum transport and to the variation of global model parameters. We further test uncertainties connected to details of the neutrino interaction rates and to the initial electron fraction of the torus. In order to relate the obtained dependencies of the hydrodynamical simulations to nucleosynthesis variations and to the kilonova signal, we compute for all models the abundances of r-process elements and basic properties of the bolometric kilonova light curve.

This paper is organized as follows: In Sect.~\ref{sec:equil-cond-that} we first review the equilibrium conditions for weak interactions and corresponding $Y_e$ values and characteristic timescales and test their sensitivity to commonly used approximations. Section~\ref{sec:setup-numer-models} describes the setup of our numerical models and of the post-processing steps aiming at evaluating the nucleosynthesis yields and kilonova light curve. In Sect.~\ref{sec:results} we first summarize basic features of the torus evolution and the neutrino emission, followed by an analysis of the impact of neutrino absorption on the torus evolution and on the outflow. Moreover, we discuss the nucleosynthesis yields and the kilonova properties. In Sect.~\ref{sec:discussion} we discuss implications of our results based on a comparison with existing studies. Finally, in Sect.~\ref{sec:summary-conclusions} we summarize and conclude our study.

%%%%%%%%%%%%%%%%%%%%%%%%%%%%%%%%%%%%%%%%%%%%%%%%%%%%%%%%%%%%%%%%%%%%%%%%%%%%%%%%%%
\section{Equilibrium conditions for $Y_e$}\label{sec:equil-cond-that}

Before discussing numerical models we first review the neutrino interaction rates, equilibrium conditions, and characteristic timescales that are relevant for the evolution of the electron fraction, $Y_e$, in neutrino-cooled accretion disks.

\subsection{Neutrino emission and absorption rates}\label{sec:neutr-emiss-absorpt}

The interactions mainly responsible for changing $Y_e$ in neutrino-cooled disks are the nucleonic $\beta$-processes, namely electron capture on protons, positron capture on neutrons, electron neutrino capture on neutrons, and electron anti-neutrino capture on protons. The interaction rates corresponding to these processes are given by\footnote{We neglect phase space blocking for neutrinos and nucleons as well as other rate corrections that only play a role at larger densities ($\rho\gg 10^{12}\,$g\,cm$^{-3}$) than typically encountered in neutrino-cooled accretion disks.} \citep{Bruenn1985, Horowitz2002a}:
\begin{subequations}\label{eq:betarates}
\begin{align}
  \lambda_{e^-} & = K_\beta  \int_0^{\infty} \eps^2 F_{e^-}(\eps_+)\eps_+^2 
  \sqrt{1  - \left(\frac{m_e c^2}{\eps_+}\right)^2} \mathrm{d}\eps  \, ,\label{eq:beta1}\\
  \lambda_{e^+} & = K_\beta  \int_{\eps_0}^{\infty}  \eps^2 F_{e^+}(\eps_-)\eps_-^2 
  \sqrt{1  - \left(\frac{m_e c^2}{\eps_-}\right)^2} \mathrm{d}\eps  \, , \label{eq:beta2}\\
  \lambda_{\nu_e} & = K_\beta  \int_0^{\infty}  \eps^2
  F_{\nu_e}(\eps)(1-F_{e^-}(\eps_+))\eps_+^2
  \sqrt{1  - \left(\frac{m_e c^2}{\eps_+}\right)^2} \mathrm{d}\eps  \, , \label{eq:beta3}\\
  \lambda_{\bar\nu_e} & = K_\beta  \int_{\eps_0}^{\infty} \eps^2
  F_{\bar\nu_e}(\eps)(1-F_{e^+}(\eps_-))\eps_-^2
  \sqrt{1  - \left(\frac{m_e c^2}{\eps_-}\right)^2} \mathrm{d}\eps  \, ,\label{eq:beta4}
\end{align}
\end{subequations}
where $F_x(\eps)$ is the distribution function of particle $x$ at energy $\eps$ integrated over solid angles in momentum space, $c$ the speed of light, $m_e$ the electron mass, $\eps_0=Q_{np}+m_e c^2$ with $Q_{np}$ being the neutron-proton mass difference, $\eps_\pm=\eps\pm Q_{np}$, and $K_\beta^{-1}=(1506\,\mathrm{s})(m_e c^2)^5$. The composition of the gas and its thermodynamic properties enter the rates through the distribution functions $F_{e^\pm}$. The effects of weak magnetism and nucleon recoil can additionally be taken into account by multiplying the integrands in Eq.~(\ref{eq:betarates}) by correction factors $R_{\nu_e/\bar\nu_e}^{\mathrm{wm}} (\eps)$ (see \citealp{Horowitz2002a} for explicit expressions). With the above rates the evolution equation of $Y_e$ for a Lagrangian fluid element reads:
\begin{align}\label{eq:yeevo}
  \frac{\dd Y_e}{\dd t} = (\lambda_{e^+}+\lambda_{\nu_e}) Y_n - (\lambda_{e^-}+\lambda_{\bar\nu_e})Y_p \, ,
\end{align}
where $Y_{n/p}=n_{n/p}/n_B$ is the number of free neutrons/protons relative to the total number of baryons. In a gas consisting only of free neutrons and protons one has $Y_{p}=Y_e$ and $Y_n=(1-Y_e)$, whereas in a gas composed of nuclei in nuclear statistical equilibrium (NSE) $Y_{n/p}$ are functions of density, $\rho$, temperature, $T$, and $Y_e$. Setting $\dd Y_e/\dd t =0$ in Eq.~(\ref{eq:yeevo}),
\begin{align}\label{eq:yeeq}
  (\lambda_{e^+}+\lambda_{\nu_e}) Y_n - (\lambda_{e^-}+\lambda_{\bar\nu_e})Y_p\Bigg|_{\rho,T,Y_e^{\mathrm{eq}}} = 0 \, ,
\end{align}
defines an equilibrium value, $Y_e^{\mathrm{eq}}$, that would asymptotically be reached by a fluid element with a given density and temperature and exposed to a given neutrino field. The characteristic timescale on which $Y_e$ approaches $Y_e^{\mathrm{eq}}$ can be estimated as
\begin{align}\label{eq:taubeta}
 \tau_\beta = \frac{1}{Y_p(\lambda_{e^-}+\lambda_{\bar\nu_e})+Y_n(\lambda_{e^+}+\lambda_{\nu_e})}  \, .
\end{align}
Anywhere along a fluid trajectory, weak interactions drive $Y_e$ to the local $\yeeq$ on a local timescale $\tau_\beta$. Once $\tau_\beta$ becomes longer than the expansion timescale $\tau_{\mathrm{exp}}\sim\rho/\dot{\rho}\sim r/v_r$ (with $r$ and $v_r$ being the radius and radial velocity of the fluid element) in an expanding outflow, $Y_e$ effectively remains constant, i.e. it freezes out.

\begin{figure*}
  \centering
  \includegraphics[trim= 0 30 0 10,clip,scale=0.65]{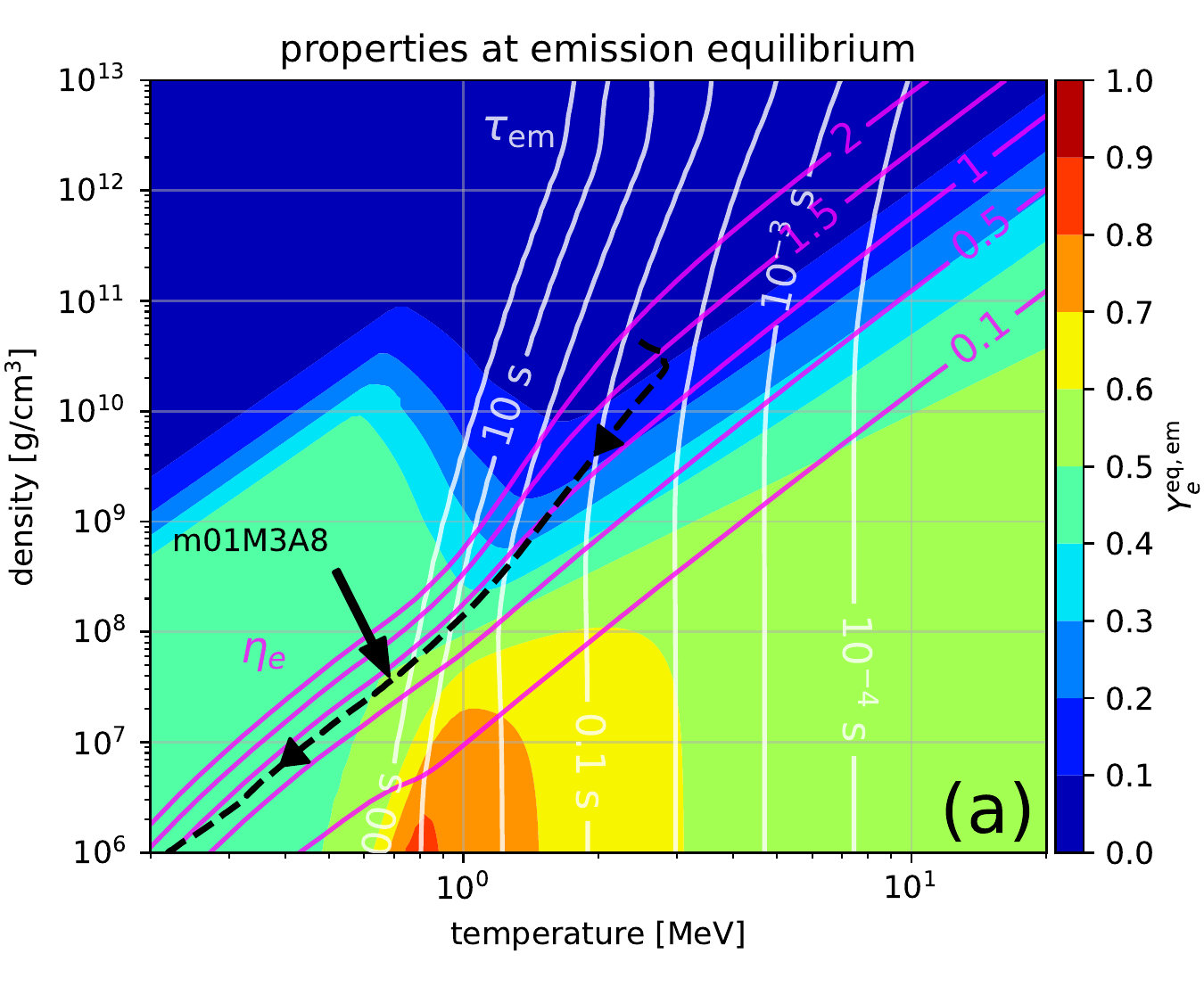}
  \includegraphics[trim=20 30 0 10,clip,scale=0.65]{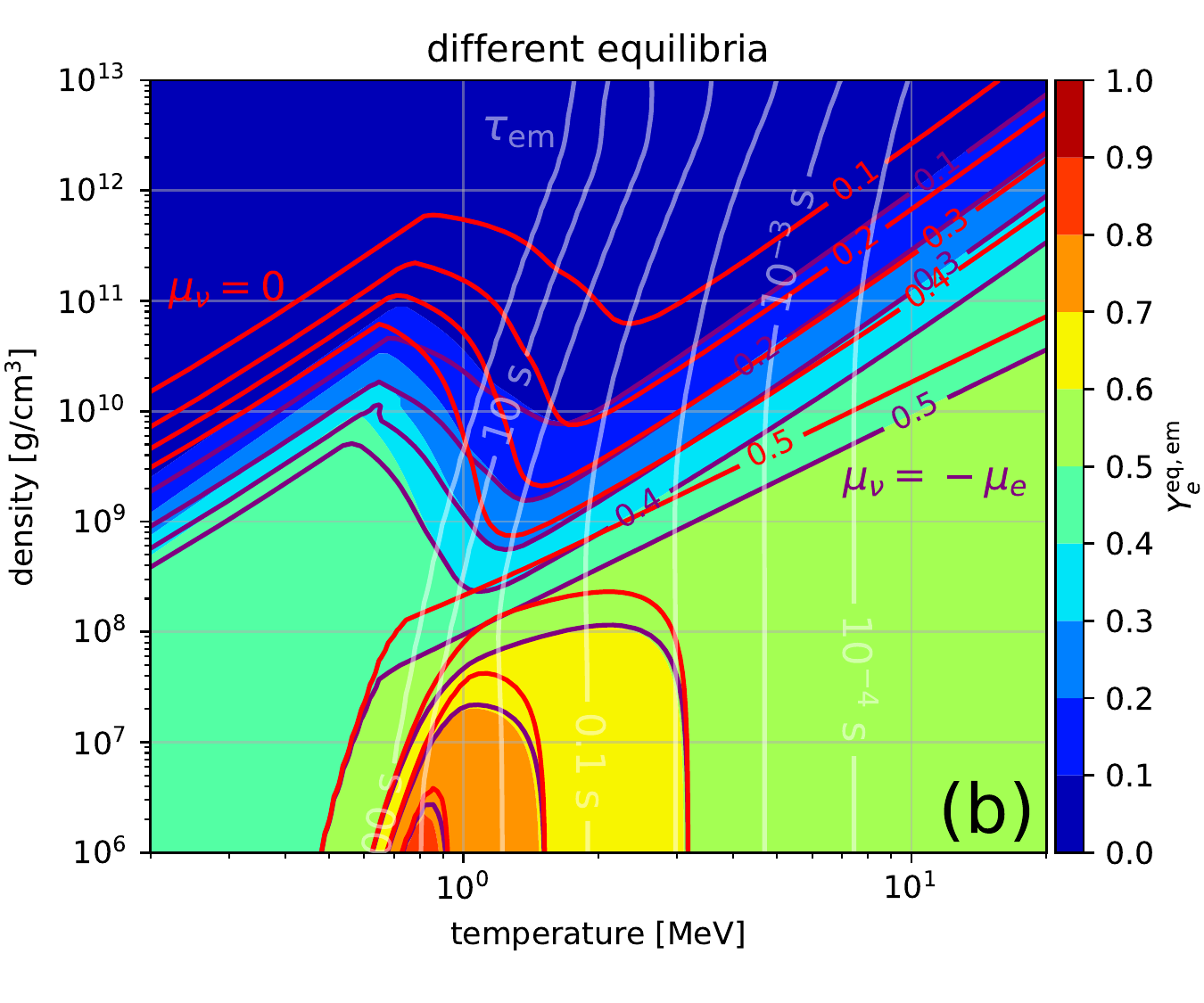}
  \includegraphics[trim= 0 30 0 10,clip,scale=0.65]{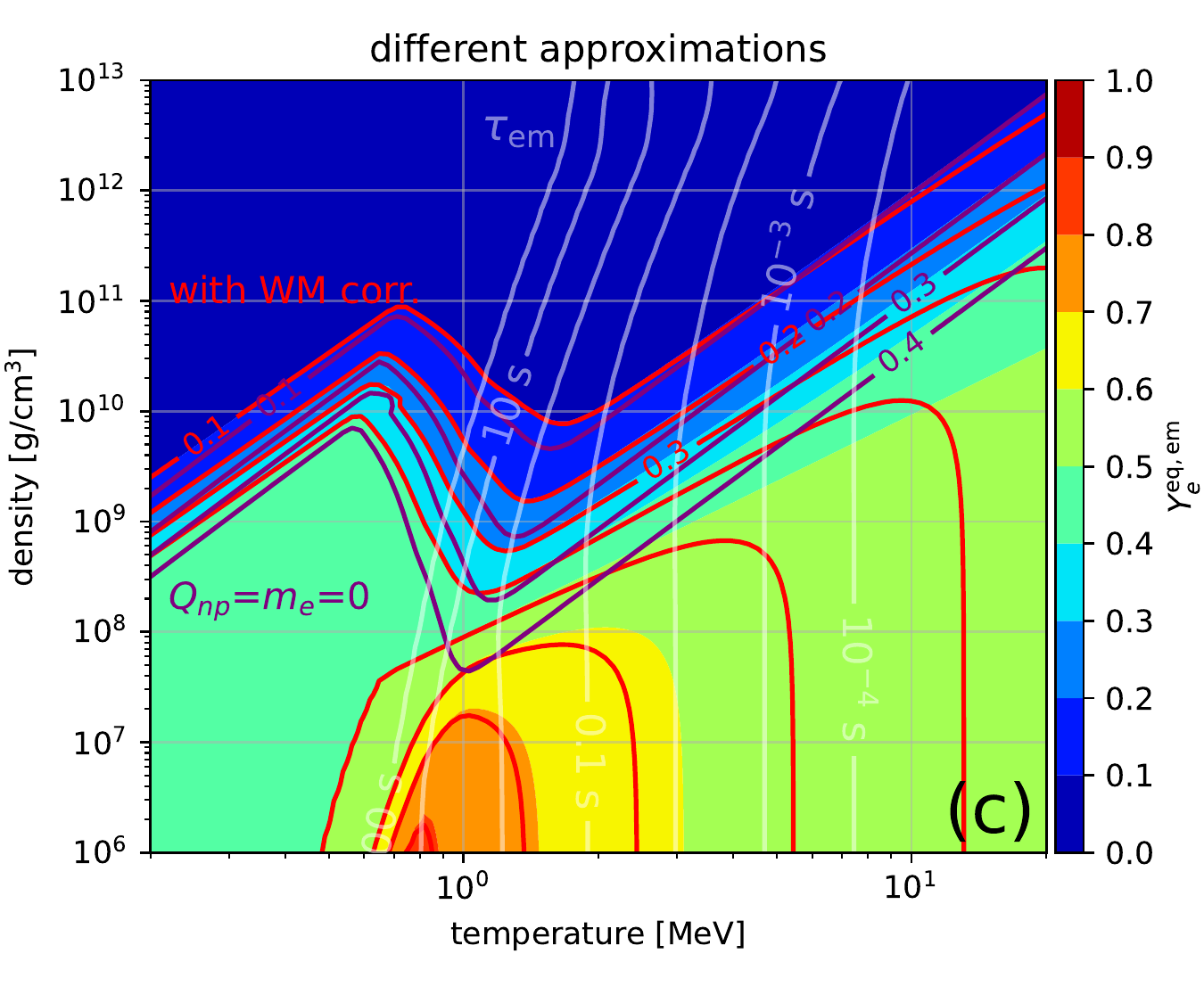}
  \includegraphics[trim=20 30 0 10,clip,scale=0.65]{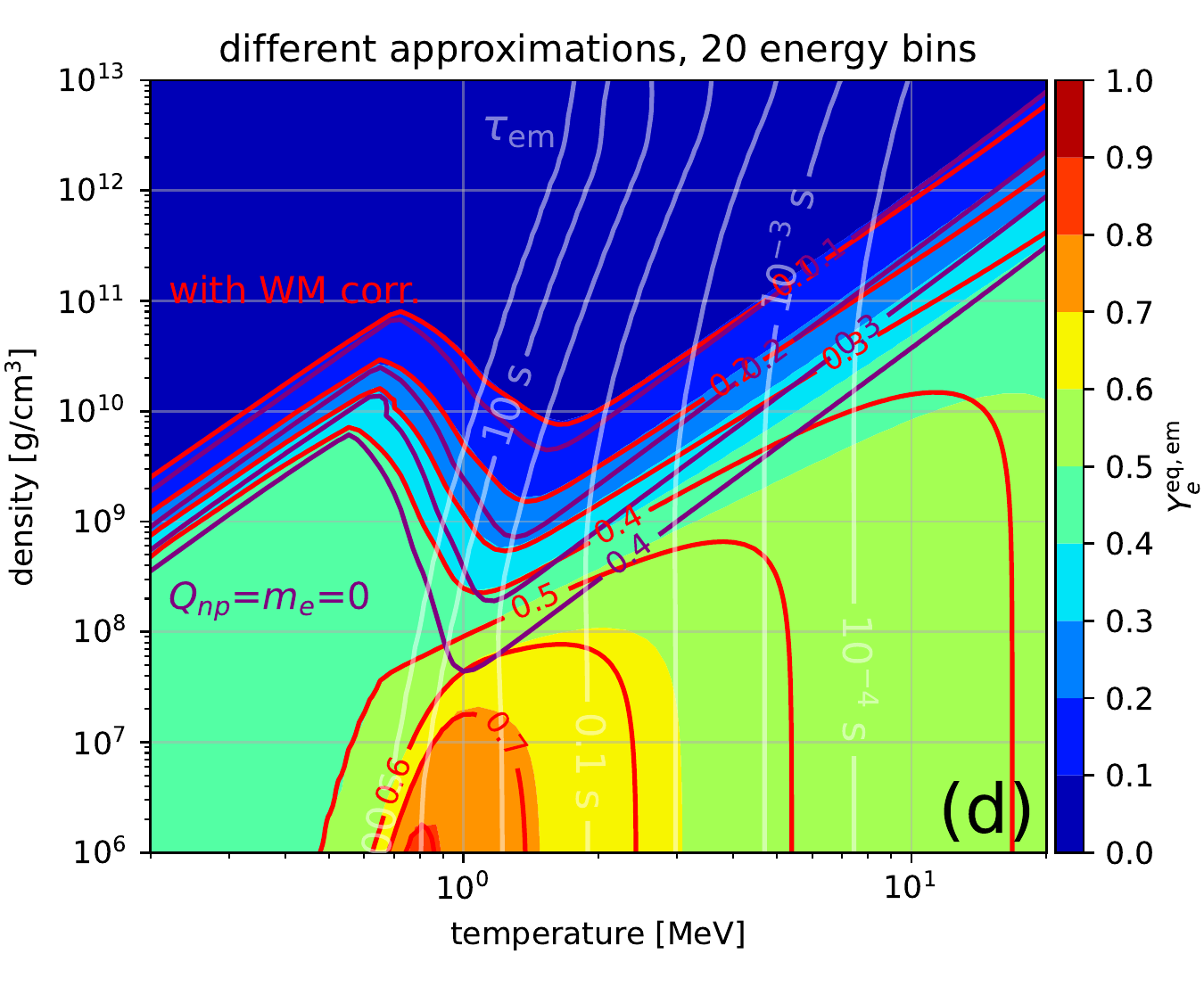}
  \includegraphics[trim= 0 10 0 10,clip,scale=0.65]{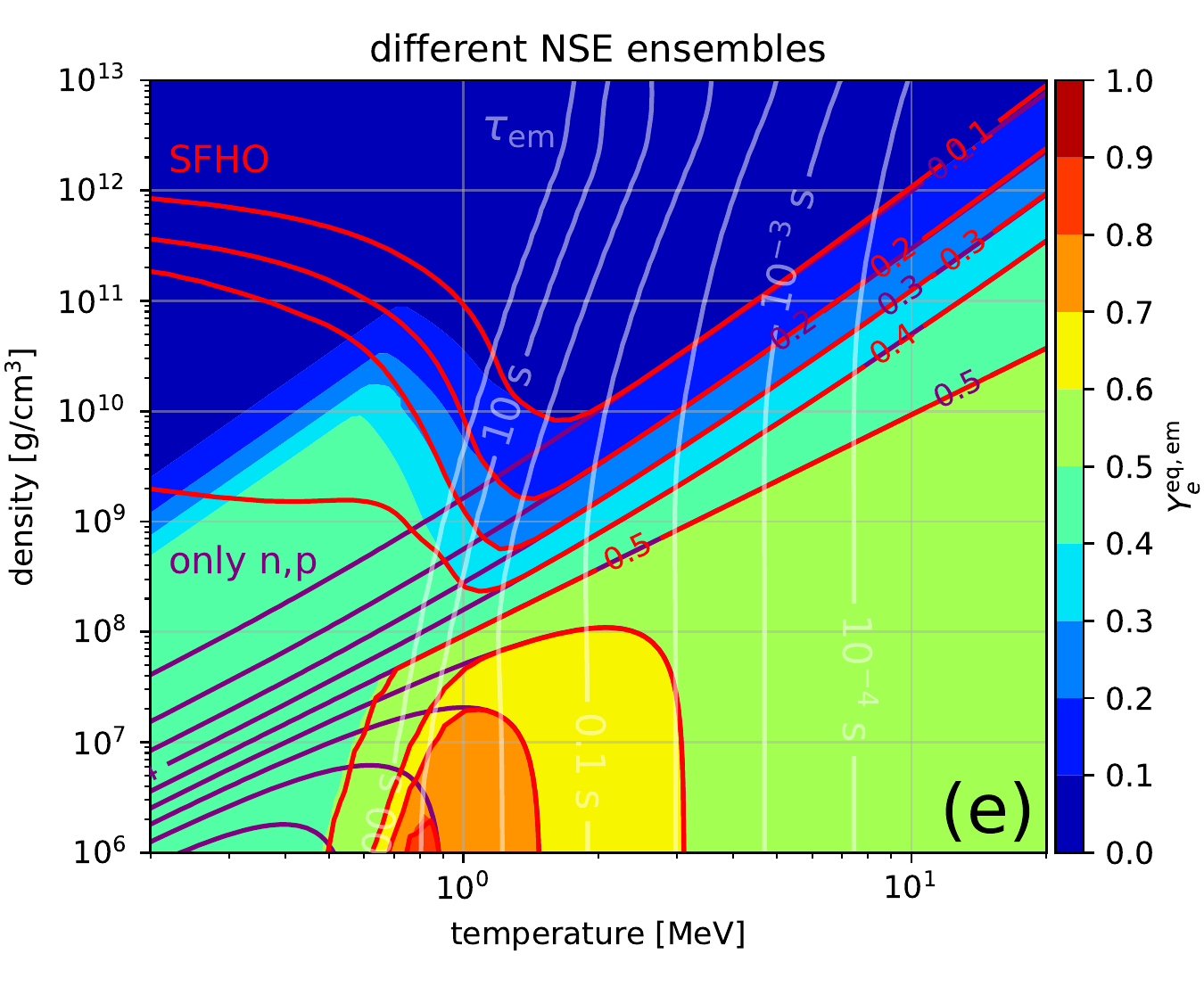}
  \includegraphics[trim=20 10 0 10,clip,scale=0.65]{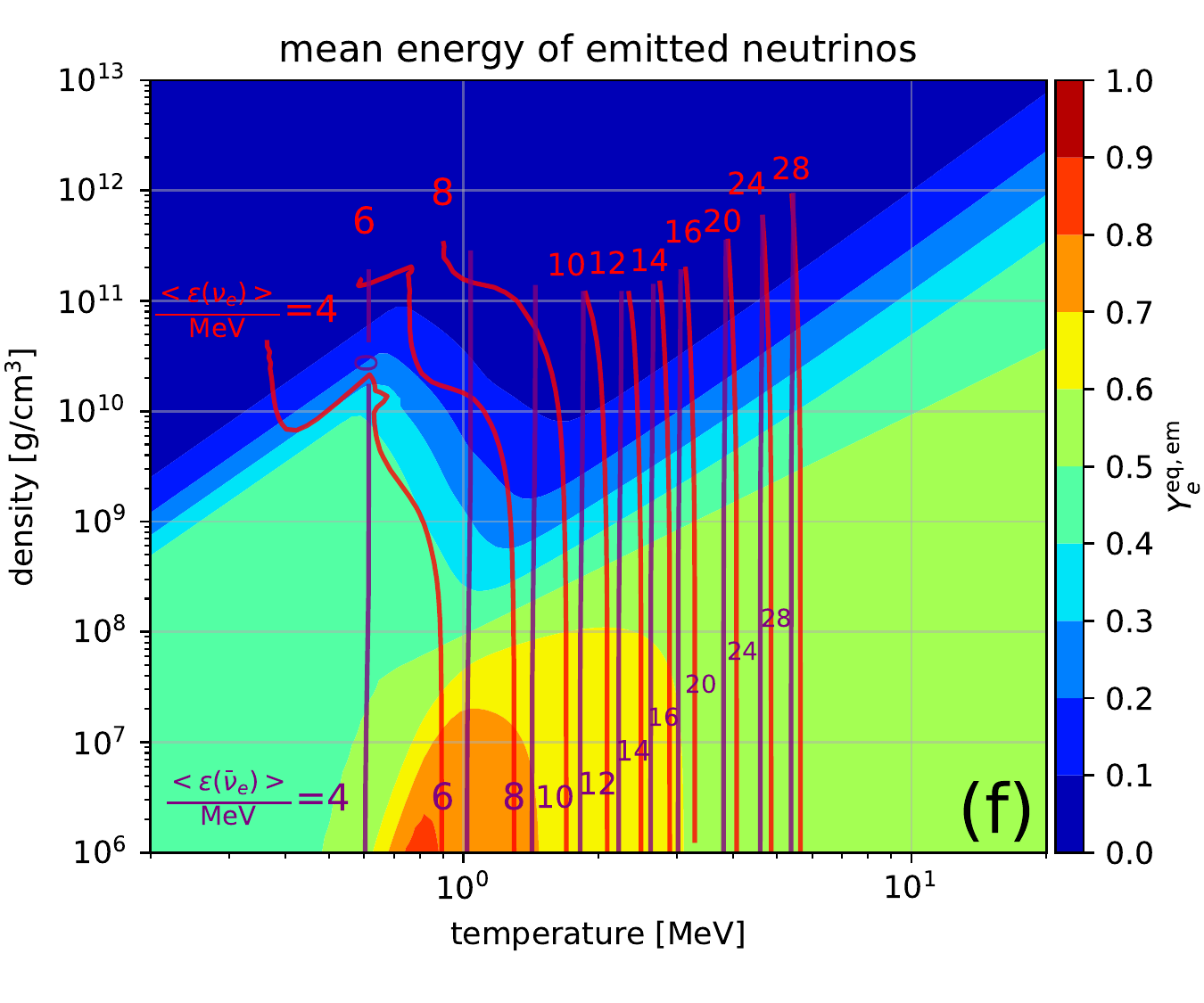}
  \caption{Properties connected to the kinetic emission equilibrium, which is established once the rate of $p+e^- \rightarrow n+\nu_e$ equals that of $n+e^+\rightarrow p+\bar\nu_e$. The color map in all panels illustrates $\yeeqem$ defined by Eqs.~(\ref{eq:beta1}),~(\ref{eq:beta2}), and~(\ref{eq:yeeqem}) and the 4-species NSE composition employed in our numerical simulations. \emph{Panel (a):} characteristic neutrino emission timescale, $\tauem$ (white lines), electron degeneracy parameter, $\eta_e$ (purple lines), and average density-temperature evolution of a fiducial numerical model (dashed black line); \emph{(b):} $Y_e$ corresponding to $\mu_{\nu}=-\mu_e$ (purple lines) and to $\mu_\nu=0$ (red lines); \emph{(c):} $\yeeqem$ computed with weak-magnetism and recoil corrections (red lines) as well as using the simplification $Q_{np}=m_e=0$ (purple lines); \emph{(d):} same as panel (c) but the color map and lines are obtained using the coarser neutrino energy grid that is employed in numerical simulations of this study; \emph{(e):} $\yeeqem$ resulting with the NSE composition of the SFHO EOS (red lines) and for a pure neutron-proton gas (purple lines); \emph{(f):} the mean energies of neutrinos, $\langle\eps\rangle$, emitted from a gas with the density, temperature, and $Y_e=\yeeqem$ given at each point (where some regions less relevant to the freeze out are neglected). All $Y_e$ contours show values of 0.1, 0.2, etc. from top to bottom. Since the dynamical timescales of outflows in neutrino-cooled disks are typically no longer than $\sim 1\,$s, the region left of the $\tau_{\mathrm{em}}=100\,$s contour, where approximately $T\la 1\,$MeV, is irrelevant to our discussion.} \label{fig:yeeq1}
\end{figure*}	

\subsection{Limiting cases of $Y_e^{\mathrm{eq}}$}\label{sec:limit-cases-y_em}

In what follows we will briefly discuss three limiting cases of $Y_e^{\mathrm{eq}}$ and comment on the relevance of each for neutrino-cooled BH-tori.

\subsubsection{Kinetic equilibrium due to neutrino emission}\label{sec:emission-equilibrium}

Given the sub-nuclear densities and relatively low neutrino optical depths in neutrino-cooled disks, it is reasonable to assume that the bulk $Y_e$ is to a large extent determined by neutrino emission, i.e. the rates $\lambda_{e^\pm}$. In situations when neutrino absorption even becomes negligible, $\yeeq$ converges to $\yeeqem$, which is defined by
\begin{align}\label{eq:yeeqem}
  \lambda_{e^+} Y_n - \lambda_{e^-}Y_p\Bigg|_{\rho,T,Y_e^{\mathrm{eq,em}}} = 0 
\end{align}
and is a function solely of the hydrodynamic quantities (i.e. for NSE only of $\rho$ and $T$). Contours of $\yeeqem(\rho,T)$ are shown in panel (a) of Fig.~\ref{fig:yeeq1} overlaid with contours of the electron degeneracy parameter, $\eta_e$, and the characteristic timescales of neutrino emission,
%
%% \begin{subequations}\label{eq:betarates}
\begin{align}\label{eq:tauem}
  \tau_{\mathrm{em}} & = \frac{1}{Y_p\lambda_{e^-}+Y_n\lambda_{e^+}}  \nonumber\\
  & \sim K_\beta \,(k_B T)^{-5} (\mathcal{F}(\eta_e)+\mathcal{F}(-\eta_e))^{-1}  \nonumber\\
   & \sim 52\,\left(\frac{T}{1\,\mathrm{MeV}}\right)^{-5} \mathrm{s}\, .
\end{align}
%% \end{subequations}
%
As pointed out by \citet{Liu2010a} a very good approximation to $\yeeqem$, at least whenever nuclei are absent, can be recovered directly from the equation-of-state (EOS) table by exploiting the condition
\begin{align}\label{eq:yeeqliu}
  \mu_\nu\equiv \mu_p -\mu_n +\mu_e = -\mu_e
\end{align}
for the chemical potentials $\mu_i$ of species $i$\footnote{The condition $\mu_\nu=-\mu_e$ follows from the consideration that equal rates of $p + e^- \rightarrow n + \nu_e$ and $n + e^+ \rightarrow p+\bar\nu_e$ define a kinetic equilibrium, for which $\mu_p\dot{n}_p + \mu_e\dot{n}_{e^-}  =  \mu_n\dot{n}_n - \mu_e\dot{n}_{e^+}$ and where all $\dot{n}_i$ are equal (see \citealp{Liu2010a} for more details). We furthermore stress that the quantity $\mu_\nu$ should only be interpreted as neutrino chemical potential if neutrinos are thermalized, otherwise it is just a placeholder for $\mu_p -\mu_n +\mu_e$.}. Contours of $Y_e$ resulting from Eq.~(\ref{eq:yeeqliu}) are plotted as purple lines in panel (b) of Fig.~\ref{fig:yeeq1}).

Several previous studies have discussed the emission equilibrium defined by Eq.~(\ref{eq:yeeqem}) \citep[e.g.][]{Beloborodov2003,Metzger2008c,Arcones2010,Fujibayashi2020a}. The basic notion for the interpretation of $Y_e$ in neutrino-cooled disks is that during their expansion and cooling fluid elements sitting in and being released from the torus travel from top right in the $\rho-T$ domain (where $\yeeqem \ll 0.5$ due to electron degeneracy) to the bottom left region (where $\yeeqem \ga 0.5$), while locally driving $Y_e$ to $\yeeq(\rho,T)$ until $Y_e$ freezes out somewhere near the $\tauem\sim 0.01\ldots 1\,$s contours, i.e. when $\tauem$ starts exceeding the expansion timescale. As first pointed out by \citet{Chen2007}, the torus remains mildly degenerate during its expansion, i.e. with electron-degeneracies $\eta_e\sim$0.5-2, due to a self-regulating interplay between viscous heating and neutrino cooling. However, while the basic evolution of $Y_e$ in simulations of BH-tori can be explained by $\yeeqem$, corrections due to the presence of neutrinos have not been examined so far.

\subsubsection{Kinetic equilibrium due to neutrino absorption}\label{sec:absorption-equilibrium}

In the opposite limiting case when neutrino absorption dominates neutrino emission, such as in neutrino-driven winds, $\yeeq$ will be given by $\yeeqabs$, which fulfills
\begin{align}\label{eq:yeeqabs}
  \lambda_{\nu_e}Y_n - \lambda_{\bar\nu_e} Y_p\Bigg|_{\rho,T,Y_e^{\mathrm{eq,abs}}} = 0 \, ,
\end{align}
and the corresponding absorption timescale is
\begin{align}\label{eq:tauabs}
  \tau_{\mathrm{abs}} & = \frac{1}{Y_n\lambda_{\nu_e}+Y_p\lambda_{\bar\nu_e}}  \, .
\end{align}
The electron fraction in absorption equilibrium, $\yeeqabs$, depends mainly (though not solely) on the number densities and mean energies of both neutrino species. Assuming a pure nucleon gas, $Y_e^{\mathrm{eq,abs}}$ is given by 
\begin{align}\label{eq:yeeqabs2}
  Y_e^{\mathrm{eq,abs}} & = \frac{\lambda_{\nu_e}}{\lambda_{\nu_e}+\lambda_{\bar\nu_e}} \, ,
\end{align}
which can further be approximated by
\begin{align}\label{eq:yeeqabs3}
   Y_e^{\mathrm{eq,abs}} & \sim \left(1+\frac{\langle\eps_{\bar\nu_e}^2\rangle n_{\bar\nu_e}}{\langle\eps_{\nu_e}^2\rangle n_{\nu_e}} \right)^{-1} \, , \nonumber\\
    & \sim \left(1+\frac{\langle\eps_{\bar\nu_e}^2\rangle L_{N,\bar\nu_e}}{\langle\eps_{\nu_e}^2\rangle L_{N,\nu_e}} \right)^{-1} \, ,
\end{align}
where $n_\nu$ and $L_{N,\nu}$ are number densities and number fluxes (or number luminosities), respectively, for neutrino species $\nu$ and the energy averages are given by $\langle \eps_\nu^2 \rangle = (\int \eps^4 F_\nu\mathrm{d}\eps)/(\int \eps^2 F_\nu\mathrm{d}\eps)$. Approximate expressions similar to those given in Eq.~(\ref{eq:yeeqabs3}) have been employed for the purpose of investigating neutrino-driven winds in numerous studies \citep[e.g.][]{Qian1996, Horowitz1999a}. The estimate in Eq.~(\ref{eq:yeeqabs3}) neglects mass corrections (i.e. $Q_{np}=m_e=0$) and ignores Pauli blocking for $e^\pm$, while the second line additionally assumes that $L_{N,\nu_e}/L_{N,\bar\nu_e} \approx n_{\nu_e}/n_{\bar\nu_e}$. In this paper we always use Eq.~(\ref{eq:yeeqabs}) for the computation of $\yeeqabs$, because all of the aforementioned assumptions are not entirely justified in the bulk of the torus. During the neutrino-dominated phase the emitted neutrino energies are relatively low (cf. Sect.~\ref{sec:how-sens-y_em} and panel (f) of Fig.~\ref{fig:yeeq1}), electrons are mildly degenerate (cf. panel (a) of Fig.~\ref{fig:yeeq1}), and $n_{\nu_e}>n_{\bar\nu_e}$ may hold while at the same time $L_{N,\nu_e}<L_{N,\bar\nu_e}$ (e.g. \citealp{Wu2017a}).

In regions surrounding neutrino sources that are approximately in emission equilibrium (meaning
that $\dd Y_e/\dd t\approx 0$, i.e. the emission timescales are short compared to other timescales)
roughly the same number of $\nu_e$ and $\bar\nu_e$ neutrinos are emitted per unit of time, such that $\yeeqabs$ is typically close to 0.5.

\subsubsection{Thermodynamic equilibrium}\label{sec:thermal-equilibrium}

Finally, in the limiting case that the neutrino mean free paths become shorter than the hydrodynamic length scales, neutrinos become trapped and thermalized by the fluid and attain a Fermi-Dirac distribution that is defined solely by $\mu_\nu$ and $T$. In the extreme case that no neutrinos diffuse out from local fluid patches (typically for densities above $\rho\sim 10^{12}\,$g\,cm$^{-3}$), the total lepton fraction $Y_{l} = Y_e + Y_{\nu_e} - Y_{\bar\nu_e}$ (where $Y_i=n_i/n_B$) remains conserved (i.e. $\dd Y_l/\dd t=0$ along fluid trajectories) and $Y_e$ becomes an instantaneous function of $\rho, T$, and $Y_{l}$ \citep[see, e.g.,][for schemes making use of the concept of trapped neutrinos]{Sekiguchi2012a,Perego2016a,Ardevol-Pulpillo2019a}. In neutrino-cooled disks with maximum densities of only $\rho\sim 10^{10\ldots 12}\,$g\,cm$^{-3}$ neutrinos will, if at all, barely reach local thermodynamic equilibrium. However, the neutrino distribution may still be close to thermal, but with vanishing chemical potential, $\mu_\nu\rightarrow 0$, because the leakage of neutrinos drives the number densities down to $n_{\nu_e}-n_{\bar\nu_e}\ll n_{e^-} - n_{e^+}$ \citep[see, e.g.,][]{Ruffert1997, Beloborodov2003}. The condition
\begin{align}\label{eq:yeeqmu}
  \mu_\nu\big|_{\rho,T,Y_e^{\mathrm{eq,\mu_\nu=0}}} = 0
\end{align}
defines another\footnote{The apparent tension with the literature of cold neutron stars \citep[e.g.][]{Yakovlev2001a}, where no distinction is being made between $\yeeqem$ and $\yeeqmu$, can be resolved by realizing that both quantities become identical in the zero-temperature limit.} equilibrium value, namely $\yeeqmu$. Contours of $\yeeqmu(\rho,T)$ are shown in panel (b) of Fig.~\ref{fig:yeeq1} (red lines), revealing that $\yeeqmu$ exceeds $\yeeqem$ by about $\sim 0.1$ in the relevant regions. Given that neutrino-cooled disks provide conditions mainly in the transition region between the optically thin and optically thick regime, $\yeeqmu$ can thus be used as a quantity to roughly estimate (or bracket, together with $\yeeqem$ representing the opposite limiting case) the impact of neutrino absorption. This interpretation is supported by our simulations (cf. Sect.~\ref{sec:aver-therm-prop}), where we find that $\yeeqem<\yeeq\sim Y_e < \yeeqmu$ in the torus at early times during which neutrino absorption is efficient.

\subsection{Sensitivity of $Y_e^{\mathrm{eq,em}}$ to commonly employed approximations}\label{sec:how-sens-y_em}

Since $\yeeqem$ is a proxy for the $Y_e$ attained in the torus, one can assess the impact of certain modeling approximation on nucleosynthesis conditions in the ejecta without performing any simulations simply by checking their influence on $\yeeqem$.

The first assumption to test is that of neglecting corrections to $\lambda_{e^\pm}$ of Eq.~(\ref{eq:betarates}) associated with the finite electron mass and the neutron-proton mass difference by setting $Q_{np}=m_e=0$. Testing this sensitivity is motivated by the fact that many existing disk and merger models based on grey neutrino leakage schemes employed this approximation (as originally suggested by \citealp{Ruffert1996a}) in order to reduce the complexity of the integrals and therefore the computational demands. As the purple lines in panel~(c) of Fig.~\ref{fig:yeeq1} show, this simplification reduces $\yeeqem$ quite considerably compared to its original value, namely by about $0.05-0.1$ in the region $\rho\sim 10^{8\ldots 10}\,$\gcm and $T\sim 1\ldots 3\,$MeV.

Another correction to the emission rates that is worth checking is that associated with weak magnetism and nucleon recoil, by including the corresponding correction factors presented in \citet{Horowitz2002a} in the integrands of $\lambda_{e^\pm}$. This correction has been studied so far only in the context of neutrino-driven winds, where it was found to be responsible for increasing $Y_e$ in the absorption-dominated regime by about $\sim 10-20\,\%$ \citep{Horowitz1999a, Pllumbi2015a, Goriely2015}. In our case we are instead interested in the emission equilibrium and find that weak magnetism corrections shift $\yeeqem$ towards lower values (because it reduces the absorption cross section of positrons), but only by a very small amount in the regions relevant to neutrino-cooled disks (cf. red lines in panel~(c) of Fig.~\ref{fig:yeeq1}). Opposite to the previously discussed corrections associated with $Q_{np}$ and $m_e$, the impact of the weak-magnetism correction grows with temperature and therefore with the mean energy of emitted neutrinos, which is plotted for $\nu_e$ and $\bar\nu_e$ assuming emission equilibrium in panel~(f) of Fig.~\ref{fig:yeeq1}.

Since we are about to perform energy-dependent neutrino transport simulations using a limited number of energy zones, we should ensure that the transport energy grid is able to reproduce the values of $\yeeqem$ as accurately as possible. While in all other panels quantities related to $\yeeqem$ have been obtained using a very fine energy grid for the integrals appearing in Eq.~(\ref{eq:betarates}), in panel~(d) of Fig.~\ref{fig:yeeq1} we confirm that our transport grid consisting of 20 energy bins (see Sect.~\ref{sec:numerical-setup} for more details of the grid) is able to reproduce very well the quantities plotted in panel~(c).

The last question we address is whether the equilibrium $Y_e$ in BH-tori is sensitive to the ensemble of nuclear species taken into account in the EOS. In panel~(e) of Fig.~\ref{fig:yeeq1}, we therefore compare $\yeeqem$ resulting for a pure nucleon gas (purple lines), for the SFHO EOS (red lines, \citealp{Steiner2013}) that contains a large number of nuclear species, and a 4-species EOS ($n,p,\alpha$ plus one heavy nucleus, color map in Fig.~\ref{fig:yeeq1}). The 4-species EOS, used in all remaining panels of Fig.~\ref{fig:yeeq1}, will be employed also for the numerical models in the remainder of this study. Figure~\ref{fig:yeeq1} reveals that noticeable differences between the three cases of NSE ensembles appear only for emission timescales $\tauem \ga 1-10\,$s, and in particular between SFHO and our 4-species EOS only for $\tauem > 10\,$s. Since the freeze out of $Y_e$ is expected to occur already when $\tauem \la 1$\,s, we conclude that our 4-species EOS (as probably also the 3-species EOSs used by, e.g., \citealp{Siegel2018c}) should yield sufficiently accurate results for the freeze-out value of $\yeeqem$.

%%%%%%%%%%%%%%%%%%%%%%%%%%%%%%%%%%%%%%%%%%%%%%%%%%%%%%%%%%%%%%%%%%%%%%%%%%%%%%%%%%
\section{Setup of our study}\label{sec:setup-numer-models}

After discussing basic aspects of the $Y_e$ evolution, we now describe the setup of our numerical study of neutrino-cooled BH-tori.

\subsection{Hydrodynamic simulations}

\subsubsection{Investigated models}\label{sec:motiv-invest-models}

\newlength{\mytabcolsep}
\setlength{\mytabcolsep}{\tabcolsep}
\setlength{\tabcolsep}{4pt}
\begin{table*}
  \centering
  \caption{Properties of investigated numerical models. The columns provide from left to right: model name, initial torus mass, BH mass, BH spin, viscous $\alpha$-parameter, the treatment of turbulent angular momentum transport (where std. $\alpha$-vis. or $\lturb$=const. vis. means that Eq.~(\ref{eq:viseta1}) or Eq.~(\ref{eq:viseta2}) is used,  respectively), inclusion of $Q_{np}$ and $m_e$ in $\beta$-interaction rates, inclusion of weak magnetism corrections, initial electron fraction, final simulation time, percentage of initial torus mass left on the grid at final simulation time, and dimensionality of the simulation. The suffix ``-no$\nu$'' in parenthesis denotes that a separate model was evolved without neutrino absorption but otherwise similar properties. Entries not in (in) parenthesis refer to models evolved with (without) neutrino absorption.}
  \label{table_models}
  \begin{center}
  \begin{tabularx}{\textwidth}{lccccccccccc}
    %%% in Emacs use M-x align-current to align
    \hline
    model                        & $m_{\mathrm{tor}}^0$ & $M_{\mathrm{BH}}$ & $A_{\mathrm{BH}}$ & $Y_e$($t$=0) & $\alpha_{\mathrm{vis}}$ & viscosity           & mass corr. & weak magn. & $t^{\mathrm{fin}}$ & $m_{\mathrm{tor}}^{\mathrm{fin}}/m_{\mathrm{tor}}^0$ & dimensions \\
    name                         & [$M_\odot$]          & [$M_\odot$]       &                   &              &                         & treatment            & included?  & included?  & [s]                & [$\%$]                                               &            \\
    \hline                                                                                                                                                                                                                                                                                   
    m01M3A8(-no$\nu$)            & 0.01                 & 3                 & 0.8               & 0.5          & 0.06                    & std. $\alpha$-vis.   & yes        & no         & 10 \,(10)          & $<$1  \,($<$1)                                       & 2D \\
    m1M3A8(-no$\nu$)             & 0.1                  &                   &                   &              &                         &                      &            &            & 10 \,(10)          & $<$1  \,($<$1)                                       & \\
    m001M3A8(-no$\nu$)           & 0.001                &                   &                   &              &                         &                      &            &            & 10 \,(10)          & $<$1  \,($<$1)                                       & \\
    m01M5A8(-no$\nu$)            & 0.01                 & 5                 &                   &              &                         &                      &            &            & 10 \,(10)          & $<$1  \,($<$1)                                       & \\
    m01M10A8(-no$\nu$)           &                      & 10                &                   &              &                         &                      &            &            & 20 \,(20)          & $<$1  \,($<$1)                                       & \\
    m01M3A4(-no$\nu$)            &                      & 3                 & 0.4               &              &                         &                      &            &            & 10 \,(10)          & $<$1  \,($<$1)                                       & \\
    m01M3A9(-no$\nu$)            &                      &                   & 0.9               &              &                         &                      &            &            & 10 \,(10)          & $<$1  \,($<$1)                                       & \\
    m01M3A8-$\alpha$03(-no$\nu$) &                      &                   & 0.8               &              & 0.03                    &                      &            &            & 10 \,(10)          & $<$1  \,($<$1)                                       & \\
    m01M3A8-$\alpha$1(-no$\nu$)  &                      &                   &                   &              & 0.1                     &                      &            &            & 10 \,(10)          & $<$1  \,($<$1)                                       & \\
    m01M3A8-vis2(-no$\nu$)       &                      &                   &                   &              & 0.05                    & $\lturb$=const. vis. &            &            & 20 \,(20)          & 5.7   \,(6.81)                                       & \\
    m01M3A8-mhd(-no$\nu$)        &                      &                   &                   &              & --                      & MHD                  &            &            & 2.1\,(2.1)         & 12.5  \,(15.0)                                       & 3D \\
    m01M3A8-noQm(-no$\nu$)       &                      &                   &                   &              & 0.06                    & std. $\alpha$-vis.   & no         &            & 10 \,(10)          & $<$1  \,($<$1)                                       & 2D \\
    m01M3A8-wm                   &                      &                   &                   &              &                         &                      & yes        & yes        & 10                 & $<$1                                                 & \\
    m01M3A8-ye01(-no$\nu$)       &                      &                   &                   & 0.1          &                         &                      &            & no         & 10 \,(10)          & $<$1    \,($<$1)                                     & \\
    \hline
  \end{tabularx}
\end{center}
\end{table*}

We simulate BH-accretion disks with the code AENUS-ALCAR \citep{Obergaulinger2008a, Just2015b} that handles neutrino transport in the M1 approximation and solves the (viscous or magneto-) hydrodynamic and transport equations on a spherical polar grid using Riemann-solver based finite-volume methods. The numerical methods for all viscous models (cf. Sect.~\ref{sec:init-cond-angul}) are exactly the same as those in \citet{Just2015a} unless stated explicitly otherwise. We summarize the evolution equations in Appendix~\ref{sec:evolv-equat-hydr}. For neutrino interactions, we include the $\beta$-processes (with rates given by $\lambda_i$ of Eqs.~(\ref{eq:betarates})) as well as iso-energetic scattering with nucleons and nuclei \citep{Bruenn1985}. Justified by the relatively low temperatures and densities encountered in neutrino-cooled disks and by the dominance of the included processes under such conditions, we neglect all other types of neutrino interactions, such as channels that produce $\mu$- and $\tau$-neutrinos. We do not take into account general relativistic effects, except that we use the pseudo-Newtonian gravitational potential by \citet{Artemova1996} to approximately incorporate effects associated with the innermost stable circular orbit (ISCO) and to mimic the shrinking of the ISCO with increasing BH spin.

One of the main goals of this study is to characterize the impact of neutrino absorption and quantify its importance relative to neutrino emission in determining $Y_e$ in the torus and the ejected material. To this end we conduct for most models two simulations, one with full M1 transport (i.e. including emission and absorption) and one neglecting neutrino absorption as well as any other neutrino reactions except neutrino emission. In doing so we aim at bracketing the maximum error that could be encountered when neglecting absorption or when treating absorption with more simplistic neutrino schemes. Although the M1 scheme used in this study is not as accurate as a Boltzmann scheme that solves the six-dimensional neutrino transfer equations, it is an actual transport scheme -- in the sense that it solves time-dependent conservation equations for the neutrino energy density and flux density -- and is consistent with the transfer equations in the diffusion and free streaming limits. The M1 scheme has shown excellent agreement with a ray-by-ray\footnote{The ray-by-ray approximation assumes that the radiation field is axisymmetric around each radial coordinate line in a spherical polar coordinate system. This approximation was found to be well justified in 3D CCSN simulations \citep{Glas2019a}, whereas in axisymmetry \citep{Skinner2016, Just2018a} it may artificially promote shock runaway.} Boltzmann solver (VERTEX, \citealp{Rampp2002, Buras2006}) for describing core-collapse supernovae \citep{Just2018a}. Its accuracy in the context of merger remnants has only been tested for time independent configurations \citep{Just2015a, Foucart2018b, Sumiyoshi2021a} or short-term simulations \citep{Foucart2020a}, where no serious shortcomings were found.

Table~\ref{table_models} summarizes the set of investigated models. Our fiducial reference model, m01M3A8, is chosen to have a BH mass of $M_{\mathrm{BH}}=3\,\Msol$, BH spin parameter of $A_{\mathrm{BH}}=0.8$, and a relatively low initial torus mass, $m^0_{\mathrm{tor}}=0.01\,\Msol$. By using such a low torus mass our results will bracket the impact of absorption from below, i.e. all disks with larger masses but otherwise same parameters can be expected to show an even stronger dependence on absorption. Moreover, another reason for exploring the low-mass regime is that low-mass tori may represent, at least approximately, BH-disks formed in the core of collapsars \citep{Siegel2019b, Miller2020s}.

In order to test the sensitivity of the results with respect to global parameters and modeling assumptions, we vary the initial disk mass (indicated by number following ``m'' in the model name), BH mass and spin (``M'' and ``A'' in model name, respectively), and the treatment of turbulent angular momentum transport (as described in Sect.~\ref{sec:init-cond-angul}). Motivated by the discussion in Sect.~\ref{sec:how-sens-y_em}, we also test (model m01M3A8-noQM) the impact of neglecting mass corrections, i.e. setting $Q_{np}=m_e=0$ in the $\beta$-processes, Eqs.~(\ref{eq:betarates}), and of including weak magnetism and recoil corrections\footnote{The reason why our fiducial model does not already include weak magnetism corrections is not related to any physics argument. After including weak-magnetism corrections at a later stage of this project, we deemed it unnecessary to repeat a large number of simulations that had already been performed without weak magnetism corrections, in particular considering that the quantitative impact of this correction is rather small.} (model m01M3A8-wm).

We set up the initial torus as equilibrium configuration with constant specific angular momentum (similarly as in \citealp{Just2015a, Fernandez2013b, Fujibayashi2020a, Siegel2019b}). We fix the initial specific entropy at $8\,k_B$ per baryon, put the location of the initial density maximum at a radius of $40\,\mathrm{km}\times (M_{\mathrm{BH}}/(3\,\Msol))$, and set the initial electron fraction, $Y_e^0$, to a constant value everywhere, namely to $Y_e^0=0.5$ for all except one model, for which we use $Y_e^0=0.1$ in order to test the sensitivity. For NS-BH mergers and probably for most NS-NS configurations, the value of 0.5 is an overestimation, while it is a more realistic choice for collapsar engines. However, we use it here deliberately for most models in order to erase any possible contribution from the initial condition in producing r-process material and to examine which and how many r-process elements can be generated self-consistently only by the disk. Thus, the final r-process abundances resulting from models with $Y_e^0=0.5$ can in that respect be considered as lower limits if the disk models are interpreted as remnants of compact-object mergers.

\subsubsection{Angular momentum transport}\label{sec:init-cond-angul}

Most of our models are run in axisymmetry and employ the well-known $\alpha$-viscosity scheme \citep{Shakura1973} to mimic turbulent angular momentum transport. We employ for the dynamic viscosity coefficient the prescription
\begin{align}\label{eq:viseta1}
  \eta_{\mathrm{vis}} = \alpvis \rho c_i \frac{c_i}{\Omega_K}
\end{align}
(where the isothermal sound speed $c_i=\sqrt{P/\rho}$ with the gas pressure $P$ and the Keplerian angular velocity $\Omega_K$), which we call the ``standard $\alpha$-viscosity'' prescription here, because it was already used in a large number of previous disk studies \citep[e.g.][]{Narayan1994, Igumenshchev2000, Setiawan2004, Fernandez2013b, Just2015a}. We only take into account the $r\phi$- and $\theta\phi$-components of the viscous stress tensor, because we only intend to parametrize the turbulent stresses related to differential rotation. In most cases we use $\alpvis=0.06$, while we also added models with lower and higher values of $\alpvis$ (cf. models with suffix $\alpha03$ and $\alpha1$, respectively, in Table~\ref{table_models}).

Since the $\alpha$-viscosity scheme is nothing but a mean-field parametrization of unresolved small-scale physics based on heuristic dimensional arguments, one is of course left with certain freedom for defining $\eta_{\mathrm{vis}}$. \citet{Fujibayashi2020a} made use of this liberty and argued that the characteristic length scale $\lturb\sim c_i/\Omega_K$ implicitly assumed in the prescription of Eq.~(\ref{eq:viseta1}) may be overestimated at large radii. They instead suggested a different formulation of the $\alpha$-viscosity for which $\lturb$=const. and
\begin{align}\label{eq:viseta2}
  \eta_{\mathrm{vis}} = \alpvis \rho c_s \lturb
\end{align}
(with the adiabatic sound speed $c_s=\sqrt{\gamma P/\rho}$). In order to test the sensitivity with respect to the viscosity prescription we implemented this treatment in model m01M3A8-vis2 and its pendant neglecting neutrino absorption, using $\lturb=9\,$km together with $\alpvis=0.05$ (i.e. the same values as used in \citealp{Fujibayashi2020a}).

The third, and most consistent option of treating angular momentum transport is to evolve MHD models, starting with a pre-defined magnetic field configuration; cf. models m01M3A8-mhd(-no$\nu$). Moreover, we also switch from a Newtonian axisymmetric description to a special relativistic three-dimensional (3D) evolution (see Appendix~\ref{sec:evolv-equat-hydr} for the evolved equations), which is necessary to overcome time-step limitations (due to unlimited Alfven speeds in Newtonian MHD) and the anti-dynamo theorem \citep{Moffatt1978}, respectively. The initial torus in the MHD models has the same properties as for the viscous models, but is now endowed with a single poloidal magnetic-field loop, of which the magnetic field strength is chosen such that the ratio of mass-weighted average of gas pressure to mass-weighted average of magnetic pressure is $\beta_{\mathrm{init}}\approx 200$. In order to distribute the initial magnetic field as uniformly as possible in the torus, the magnetic-field configuration is constructed in the same fashion (apart from a higher value of  $\beta_{\mathrm{init}}$ in our case) as described in \citet{Fernandez2019b}. We note that the initial magnetic field configuration can have a non-negligible leverage on the torus evolution and ejecta properties \citep{Beckwith2008,Christie2019a}, while additionally also the demand on the grid resolution is higher in MHD models than in viscous disks. However, given the huge computational expenses to perform the special relativistic 3D MHD simulations, we were unable to address these aspects in the current study. For the same reason, we could only evolve these models until $\sim 2\,$s (compared to evolution times of $\geq 10\,$s for the viscous models). Nevertheless, to our knowledge the MHD simulations presented here are the longest performed so far using genuine neutrino transport (i.e. solving time-dependent conservation equations for neutrinos).

\subsubsection{Numerical setup}\label{sec:numerical-setup}

The numerical setup is as follows: For the viscous (MHD) models the radial domain is discretized between $r_{\mathrm{min}}$ and $2\times 10^4$\,km ($1\times 10^4\,$km) using a logarithmic grid with $N_r=400$ (320) zones, where $r_{\mathrm{min}}$ is approximately given by the arithmetic average between the ISCO radius and the radius of the event horizon (e.g. $r_{\mathrm{min}}\approx 10\,$km for $\MBH=3\,\Msol$ and $\ABH=0.8$). All viscous models are evolved with a polar grid of 160 zones distributed uniformly between $\theta=0$ and $\pi$. The 3D MHD models make use of 128 polar zones, of which 126 zones are distributed non-uniformly between $\pi/7<\theta<6\pi/7$ in the form suggested by \citet{Sc-adowski2015a} with a cell width increasing from $\Delta\theta \approx 0.43^\circ$ at the equator to $\approx 4^\circ$ close to the pole, and the two remaining zones attach both ends of the grid to the polar axis. The azimuthal coordinate, $\phi$, in the MHD models runs from 0 to $\pi/2$ with 32 uniform zones, while we assume $\pi/2$-periodicity to fill the remaining quadrants. This assumption, which is not uncommon in the existing literature \citep[e.g.][]{DeVilliers2003b} and was necessary in order to keep our simulations computationally feasible, precludes a consistent evolution of the lowest order modes in azimuthal direction, which may contribute to turbulent angular momentum transport, though probably not on a dominant level. The energy space for neutrinos is discretized logarithmically between 0 and 80\,MeV by 20 (12) zones for the viscous (MHD) models. In order to prevent ill-defined zero-density regions, we apply a floor value for the density that is constant at radii lower than $1000\,$km (between 1-100\,g\,cm$^{-3}$ depending on the model) and beyond that radius decreases roughly as $r^{-3}$. For the MHD models we additionally enforce the condition that the magnetic energy should never be larger than 50 times the rest-mass energy in a given cell, where both energies are measured in the comoving frame.

\subsection{Outflow trajectories and nucleosynthesis post-processing}\label{sec:outfl-traj-nucl}

In order to analyze the outflow and obtain the nucleosynthesis yields we extract fluid trajectories from the hydrodynamical simulations and post-process the trajectories in a separate step. We consider as ejecta all material beyond a radius of $r=10^4\,$km, i.e. we do not apply a criterion to check whether material is gravitationally unbound. In all models performed in axisymmetry and employing a viscosity scheme, we use the available output data from the simulation to numerically integrate particle trajectories backward in time starting at a radius of $r=10^9\,$cm and at pre-defined times and polar angles for a total number of about 600-1100 trajectories per model. For the 3D MHD models we instead distribute $5\times 10^4$ tracer particles of equal mass in the initial disk and advect their locations during the simulation while recording their thermodynamic properties. As a cross check, we verified that our finite set of representative trajectories yields very similar mass distributions versus electron fraction, entropy, or velocity at a fixed radius as those resulting when integrating the mass fluxes at a fixed radius using all output data written throughout the simulation. The agreement suggests that statistical errors due to insufficient sampling of the ejecta is not a major source of uncertainty.

The r-process nucleosynthesis is calculated by post-processing the thus obtained trajectories of the ejected matter. The density and temperature evolution is taken directly from the hydrodynamical trajectories. The initial composition is determined by nuclear statistical equilibrium when the density has dropped below the drip density  and matter has cooled below $T=10^{10}$~K during the expansion. As soon as the temperature has fallen below 10~GK, further changes of the composition are followed by a full network calculation including all 5000 nuclear species from protons up to $Z=110$ that lie between the valley of $\beta$-stability and the neutron-drip line  \cite{Goriely2015a}. All charged-particle fusion reactions as well as their reverse reactions on light elements up to Th isotopes are included in addition to radiative neutron captures and photodisintegrations on all species up to $Z=110$ isotopes. The reaction rates on light species are taken from the NETGEN library, which includes all the latest compilations of experimentally determined reaction rates \citep{Xu2013}. Experimentally unknown reactions are estimated with the TALYS code \citep{Koning2005,Goriely2008} on the basis of the Skyrme Hartree-Fock-Bogolyubov (HFB) nuclear mass model, HFB-21 \citep{Goriely2010}. On top of these reactions, $\beta$-decays as well as $\beta$-delayed neutron emission probabilities are also included, the corresponding rates being taken from the relativistic mean-field plus ramdom-phase-approximation calculation \citep{Marketin2016a}.

All fission rates, i.e. the neutron-induced, photo-induced, $\beta$-delayed and spontaneous fission rates, are estimated on the basis of the HFB-14 fission paths \citep{Goriely2007, Goriely2015a} and the nuclear level densities within the combinatorial approach \citep{Goriely2008} are obtained with the same single-particle scheme and pairing strength. The neutron-induced fission rates are calculated with the TALYS code for all nuclei  with $90 \le Z \le 110$ \citep{Goriely2009}. Similarly, the $\beta$-delayed and spontaneous fission rates are estimated with the same TALYS fission barrier penetration calculation. The $\beta$-delayed fission rate takes into account the full competition
between the fission, neutron and photon channels, weighted by the population probability given by the $\beta$-decay strength function \citep{Kodama1975}. The fission fragment yield distribution is estimated with the renewed statistical scission-point model based on microscopic ingredients, the so-called SPY model, as described in \citet{Lemaitre2019a}.

\subsection{Estimation of kilonova light curves}\label{sec:estim-kilon-lightc}

The possibility to observe the torus ejecta and r-process yields directly by means of the emitted kilonova opens up prospects of quantitatively determining the global parameters of the merging binary, e.g. the post-merger torus mass, its composition etc. However, many dependencies still remain unexplored that establish the link between the input physics of hydrodynamic models and the resulting kilonova produced by the ejected material. In order to explore some basic sensitivities we employ a simple and approximate kilonova computation scheme to translate the variations of $Y_e$ and of nuclear mass fractions found for our hydrodynamic models into variations of observable features, such as the time, photospheric temperature, and bolometric luminosity at peak emission.

Depending on the intended level of consistency, various more and less sophisticated approaches have previously been used for computing the kilonova light curve based on the ejecta properties (e.g. \citealp{Li1998, Kulkarni2005, Metzger2010c, Grossman2014, Kasen2015, Perego2017a, Kawaguchi2018a, Metzger2019a,  Hotokezaka2020a, Korobkin2020a}). In our study we employ an approximate and to-our-knowledge new method that makes direct use of the mass and velocity of each outflow trajectory as well as the corresponding nucleosynthesis data (lanthanide plus actinide fraction, $\XLA \equiv \Xlan+\Xact$, and radioactive heating rate per mass, $q_{\mathrm{rad}}$). While not being as accurate as sophisticated radiative transfer schemes, our method has the advantage over several existing approximate schemes that it captures the detailed dependence of mass density, heating rate, and $\XLA$ on the velocity coordinate, $v$, instead of assuming idealized functions. In the following we outline the method only briefly, while explicit equations are provided in Appendix~\ref{sec:meth-comp-kilon}. We assume spherical symmetry, homology (i.e. $r\propto v$), and employ the grey approximation for the light curve, such that the only coordinate degrees of freedom are the expansion time and ejecta velocity. The velocity space is discretized by about $N_v\sim 100$ zones and each zone is filled with ejecta particles based on their velocity at the extraction radius of $10^4\,$km. A mass-weighted integration of the corresponding quantities then yields $m(v), \XLA(v,t),$ and~$\qheat(v,t)$, where $m(v)$ equals the ejecta mass with velocity greater than $v$ and $\qheat\equiv \epsilon_{\mathrm{therm}} q_{\mathrm{rad}}$ is the effective heating rate including the thermalization factor $\epsilon_{\mathrm{therm}}$ that is obtained from interpolation of values provided for case ``Random'' in Table~1 of \citet{Barnes2016a}. We finally need opacities, $\kappa$, in order to compute a bolometric light curve from these data. Lacking opacity data based on detailed atomic models \citep[e.g.][]{Kasen2013, Tanaka2020a}, we approximate $\kappa$ as a function of $\XLA$ and $T$, which is calibrated by fitting kilonova light curves from \citet{Kasen2017a}; see Appendix~\ref{sec:meth-comp-kilon} for more details. Although this simplified prescription is a crude approximation to atomic opacities, it captures the main effect that we are interested in, namely the correlation between $\kappa$ and the lanthanide content of the ejecta. After preparation of the data in the described manner, we solve a two-moment system of conservation equations for the energy density and flux density of photons using the M1 closure.

%%%%%%%%%%%%%%%%%%%%%%%%%%%%%%%%%%%%%%%%%%%%%%%%%%%%%%%%%%%%%%%%%%%%%%%%%%%%%%%%%%
\section{Results}\label{sec:results}

\setlength{\tabcolsep}{4mm}
\begin{table*}
  %% \centering
  \caption{Properties characterizing weak interactions in the disk: Minimum of average $Y_e$, maximum optical depth at a neutrino energy of $20\,$MeV along equator, estimated times when emission ($t_{\mathrm{em}}$) and absorption ($t_{\mathrm{abs}}$) become irrelevant, mass accretion rates onto the BH at times $t_{\mathrm{em}}$ and $t_{\mathrm{abs}}$, percentage of accreted rest-mass energy radiated in the form of neutrinos until $t^{\mathrm{fin}}$, mean energy of radiated electron neutrinos ($\nu_e$) and electron antineutrinos ($\bar\nu_e$) computed as $\overline{\langle\epsilon\rangle}_{\mathrm{rel}}^{\nu}=L_\nu/L_{\nu,N}$ where $L_\nu$ and $L_{\nu,N}$ are measured at $r=500\,$km in the laboratory frame, mean energy of all absorbed neutrinos (i.e. neutrinos captured by protons or neutrons) defined by Eq.~(\ref{eq:epsabs}). Entries not in (in) parenthesis refer to models evolved with (without) neutrino absorption.}
  \label{table_weak}
  \begin{center}
  \begin{tabularx}{\textwidth}{lccccccc}
    %%% in Emacs use M-x align-current to align
    \hline
    model              & $\yemin$      & $\tau_{\mathrm{opt}}^{\mathrm{max}}(20\,\mathrm{MeV})$ & $t_{\mathrm{em/abs}}$ & $\dot{M}_{\mathrm{BH}}(t_{\mathrm{em/abs}})$ & $\bar\eta_\nu$  & $\overline{\langle\epsilon\rangle}_{\mathrm{rel}}^{\nu_e/\bar\nu_e}$ & $\overline{\langle\epsilon\rangle}_{\mathrm{abs}}^{\nu_e/\bar\nu_e}$ \\
    name               & w/ (w/o) abs. & w/ (w/o) abs.                                          & [ms]                  & [$10^{-2}M_\odot s^{-1}$])                   & [$\%$]          & [MeV]                                                                & [MeV]                                                                \\

 \hline                                                                                                             
    m01M3A8            & 0.227 (0.153) & 14.0 (11.1)                                            & 105/ 60               & 1.0/3.2                                      & 4.2             & 14.2/17.5                                                            & 23.6/28.1                                                            \\
    m1M3A8             & 0.247 (0.073) & 47.1 (106 )                                            & 175/135               & 5.2/8.6                                      & 5.2             & 13.7/17.1                                                            & 35.0/38.1                                                            \\
    m001M3A8           & 0.346 (0.336) & 1.44 (1.0 )                                            & 55/ 17                & 0.3/1.6                                      & 1.2             & 12.5/15.3                                                            & 18.9/23.5                                                            \\
    m01M5A8            & 0.261 (0.228) & 6.08 (7.10)                                            & 120/ 60               & 1.2/4.2                                      & 2.8             & 12.6/15.3                                                            & 19.3/23.7                                                            \\
    m01M10A8           & 0.370 (0.366) & 1.47 (1.50)                                            & 145/ 32               & 1.4/7.3                                      & 0.8             & 9.50/11.7                                                            & 14.0/17.9                                                            \\
    m01M3A4            & 0.240 (0.177) & 12.0 (17.1)                                            & 85/ 35                & 1.0/4.9                                      & 1.7             & 13.0/16.3                                                            & 21.6/26.4                                                            \\
    m01M3A9            & 0.224 (0.148) & 15.4 (23.4)                                            & 110/ 72               & 1.1/2.7                                      & 6.3             & 14.8/18.1                                                            & 24.9/29.4                                                            \\
    m01M3A8-$\alpha$03 & 0.176 (0.118) & 17.2 (11.9)                                            & 235/130               & 0.5/1.6                                      & 4.6             & 12.9/15.6                                                            & 20.7/25.0                                                            \\
    m01M3A8-$\alpha$1  & 0.263 (0.181) & 11.9 (19.0)                                            & 62/ 35                & 1.8/5.2                                      & 3.9             & 15.3/18.9                                                            & 26.3/31.0                                                            \\
    m01M3A8-vis2       & 0.223 (0.162) & 10.6 (15.2)                                            & 250/ 75               & 0.2/1.8                                      & 5.2             & 14.0/16.8                                                            & 22.9/28.9                                                            \\
    m01M3A8-mhd        & 0.150 (0.104) & 15.4 (20.3)                                            & 254/ 51               & 0.3/3.1                                      & 2.5             & 12.3/15.0                                                            & 19.3/26.0                                                            \\
    m01M3A8-noQm       & 0.202 (0.135) & 14.5 (11.1)                                            & 110/ 60               & 0.9/3.3                                      & 4.2             & 14.8/17.0                                                            & 23.9/27.9                                                            \\
    m01M3A8-wm         & 0.229         & 13.7                                                   & 105/ 60               & 0.9/3.1                                      & 4.3             & 14.3/17.6                                                            & 23.9/27.5                                                            \\
    m01M3A8-ye01       & 0.100 (0.088) & 14.5 (20.8)                                            & 105/ 62               & 1.0/3.1                                      & 4.0             & 14.8/17.4                                                            & 24.0/28.1                                                            \\
    \hline
  \end{tabularx}
\end{center}
\end{table*}

\begin{figure*}
  \centering
  \includegraphics[width=0.99\textwidth]{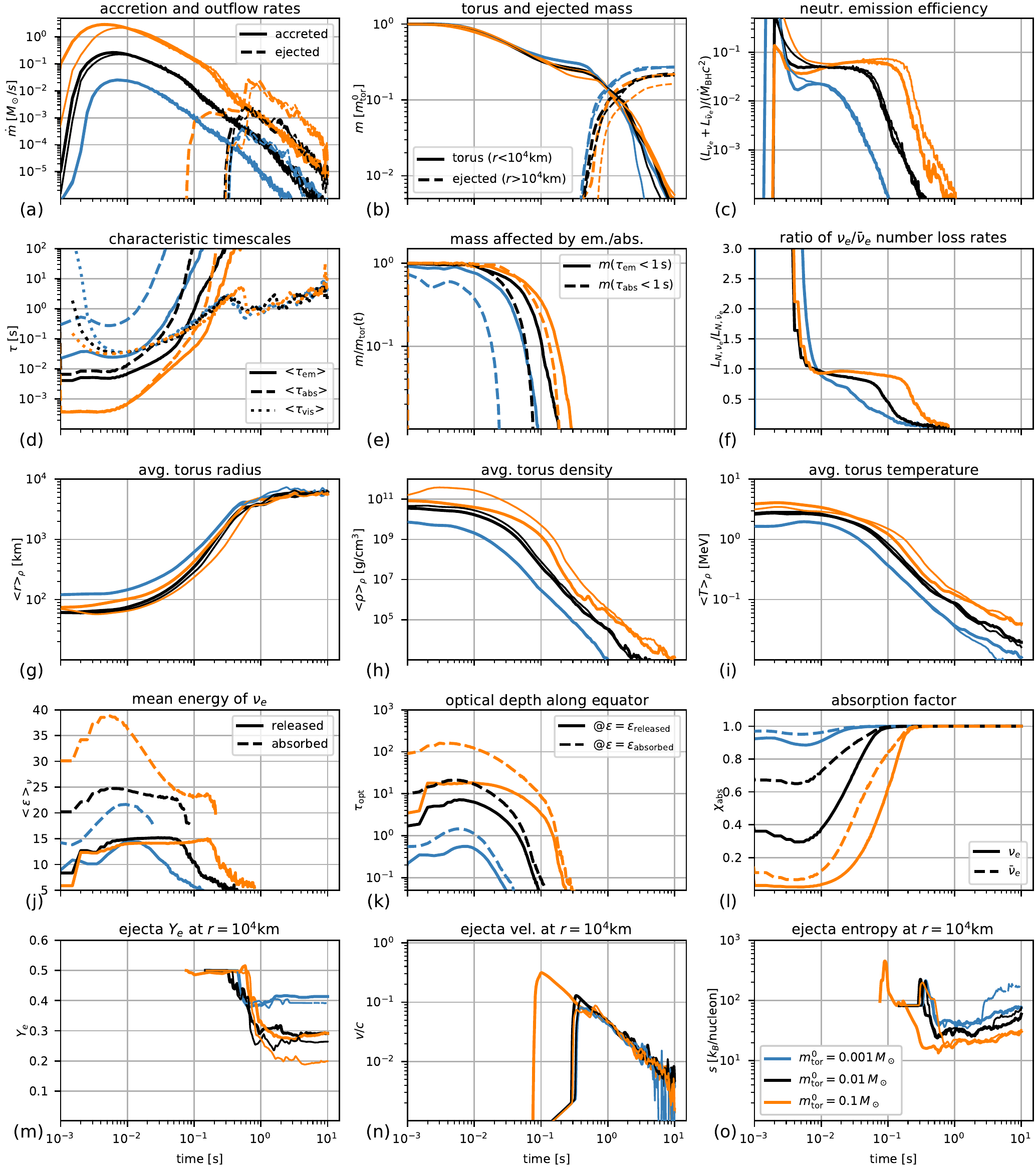}
  \caption{Time evolution of global properties for models m001M3A8 (blue lines), m01M3A8 (black lines), and m1M3A8 (orange lines). Thin lines of same color denote the corresponding ``no$\nu$'' models ignoring neutrino absorption. \emph{Panel~(a):} mass-accretion rates onto the BH and ejecta mass-flux rates at $r=10^4\,$km; \emph{(b):} mass of torus (i.e. at $r<10^4\,$km) and ejected material (i.e. at $r>10^4\,$km) ; \emph{(c):} total neutrino luminosity (measured at $r=500\,$km in the laboratory frame) divided by rest-mass energy accretion rate onto the BH; \emph{(d):} characteristic timescales of emission, Eq.~(\ref{eq:tauemtorus}), absorption, Eq.~(\ref{eq:tauabstorus}), and viscous expansion, Eq.~(\ref{eq:tauvistorus}); \emph{(e):} fraction of the torus mass for which $\tauem<1\,$s (solid lines) and $\tauabs<1\,$s (dashed lines); \emph{(f):} ratio of $\nu_e$ over $\bar\nu_e$ number fluxes measured at $r=500\,$km in the laboratory frame; \emph{(g),~(h),~(i):} mass-weighted averages (cf. Eq.~(\ref{eq:masswavg})) of radial coordinate, density, and temperature, respectively; \emph{(j):} mean energy of released $\nu_e$ as measured at $r=500\,$km (solid lines, cf. Eq.~(\ref{eq:epsrel})) and of absorbed $\nu_e$ (dashed lines, cf. Eq.~(\ref{eq:epsabs})); \emph{(k):} the optical depth along the equator for the energies shown in panel~(j); \emph{(l):} absorption factor, Eq.~(\ref{eq:absfac}); \emph{(m),~(n),~(o):} mass-weighted surface average of outflow material crossing the sphere at $r=10^4\,$km of $Y_e$, velocity, and entropy per baryon, respectively (computed as $\langle S \rangle = (\int \rho S \dd\Omega)/(\int\rho\dd\Omega)|_{r=10^4\,\mathrm{km}}$ for quantity $S$).} \label{fig:globdat_mtorus}
\end{figure*}

\begin{figure*}
  \centering
  \includegraphics[width=0.99\textwidth]{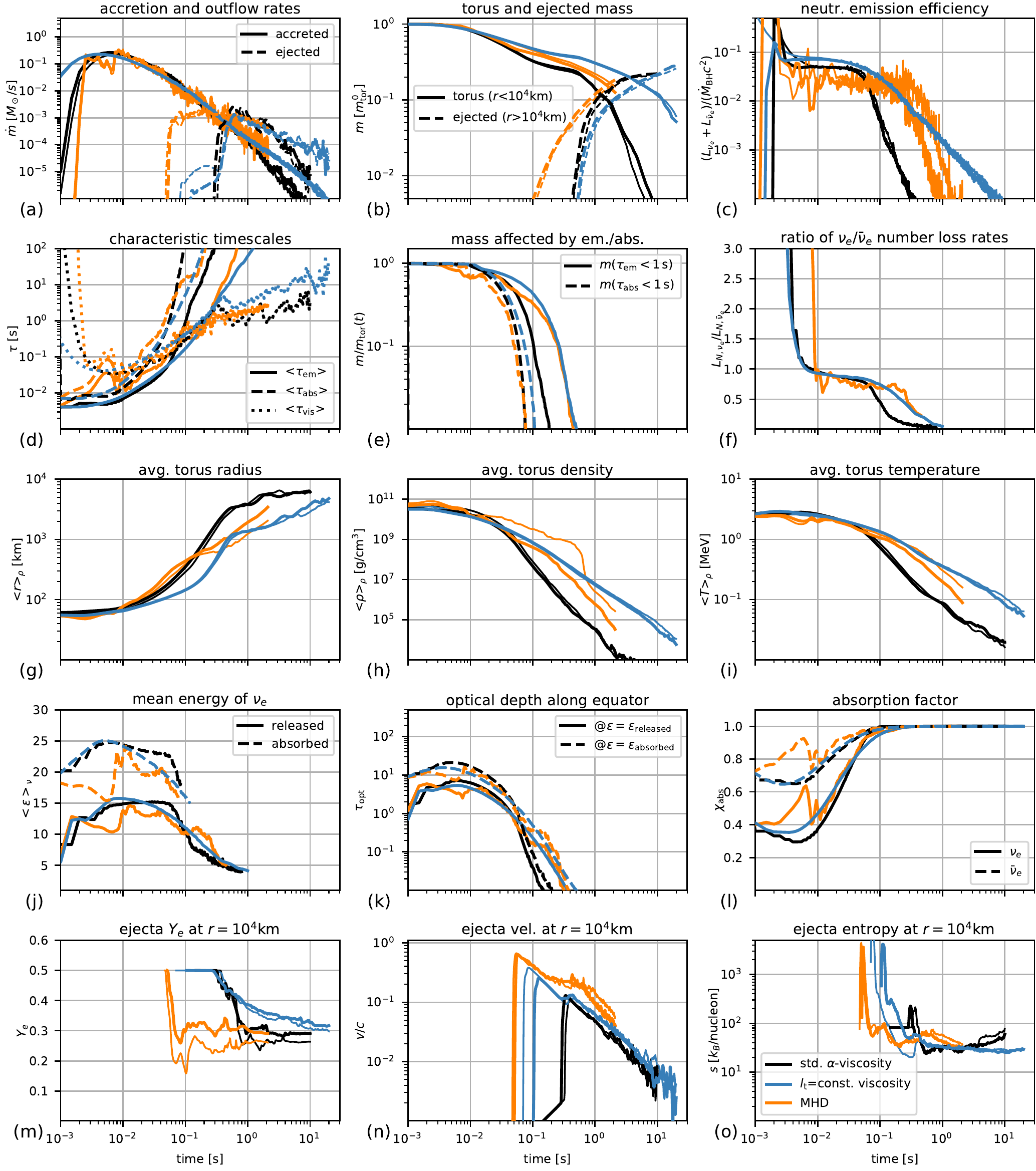}
  \caption{Same as Fig.~\ref{fig:globdat_mtorus} but for the series of models with different treatments of angular momentum transport, m01M3A8 (black lines), m01M3A8-vis2 (blue lines), and m01M3A8-mhd (orange lines).}
  \label{fig:globdat_vis}
\end{figure*}	

\subsection{Basic features}\label{sec:general-features}

\subsubsection{Neutrino emission and weak freeze out}\label{sec:neutr-emiss-weak}

We first review, mostly on the basis of the fiducial models m01M3A8(-no$\nu$), some generic features of the neutrino emission and the weak interaction freeze out. Although many of the following aspects are not entirely new (e.g. \citealp{Metzger2008c, Fernandez2013b, Just2015a}), we briefly summarize them here and introduce quantities, which will subsequently serve as diagnostic tools to assess the impact of neutrino absorption.

Figures~\ref{fig:globdat_mtorus} and~\ref{fig:globdat_vis} show several global quantities as functions of time, Fig.~\ref{fig:evolphases} schematically illustrates the conditions for weak interactions in the disk, and Figs.~\ref{fig:contours_m01}, \ref{fig:contours_m1}, and~\ref{fig:contours_m01_mhd} provide contours of density, $Y_e$, and temperature, as well as of neutrino emission timescales and absorption timescales for different times of evolution. Here and in the following we consider as the torus all material that is located between the inner radial boundary of the computational domain and the radius $r_1 \equiv 10^4\,$km. We measure luminosities of neutrino energy and number at $r=500\,$km in the laboratory frame. Mass-weighted averages of any quantity $X$ are computed as:
\begin{align}\label{eq:masswavg}
  \langle X\rangle_\rho = \frac{\int_{r<r_1}\rho X\mathrm{d}V} {\int_{r<r_1}\rho \mathrm{d}V} \, .
\end{align}

The torus evolution can roughly be divided into two phases \citep{Metzger2008c}: In the first, neutrino-dominated phase neutrino cooling is efficient in removing heat from viscous processes, i.e. the neutrino emission timescale, estimated as
\begin{align}\label{eq:tauemtorus}
  \langle\tau_{\mathrm{em}}\rangle = \frac{\int_{r<r_1}n_B\mathrm{d}V} {\int_{r<r_1}\dot{n}_{\mathrm{em}}\mathrm{d}V}
\end{align}
(where $\dot{n}_{\mathrm{em}} = \lambda_{e^-}n_p+ \lambda_{e^+}n_n$ with $\lambda_i$ from Eq.~(\ref{eq:betarates}) and proton-/neutron number densities $n_{p/n}$) is shorter than the viscous timescale, computed as
\begin{align}\label{eq:tauvistorus}
  \langle\tau_{\mathrm{vis}}\rangle = \frac{m_{\mathrm{tor}}}{\dot{m}_{\mathrm{tor}}}
\end{align}
(where $m_{\mathrm{tor}}$ is the total baryonic mass within the sphere of radius $r_1$); see panel~(d) in Fig.~\ref{fig:globdat_mtorus}. Moreover, during the neutrino-dominated phase fluid elements accreting onto the BH typically radiate away a sizable fraction of their rest-mass energy in the form of neutrinos, i.e.  $\eta_\nu=L_\nu/(\dot{M}_{\mathrm{BH}}c^2)\ga 1-5\,\%$ (cf. panel~(c) of Fig.~\ref{fig:globdat_mtorus}). In the subsequent, non-radiative phase, $\eta_\nu$ plunges and viscosity becomes a powerful agent of mass ejection \citep[e.g.][]{Fernandez2013b, Just2015a}.

The evolution of $Y_e$ as well as of the equilibrium values introduced in Sect.~\ref{sec:equil-cond-that}, mass averaged up to a radius of $10^4$\,km, is depicted in the top rows of Figs.~\ref{fig:torusye_time} (for models with different torus masses) and~\ref{fig:torusye_time_vis} (for models with different viscosity treatments). During the neutrino-dominated phase a large fraction of the torus is close to weak equilibrium, meaning that neutrino emission is efficient enough for $Y_e$ to roughly track its respective\footnote{Recall that $\yeeq=\yeeqem$ for models without neutrino absorption.} equilibrium value $\yeeq$. Therefore, in all models that start with an initial value of $Y_e=0.5$, the torus average, $\yeavg$ (black lines), first drops from its initial value until approximately reaching $\yeeqavg$ (green or red lines for models with or without neutrino absorption, respectively), and then rises again in an attempt to match $\yeeqavg$, which in turn increases owing to viscous expansion and a subsiding level of electron degeneracy in the torus. However, the decreasing rates of neutrino production (i.e. growing emission timescales $\tauem$) in the more and more diluting torus thwart and ultimately terminate\footnote{The late-time variations of $\yeavg$ visible in Figs.~\ref{fig:torusye_time}~and~\ref{fig:torusye_time_vis} are not caused by weak interactions but by material leaving the control volume.} the evolution of $Y_e$ in the torus.

Closely connected to the aforementioned features is the behavior of the ratio of $\nu_e$ to $\bar\nu_e$ number luminosities, $L_{N,\nu_e}/L_{N,\bar\nu_e}$, which is plotted in panel (f) of Fig.~\ref{fig:globdat_mtorus}. After a short initial phase of $L_{N,\nu_e}/L_{N,\bar\nu_e}\gg 1$, during which the torus with initially $Y_e=0.5$ deleptonizes towards its weak-equilibrium value, the ratio $L_{N,\nu_e}/L_{N,\bar\nu_e}$ remains fairly close to, but marginally below, unity during the neutrino-dominated phase. The condition $L_{N,\nu_e}\sim L_{N,\bar\nu_e}$ is basically equivalent to $\dd \yeavg/\dd t\sim 0$ and expresses the circumstance that $Y_e\sim \yeeq$ and that neutrino-emission timescales are short or comparable to the dynamical (i.e. viscous) timescales. The fact that $L_{N,\nu_e} < L_{N,\bar\nu_e}$ derives from the tendency that $Y_e$ keeps running behind the increasing values of $\yeeq$. Once the torus becomes non-radiative, the number luminosities drop, $Y_e$ decouples from $\yeeq$, and the result is a growing disparity between $Y_e$ and $\yeeq$, which is counteracted by the system with boosting the production rates of $\bar\nu_e$ relative to those of $\nu_e$. This explains the drop of $L_{N,\nu_e}/L_{N,\bar\nu_e}$ at late times.

The efficiency by which neutrino emission can change the average electron fraction of the torus can be measured by the fraction $m_{\mathrm{em}}/\mtor$ of torus material that produces neutrinos on timescales $\tauem$ shorter than a certain threshold value; here we use $\tauem<1$\,s. This quantity is plotted in panel (e) of Fig.~\ref{fig:globdat_mtorus} and it exhibits a steep decline marking the end of the neutrino-dominated phase (e.g. for model m01M3A8 at about $t \sim 100\,$ms). We therefore use this quantity to measure the freeze-out time, $t_{\mathrm{em}}$, as the time when $m_{\mathrm{em}}/\mtor$ drops below $10\,\%$. We caution, however, that $t_{\mathrm{em}}$ is only a crude estimate and that fluid elements freeze out over an extended range of time before and after $t_{\mathrm{em}}$ (see, e.g., plots of $m^{\mathrm{FO}}$ in the bottom panels of Figs.~\ref{fig:torusye_time}~and~\ref{fig:torusye_time_vis}). For our set of models, $t_{\mathrm{em}}$ lies between 50 and 300\,ms (cf. Table~\ref{table_weak}), while it is prolonged for disks with greater masses and for more massive and faster rotating BHs. In general, torus configurations with overall lower temperatures freeze out earlier. In the extreme case of low disk compactness (i.e. low values of disk mass over disk size) and therefore low temperature, the bulk of the torus may not even be able to achieve weak equilibrium (in the sense that $\yeavg\approx\yeeqavg$) to begin with. This is the case for model m01M10A8 with a 10$\,\Msol$ central BH and for model m001M3A8 having a 0.001\,$M_\odot$ torus (cf. left panel of Fig.~\ref{fig:torusye_time}, where $\yeavg$ is barely changing from its initial value of 0.5). Table~\ref{table_weak} also reveals that the freeze-out time is roughly inversely proportional to $\alpvis$, which is not surprising as $\alpvis$ essentially regulates the accretion timescale.

Table~\ref{table_weak} also lists for all models the mass accretion rates into the BH measured at $t_{\mathrm{em}}$. Most values, albeit with large scatter, lie within $\dot{M}_{\mathrm{BH}}(t_{\mathrm{em}})\sim 10^{-3}\ldots 10^{-2}\,\Msol$\,s$^{-1}$, which is broadly consistent with analytical estimates \citep[see, e.g.,][who call this value the ignition accretion rate]{Siegel2019b, De2020a}. Finally, in order to facilitate comparison of the neutrino properties observed in our models with studies using more or less sophisticated neutrino schemes, we provide in Table~\ref{table_weak} the total percentage of accreted rest-mass energy radiated away by neutrinos and the mean energies of all emitted neutrinos.

\subsubsection{Different viscosity prescription}\label{sec:diff-visc-prescr}

While the observations described in the previous section remain applicable also for models m01M3A8-vis2(-no$\nu$), several quantitative differences appear when using Eq.~(\ref{eq:viseta2}) instead of Eq.~(\ref{eq:viseta1}) to express the dynamic viscosity coefficient $\eta_{\mathrm{vis}}$. As shown in Fig.~\ref{fig:lturbcomp}, the quantity $c_i/\Omega_K$, which is employed in the conventional  $\alpha$-viscosity prescription (cf. Eq.~(\ref{eq:viseta1})) as a proxy for the length scale of turbulent eddies, $\lturb$, grows roughly linearly with radius. In contrast, $\lturb=$const.$=9\,$km is used in models m01M3A8-vis2(-no$\nu$). This mismatch between length scales implies that the impact of viscosity in the $\lturb$=const. scheme is comparable to the conventional $\alpha$-viscosity scheme only at small radii, whereas it becomes relatively weaker at larger radii. Since the torus is expanding with time (cf. the radial coordinate of the center-of-mass, $\langle r\rangle_\rho$, shown in panel (g) of Fig.~\ref{fig:globdat_vis}), the aforementioned circumstance explains why most evolutionary features agree well at early times between models m01M3A8 (black lines in Fig.~\ref{fig:globdat_vis}) and m01M3A8-vis2 (green lines), but start to diverge at later times. Approximately once the torus in model m01M3A8-vis2 reaches $\langle r\rangle_\rho\ga 100\,$km, the torus expansion decelerates, and densities and temperatures drop more slowly (see $\langle\rho\rangle_\rho$ and $\langle T\rangle_\rho$ in panels (h) and (i) of Fig.~\ref{fig:globdat_vis}, respectively). This causes neutrino emission to remain efficient for a longer time (cf. panels (c) and (e) in Fig.~\ref{fig:globdat_vis}) and to impact a greater fraction of the expanding ejecta relative to tori with a standard $\alpha$-viscosity. This is further discussed in Sect.~\ref{sec:model-dependence}.

\subsubsection{MHD prescription}\label{sec:mhd-models}

The basic evolution of the MHD models, m01M3A8-mhd(-no$\nu$), is well in agreement with that reported in previous  studies of MHD disks \citep[e.g.][]{DeVilliers2003, McKinney2014a, Siegel2018c, Fernandez2019b}: Right after the start of the simulation the poloidal fields are wound up into toroidal fields, which soon become the dominant field component in the main body of the disk. Meanwhile, the MRI starts to grow and quickly generates turbulence that fuels angular momentum transport. In the surroundings of the disk a magnetized corona forms, while the polar regions develop a funnel of strong radial $B$-field, which dominates all baryonic energies and persists until the end of the simulation. Importantly, even though the flow pattern is turbulent right from the beginning in the MHD models -- whereas it remains rather laminar in the viscous models during the neutrino-dominated phase -- the features of neutrino emission and weak freeze out described in Sect.~\ref{sec:general-features} are similarly observed for the MHD models (see, e.g., Fig.~\ref{fig:globdat_vis}). Specific features of the MHD models in comparison to the viscous models will be discussed in Sect.~\ref{sec:model-dependence}.

Comparing the MHD models with and without neutrino absorption (e.g. panel~(h) in Fig.~\ref{fig:globdat_vis} showing the average torus density) we observe that the evolution of the optically thin torus noticeably lags behind that of the full transport model. The most likely explanation for this is that the almost perfect neutrino cooling in the no-absorption model until about $t\sim 600\,$ms reduces the disk thickness so dramatically that the dominant MRI modes become numerically under-resolved. This explanation is supported by the sudden transition to a faster disk expansion once neutrino cooling has become inefficient (see, e.g., strong decline of torus density in panel~(h) at about $t\sim 600\,$ms). Figure~\ref{fig:mridx} backs this interpretation, where we compare radial profiles of the number of grid points covering the MRI wavelength (left panel) and the disk scale height (right panel). A higher grid resolution could therefore solve this issue, however, the necessary grid refinement factor may be large, and we were unable to run additional high-resolution simulations given our limited budget of computing time. The suppression of the MRI in model m01M3A8-mhd-no$\nu$ to some extent delays the ejection of material and, by doing so, facilitates ejecta with relatively high $Y_e$ (see Sect.~\ref{sec:model-dependence} for a deeper discussion of this aspect).

\subsection{Role of neutrino absorption}\label{sec:role-neutr-absorpt}

\begin{figure*}
  \centering
  \includegraphics[trim=320 0 100 0,clip,width=0.4\textwidth]{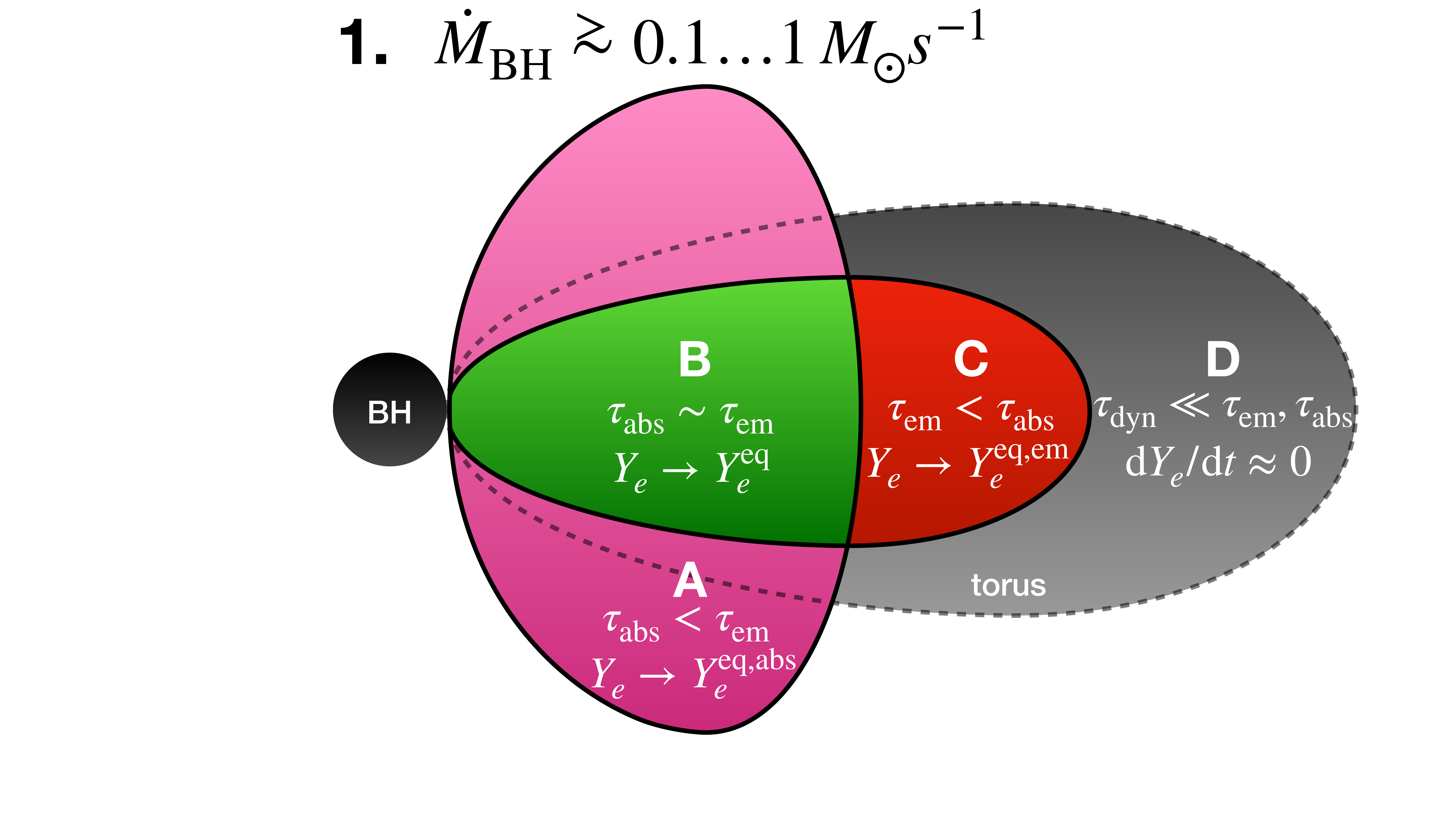}\qquad\qquad
  \includegraphics[trim=320 0 100 0,clip,width=0.4\textwidth]{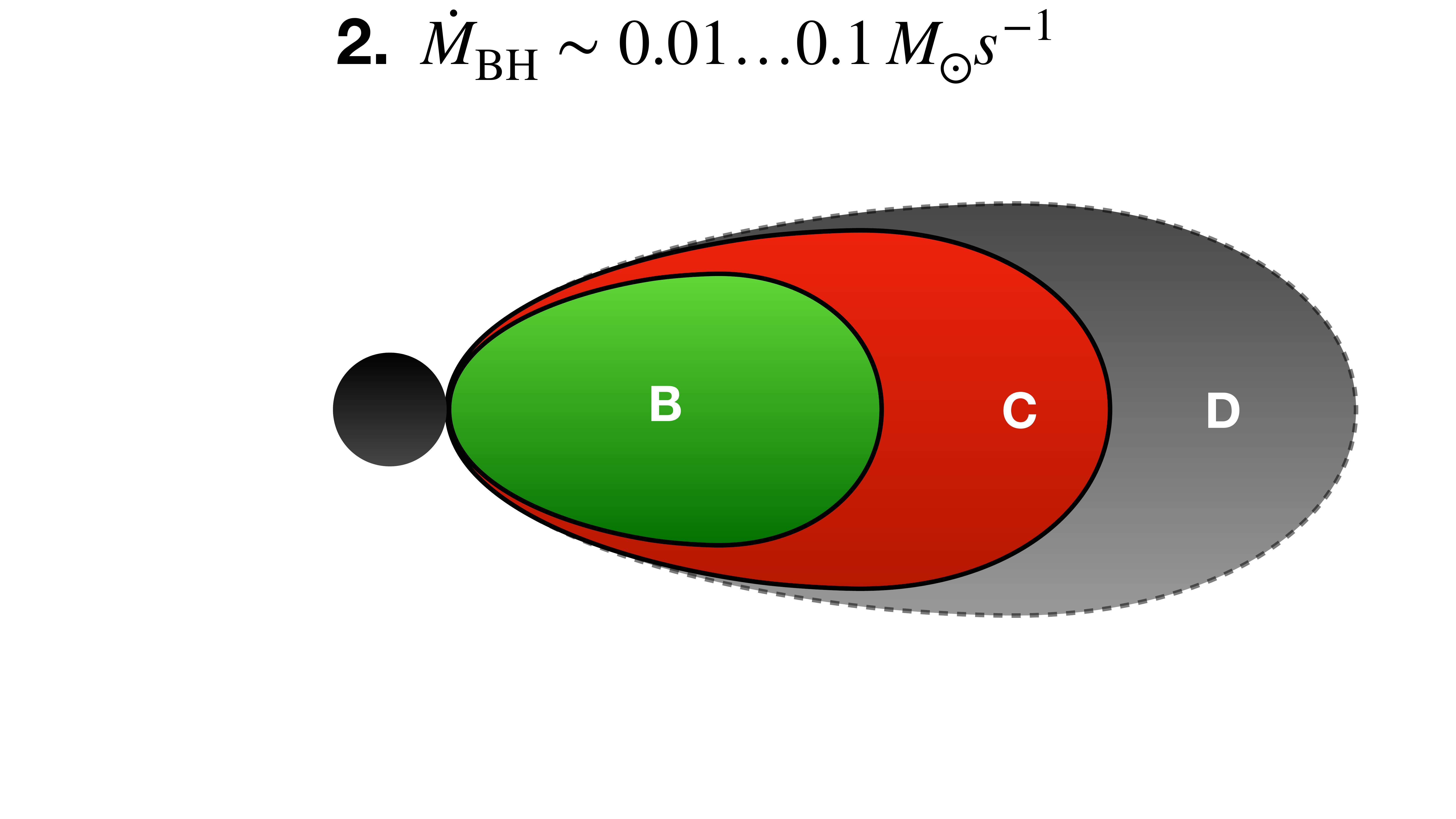}\\
  \includegraphics[trim=320 120 100 0,clip,width=0.4\textwidth]{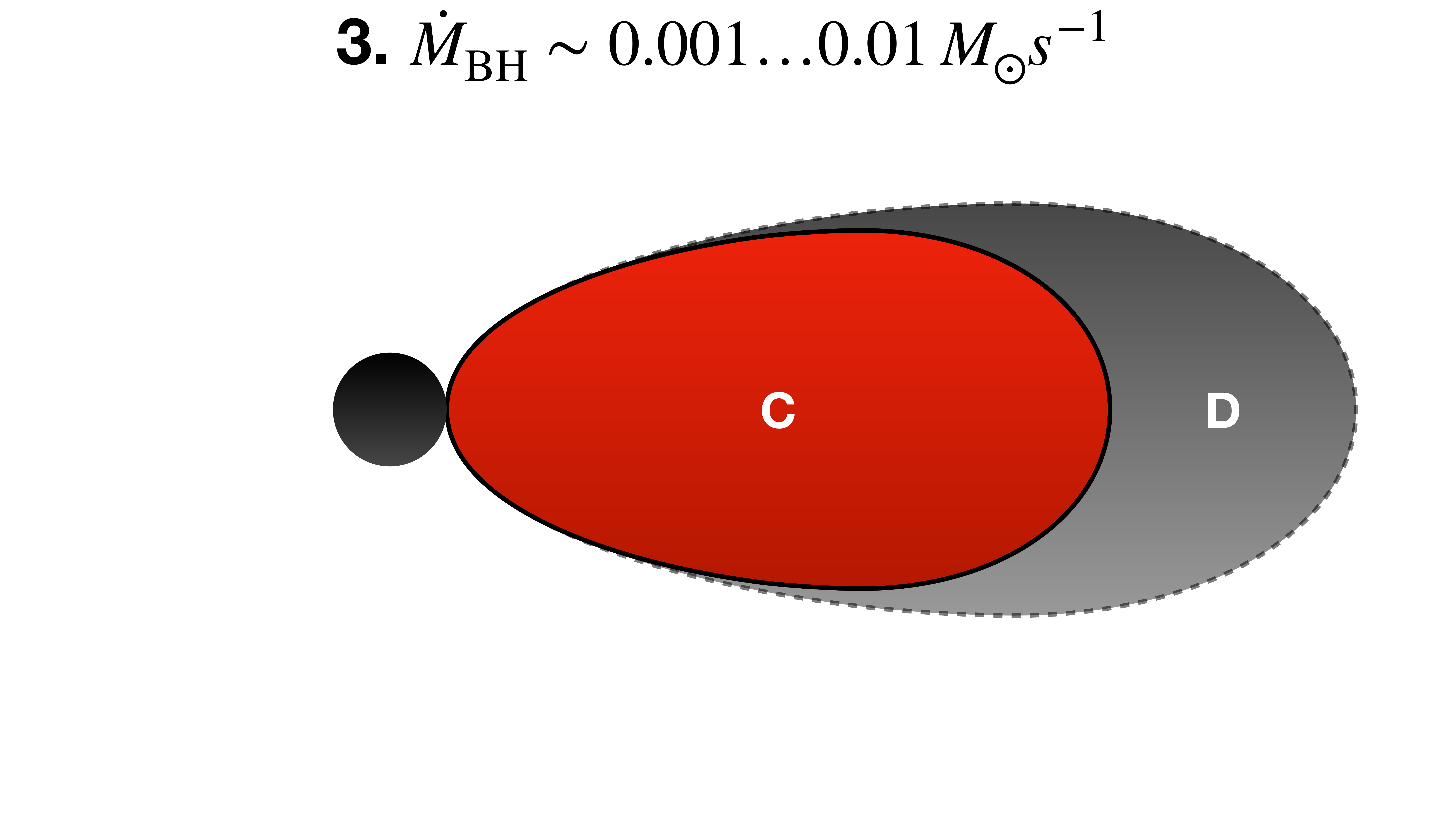}\qquad\qquad
  \includegraphics[trim=320 120 100 0,clip,width=0.4\textwidth]{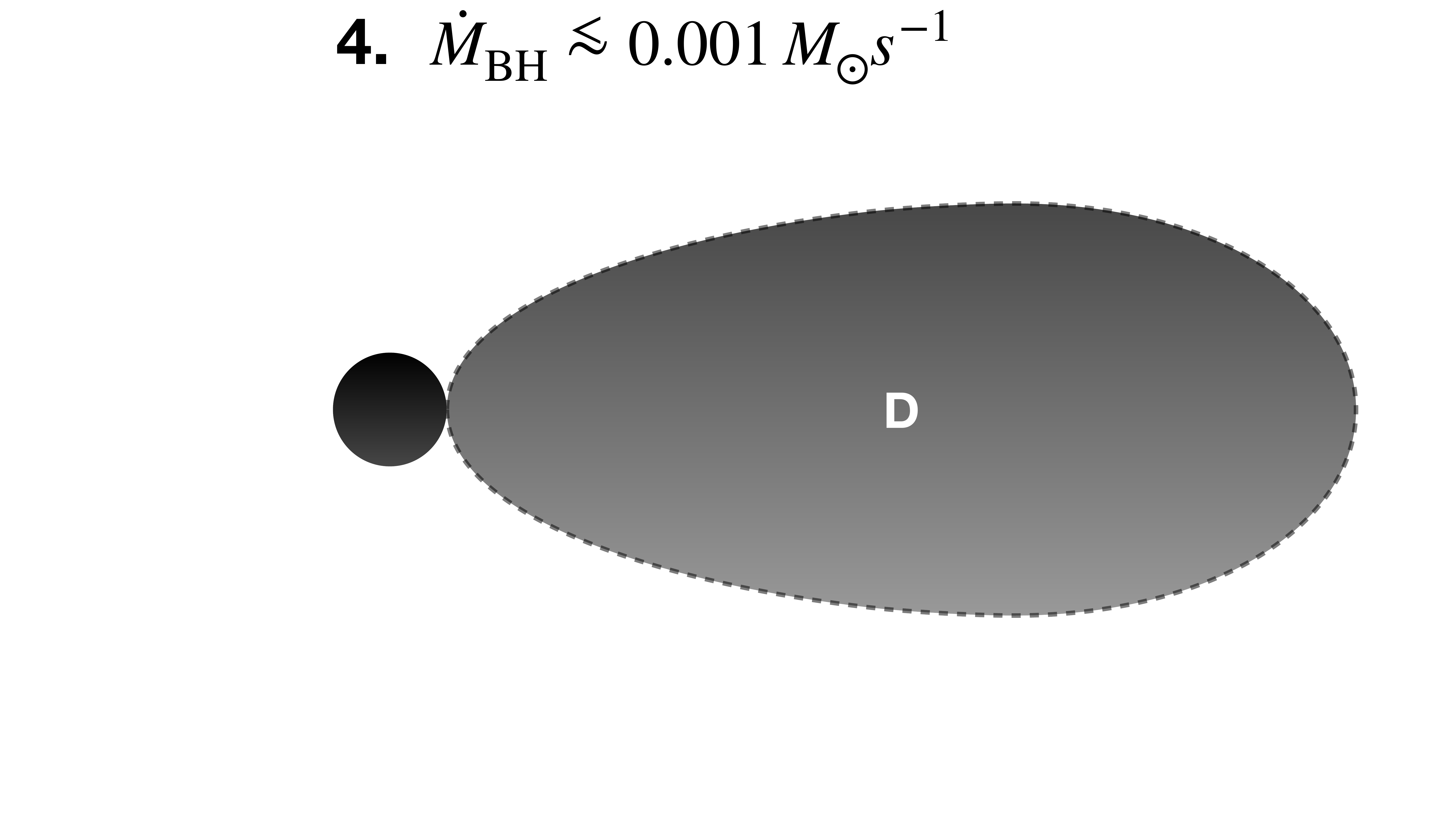}
  \caption{Schematic illustration of the characteristic weak-interaction regimes encountered in neutrino-cooled disks and their corresponding equilibrium electron fractions in dependence of the mass accretion rate onto the central BH, $\dot{M}_{\mathrm{BH}}$. The torus can roughly be divided into a region where only neutrino absorption is relevant (A), where both neutrino emission and absorption are relevant (B), where only neutrino emission is relevant (C), and where all weak interactions are inefficient (D). The cases 1 to 4 indicate different regimes of mass accretion rates onto the BH. See Sect.~\ref{sec:equil-cond-that} for the definition of the corresponding $Y_e$-equilibria and the emission/absorption timescales, as well as Sect.~\ref{sec:char-regi} for a discussion of the regions A, B, C, and D.}
  \label{fig:evolphases}
\end{figure*}

\begin{figure*}
  \centering
  \includegraphics[trim=0 62 0 0,clip,width=0.3\textwidth]{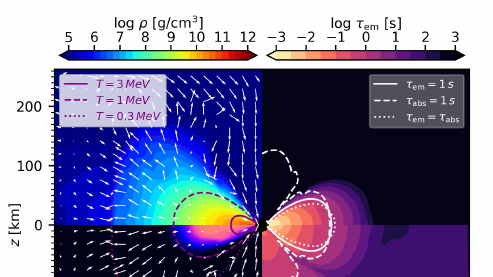}
  \includegraphics[trim=0 62 0 0,clip,width=0.3\textwidth]{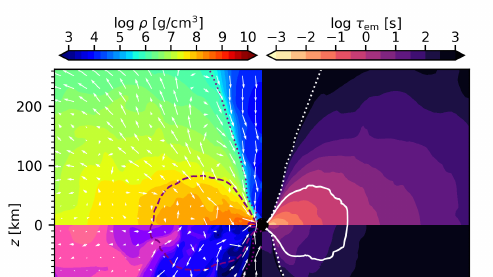}
  \includegraphics[trim=0 62 0 0,clip,width=0.3\textwidth]{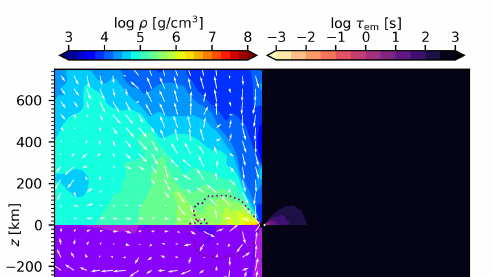}
  %% \vspace{0.2cm}\\
  \includegraphics[trim=0 0 0 37,clip,width=0.3\textwidth]{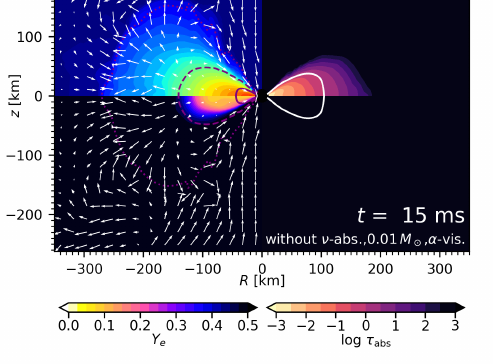}
  \includegraphics[trim=0 0 0 37,clip,width=0.3\textwidth]{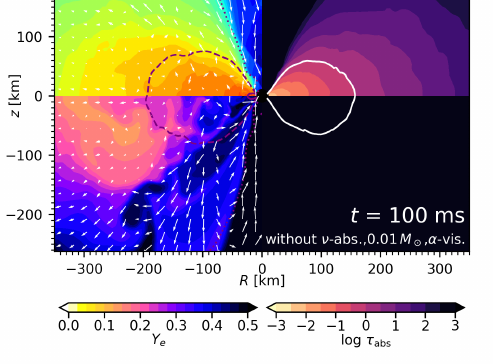}
  \includegraphics[trim=0 0 0 37,clip,width=0.3\textwidth]{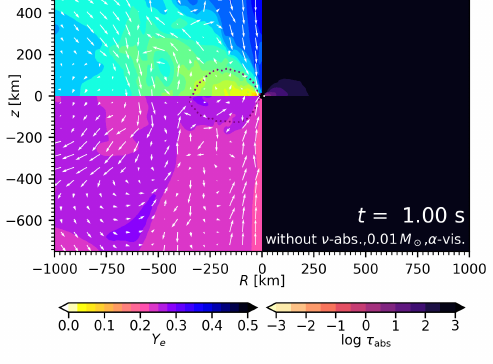}
  \caption{Snapshots of model m01M3A8 (top row) and the corresponding model without neutrino absorption, m01M3A8-no$\nu$ (bottom row). Each panel shows the density, $\rho$, electron fraction, $Y_e$, emission timescale, $\tauem$, and absorption timescale, $\tauabs$, at evolution times $t=15, 100$, and $1000\,$ms. The purple lines on the left-hand side of each box show contours of the temperature at $T=0.3, 1,$ and~$3\,$MeV, and the white lines on the right-hand side mark the surfaces of regions where the emission timescale (solid) and absorption timescale (dashed) are shorter than $1\,$s as well as the location where both timescales are equal (dotted).}
  \label{fig:contours_m01}
\end{figure*}	

\begin{figure*}
  \centering
  \includegraphics[trim=0 62 0 0,clip,width=0.3\textwidth]{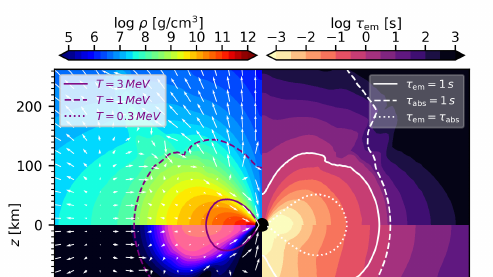}
  \includegraphics[trim=0 62 0 0,clip,width=0.3\textwidth]{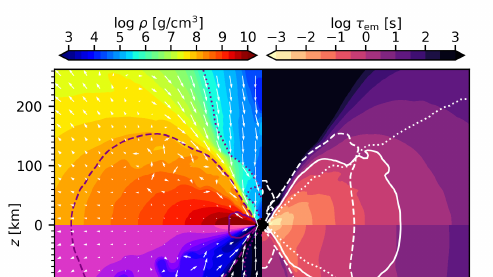}
  \includegraphics[trim=0 62 0 0,clip,width=0.3\textwidth]{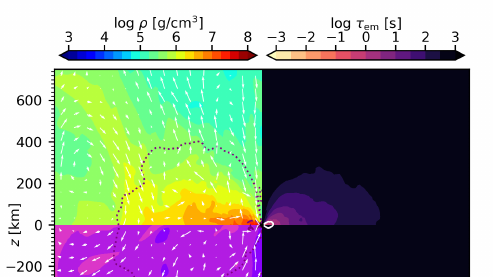}
  %% \vspace{0.3cm}\\
  \includegraphics[trim=0 0 0 37,clip,width=0.3\textwidth]{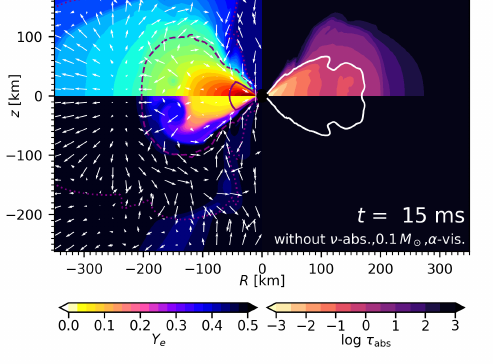}
  \includegraphics[trim=0 0 0 37,clip,width=0.3\textwidth]{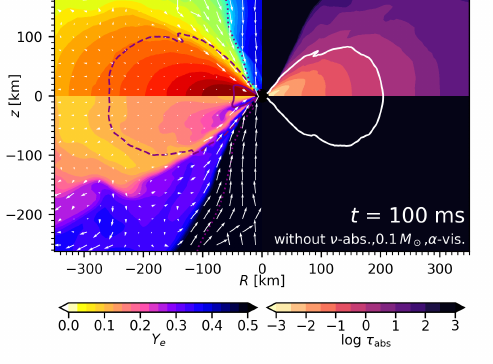}
  \includegraphics[trim=0 0 0 37,clip,width=0.3\textwidth]{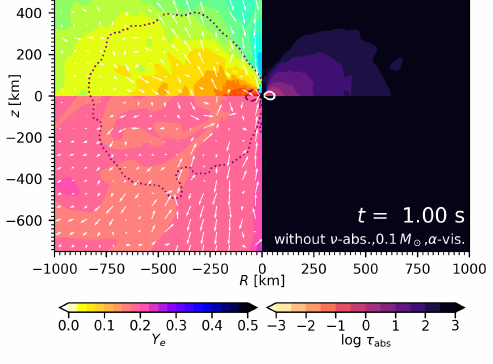}
  \caption{Same as Fig.~\ref{fig:contours_m01} but for models m1M3A8 (upper panels) and m1M3A8-no$\nu$ (lower panels) with initial torus mass of 0.1\,$\Msol$ instead of 0.01$\,\Msol$.}
  \label{fig:contours_m1}
\end{figure*}	

\begin{figure*}
  \centering
  \includegraphics[trim=0 62 0 0,clip,width=0.3\textwidth]{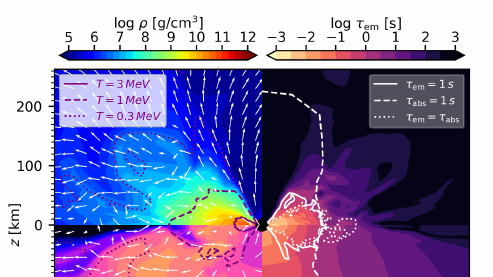}
  \includegraphics[trim=0 62 0 0,clip,width=0.3\textwidth]{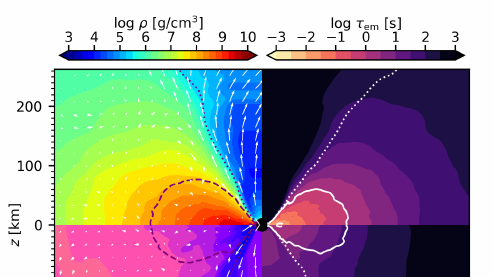}
  \includegraphics[trim=0 62 0 0,clip,width=0.3\textwidth]{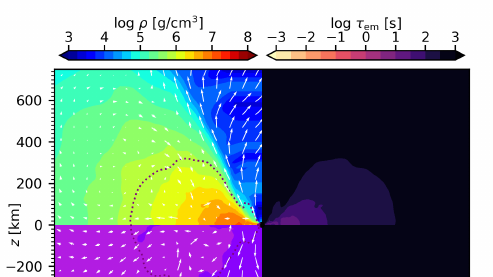}
  %% \vspace{0.3cm}\\
  \includegraphics[trim=0 0 0 37,clip,width=0.3\textwidth]{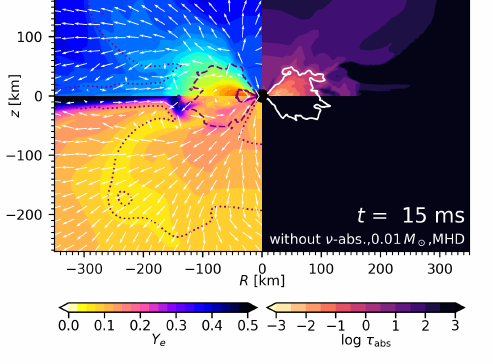}
  \includegraphics[trim=0 0 0 37,clip,width=0.3\textwidth]{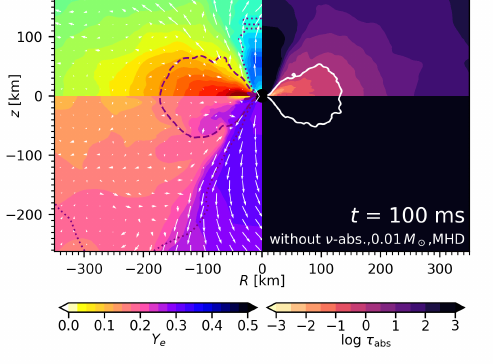}
  \includegraphics[trim=0 0 0 37,clip,width=0.3\textwidth]{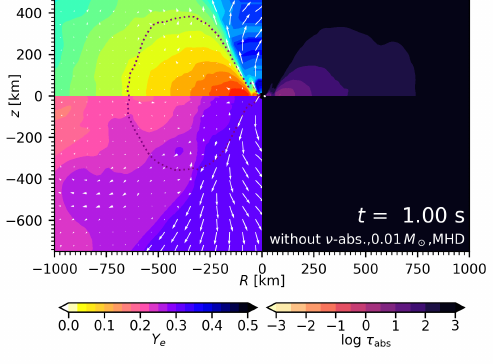}
  \caption{Same as Fig.~\ref{fig:contours_m01} but for models m01M3A8-mhd (upper panels) and m01M3A8-mhd-no$\nu$ (lower panels), where the viscosity is replaced by magnetic fields, and showing data that has been obtained by averaging along the azimuthal direction.}
  \label{fig:contours_m01_mhd}
\end{figure*}	

\begin{figure*}
  \centering
  \includegraphics[width=0.99\textwidth]{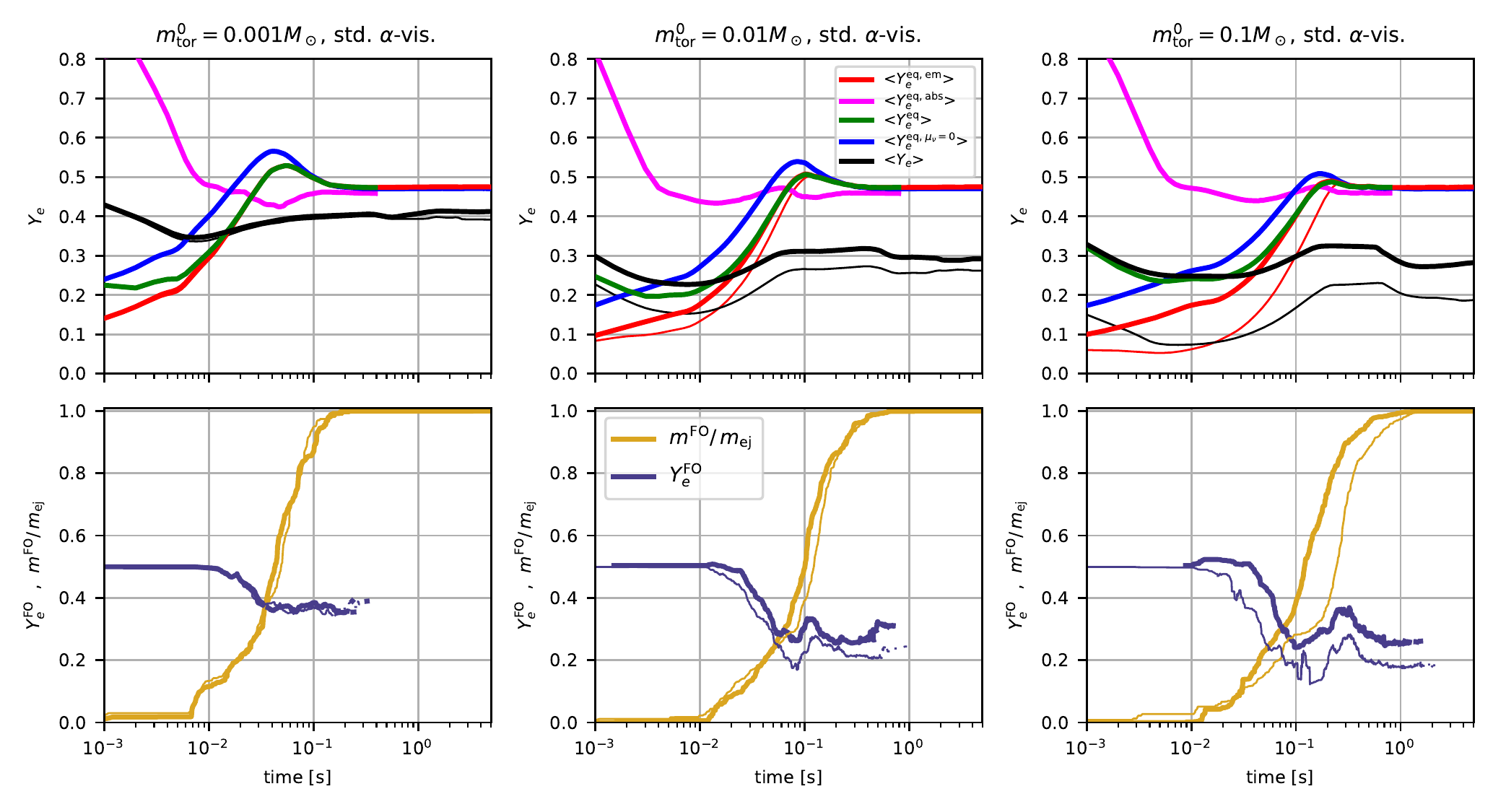}
  \caption{\emph{Top row:} Mass-weighted averages of the electron fraction, $Y_e$ (black lines), as well as the equilibrium quantities $\yeeqem$ (Eq.~(\ref{eq:yeeqem}), red lines), $\yeeqabs$ (Eq.~(\ref{eq:yeeqabs}), magenta lines), $\yeeq$ (Eq.~(\ref{eq:yeeq}, green lines)), and $\yeeqmu$ (Eq.~(\ref{eq:yeeqmu}), blue lines) for models m001M3A8(-no$\nu$) (left panel), m01M3A8(-no$\nu$) (middle panel), and m1M3A8(-no$\nu$) (right panel). Thick lines (thin lines) belong to models including (neglecting) neutrino absorption. The averages include all material within radii of $10^4\,$km. Note that the reason for $\langle Y_e\rangle$ not remaining exactly constant after weak freeze out is that torus material leaves the considered volume due to mass ejection at $10^4\,$km and mass accretion onto the BH. \emph{Bottom row:} Mass fraction of outflow material for which $Y_e$ is already frozen out, $m^{\mathrm{FO}}/m_{\mathrm{ej}}$, i.e. with weak interaction timescales longer than $10\,$s (golden lines), and average $Y_e$ of all ejecta trajectories that are freezing out during a short time window $\Delta t$ centered around the current time $t$, $Y^{\mathrm{FO}}_e$.}
  \label{fig:torusye_time}
\end{figure*}	

\begin{figure*}
  \centering
  \includegraphics[trim=199 0 0 0,clip,width=0.66\textwidth]{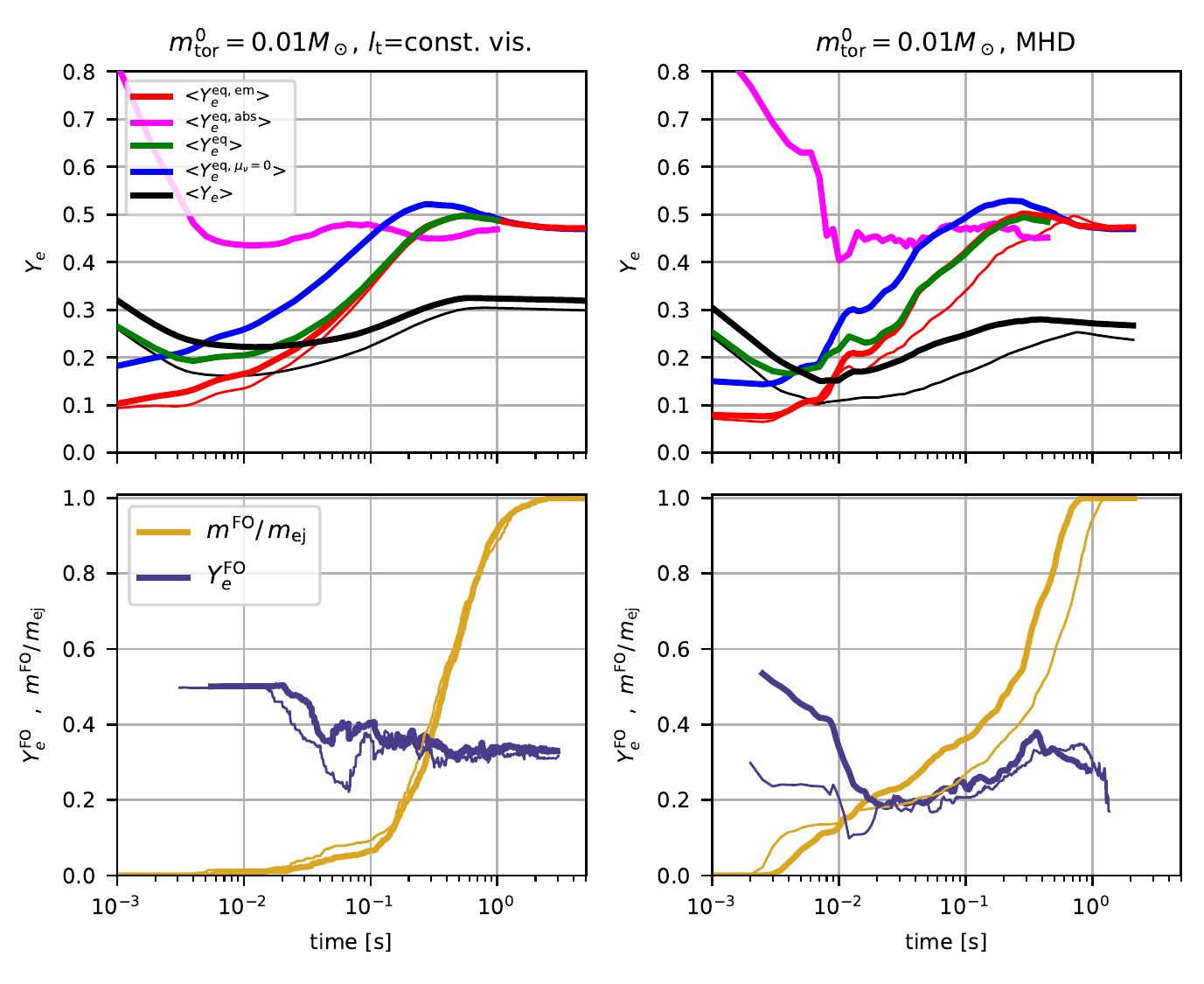}
  \caption{Same as Fig.~\ref{fig:torusye_time} but for models m01M3A8-vis2(-no$\nu$) and m01M3A8-mhd(-no$\nu$).}
  \label{fig:torusye_time_vis}
\end{figure*}	

\begin{figure}
  \centering
  \includegraphics[width=0.4\textwidth]{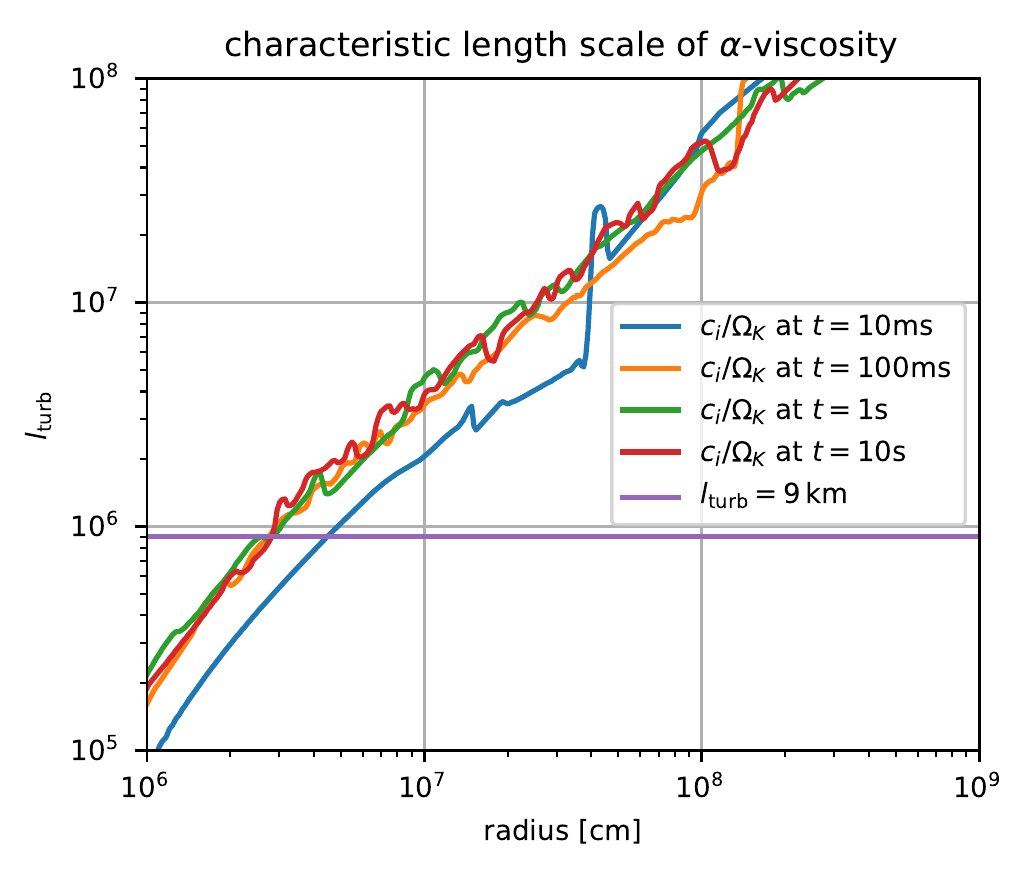}
  \caption{Comparison of $\lturb=c_i/\Omega_K$ (where $c_i=\sqrt{P/\rho}$ and $\Omega_K$ is the Keplerian angular velocity) used as characteristic length scale in the standard $\alpha$-viscosity approach with $\lturb=9\,$km used in models m01M3A8-vis2(-no$\nu$). The profiles of $\lturb=c_i/\Omega_K$ are computed along the equator for the indicated times based on the data of model m01M3A8.}
  \label{fig:lturbcomp}
\end{figure}

\begin{figure}
  \centering
  \includegraphics[width=0.48\textwidth]{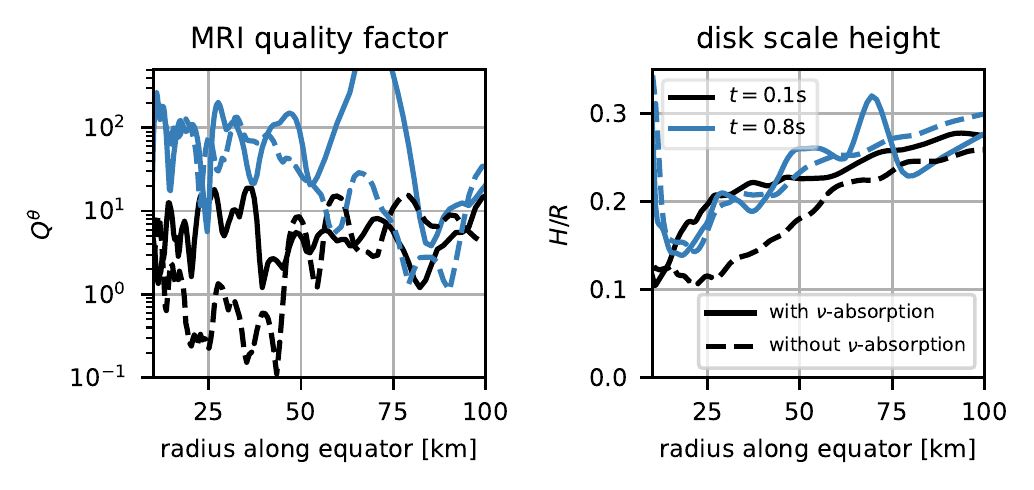}
  \caption{\emph{Left panel:} Equatorial profile of the MRI quality factor, $Q^\theta$, approximately representing the number of grid zones that resolve the wavelength of fastest MRI growth \citep[see, e.g., Eq.~(A1) in][for its computation]{Miller2019a} for the two 3D MHD models with and without neutrino absorption at two different times $t=0.1$ and 0.8\,s. \emph{Right panel:} Radial profile of the disk scale height normalized to the equatorial radius, $H/R\approx \sqrt{\left(\int_\theta \rho (\theta-\pi/2)^2\dd\theta\right)/\left(\int_\theta \rho \dd\theta\right)}$. At early times, neglecting neutrino absorption leads to a geometrically thinner disk at radii $\la 50\,$km, in which the MRI wavelength is shorter and therefore less well resolved compared to the model including neutrino absorption.}
  \label{fig:mridx}
\end{figure}

%% \begin{figure*}
%%   \centering
%%   \includegraphics[trim=320 0 100 0,clip,width=0.4\textwidth]{./phase1.pdf}\qquad\qquad
%%   \includegraphics[trim=320 0 100 0,clip,width=0.4\textwidth]{./phase2.pdf}\\
%%   \includegraphics[trim=320 120 100 0,clip,width=0.4\textwidth]{./phase3.pdf}\qquad\qquad
%%   \includegraphics[trim=320 120 100 0,clip,width=0.4\textwidth]{./phase4.pdf}
%%   \caption{Schematic illustration of the characteristic weak-interaction regimes encountered in neutrino-cooled disks and their corresponding equilibrium electron fractions in dependence of the mass accretion rate onto the central BH, $\dot{M}_{\mathrm{BH}}$. The torus can roughly be divided into a region where only neutrino absorption is relevant (A), where both neutrino emission and absorption are relevant (B), where only neutrino emission is relevant (C), and where all weak interactions are inefficient (D). The cases 1 to 4 indicate different regimes of mass accretion rates onto the BH. See Sect.~\ref{sec:equil-cond-that} for the definition of the corresponding $Y_e$-equilibria and the emission/absorption timescales, as well as Sect.~\ref{sec:char-regi} for a discussion of the regions A, B, C, and D.}
%%   \label{fig:evolphases}
%% \end{figure*}	

\begin{figure*}
  \centering
  \includegraphics[width=0.99\textwidth]{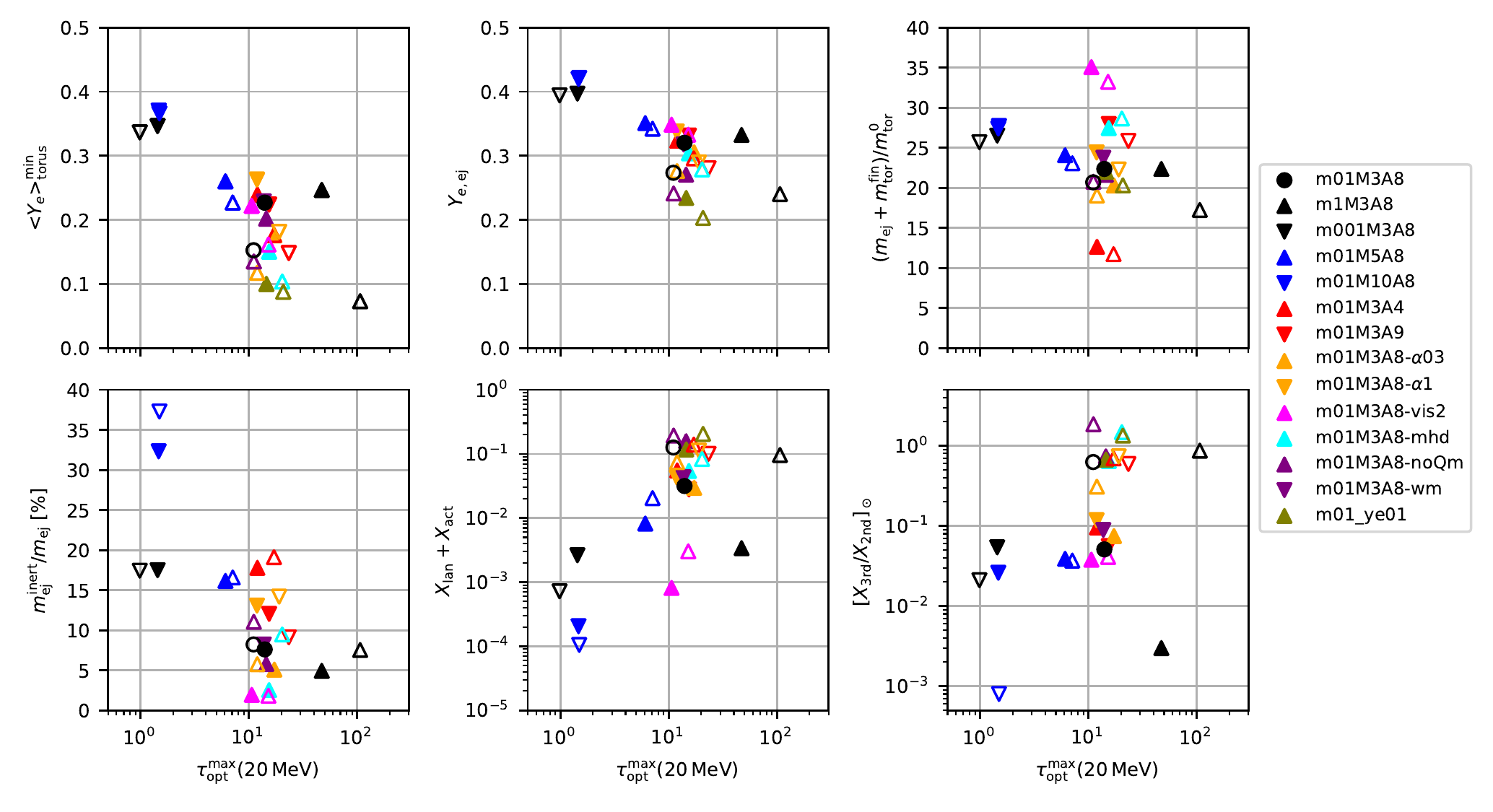}
  \caption{Global comparison of models visualizing data of Tables~\ref{table_weak},~\ref{table_ejecta}, and~\ref{table_nuc}. For each torus model we show as function of its maximum optical depth the average electron fraction of the ejected material, $Y_e^{\mathrm{ej}}$, minimum of the average electron fraction of the torus, $\yemin$, mass fraction of ejected fluid that remained unaffected by weak interactions, $m_{\mathrm{ej}}^{\mathrm{inert}}/m_{\mathrm{ej}}$, the sum of lanthanide and actinide mass fractions, $X_{\mathrm{LA}}=X_{\mathrm{lan}}+X_{\mathrm{act}}$, and the ratio of 3rd-peak to 2nd-peak abundances normalized to the solar value,  $[X(\mathrm{3rd})/X(\mathrm{2nd}]_\odot$. The mass fractions are defined in the caption of Table~\ref{table_nuc}. Filled (open) symbols refer to models including (ignoring) neutrino absorption.}
  \label{fig:tabdat}
\end{figure*}

\subsubsection{Characteristic regions}\label{sec:char-regi}

Before investigating the quantitative impact of neutrino absorption, we first have a look at some qualitative features. In Figs.~\ref{fig:contours_m01}--\ref{fig:contours_m01_mhd}, contour maps are shown of the neutrino emission timescales, $\tauem$, and absorption timescales, $\tauabs$, for models m01M3A8, m1M3A8, m01M3A8-mhd, and their corresponding ``no$\nu$'' pendants in which neutrino absorption is neglected. The conditions for weak interactions observed in our models motivate the definition of four characteristic regions in the torus, which are schematically depicted in Fig.~\ref{fig:evolphases}. The B region (green) in Fig.~\ref{fig:evolphases} can be identified in Figs.~\ref{fig:contours_m01}--\ref{fig:contours_m01_mhd} by the overlap region, where both neutrino emission (white solid lines) and absorption (white dashed lines) are efficient. This region, if present at a given time, is located around the densest parts of the torus and typically extends to radii of $r\sim 100-250\,$km. In this region the electron fraction is close to $\yeeq$ (cp. Eq.~(\ref{eq:yeeq})). During early evolution times and for high mass accretion rates, the northern and southern surface layers of the torus are surrounded by a region (denoted A in Fig.~\ref{fig:evolphases}), in which the temperatures are relatively low, such that absorption dominates neutrino emission, i.e. $\tauabs<\tauem$. Here, $Y_e$ is pushed towards $\yeeqabs$. Moreover, neutrino heating in this region may even be dynamically important and launch a thermal wind\footnote{Given the low torus masses for most of our models, a neutrino-driven wind is observed only in model m1M3A8 during a short initial phase, as can be seen when comparing the mass ejection rates in panel (a) of Fig.~\ref{fig:globdat_mtorus} between models m1M3A8 and m1M3A8-no$\nu$.}. Next, since the torus optical depth is largest in the equatorial direction, the edge of the torus around the midplane experiences relatively low neutrino absorption rates, while neutrino emission rates may still be significant. In this region (C in Fig.~\ref{fig:evolphases}) $Y_e\rightarrow \yeeqem$. Finally, region D subsumes all torus material where weak interactions take place on timescales much longer than the viscous (or expansion) timescales and $Y_e$ is basically frozen out.

The four different cases sketched in Fig.~\ref{fig:evolphases} represent different regimes of mass accretion rates, $\dot{M}_{\mathrm{BH}}$. During the evolution of a sufficiently massive torus, region A will disappear first, when $\dot{M}_{\mathrm{BH}}\la 0.1\,\Msol\,$s$^{-1}$, followed by region B once the optical depth for neutrinos falls below unity (approximately once $\dot{M}_{\mathrm{BH}}\la 0.01\,\Msol\,$s$^{-1}$). Finally, region C vanishes around the freeze-out time, $t_{\mathrm{em}}$, when $\dot{M}_{\mathrm{BH}}$ drops below $\sim 0.001\,\Msol\,$s$^{-1}$. As a result, a fixed-mass torus (such as formed in a NS merger) may never exhibit regions A, B, or C if its initial mass is too low to achieve the corresponding mass-accretion rates.

As can be seen in Fig.~\ref{fig:contours_m01_mhd}, the turbulent behavior of the MHD models strongly distorts instantaneous iso-contours and sometimes even disrupts them into multiple patches. However, these perturbations are stochastic in nature, hence, a classification based on Fig.~\ref{fig:evolphases} and described above still holds for the MHD models but should then be interpreted in the time-averaged sense, i.e. considering properties that are averaged over a few dynamical timescales.

\subsubsection{Optical depth and absorption factor}\label{sec:optical-depth}
  
Next, since to our knowledge this quantity has rarely been systematically reported so far for neutrino-cooled disks, we have a brief look at the optical depth,
\begin{align}\label{eq:tauopt}
  \tauopt(\eps) = \int \kappa_{\mathrm{abs}}(\eps)\mathrm{d}r \, ,
\end{align}
where $\kappa_{\mathrm{abs}}(\eps)$ is the total opacity as function of neutrino energy $\eps$.
While the optical depth, and therefore the neutrino emission rates, depend on the direction in which neutrinos leave the torus, we are only interested here in a typical, representative optical depth, which we choose to compute in Eq.~(\ref{eq:tauopt}) along the equatorial direction from infinity to the central BH, i.e. through the entire computational domain. A straightforward estimate of $\tauopt$ can be obtained by adopting for the neutrino energy the mean energy of released neutrinos measured far away from the torus, given by
\begin{align}\label{eq:epsrel}
  \langle\eps_{\mathrm{rel}}\rangle = \frac{L_\nu}{L_{\nu,N}} \, .
\end{align}
We use the approximate formula \citep[e.g.][]{Bruenn1985}
\begin{align}\label{eq:kappaopt}
  \kappa_{\mathrm{abs}}(\eps) \approx \frac{\sigma_0}{4 m_e c^2}(3 g_A^2+1) n_B \eps^2 \, ,
\end{align}
for calculating all optical depths in this paper (where $\sigma_0=1.761\times 10^{-44}\,$cm$^{2}$ and $g_A=1.254$). The solid lines in panels (j) and (k) of Figs.~\ref{fig:globdat_mtorus}~and~\ref{fig:globdat_vis} depict the evolution of $\eps$ and $\tauopt(\eps)$ computed in this way for various models. However, the numbers resulting in this case for $\eps$ and $\tauopt(\eps)$ are systematically underrated, because they disregard the fact that preferrably neutrinos of higher energy are absorbed, owing to the $\eps^2$ dependence of $\kappa_{\mathrm{abs}}$. Therefore, a more appropriate energy for measuring the impact of absorption is given by the average energy of all neutrinos captured by nucleons per unit of time, i.e.
\begin{align}\label{eq:epsabs}
  \langle\eps_{\mathrm{abs}}\rangle = \frac{\int_{r<r_1} \dot{e}_{\mathrm{abs}}\mathrm{d}V}{\int_{r<r_1} \dot{n}_{\mathrm{abs}}\mathrm{d}V}  \, ,
\end{align}
where $\dot{n}_{\mathrm{abs}} = \lambda_{\bar\nu_e}n_p + \lambda_{\nu_e}n_n$ and $\dot{e}_{\mathrm{abs}}$ is the corresponding energy-absorption rate that results after replacing $\eps^2$ by $\eps^3$ in the rates $\lambda_{\nu_e/\bar\nu_e}$, cf. Eq.~(\ref{eq:betarates}). The resulting changes in $\eps$ and $\tauopt(\eps)\propto \eps^2$ are quite significant, approximately a factor of 2 and 4, respectively, during the neutrino-dominated phase. For the fiducial model, m01M3A8, with a relatively low torus mass of $0.01\,\Msol$ the optical depth computed in this way exceeds 10 during the first $\sim 20\,$ms and drops below $\tauopt=1$ only after $t\approx 60\,$ms. The sharp dependence of the optical depth on the detailed neutrino energy spectrum highlights the importance of using an energy-dependent neutrino transport scheme for investigating optical-depth related effects in neutrino-cooled disks.

We provide for each model in Table~\ref{table_weak} the time-integrated mean energies of released and absorbed neutrinos as well as the maximum value of the optical depth\footnote{The reason why in Table~\ref{table_weak} we employ a fixed neutrino energy of $\eps=20\,$MeV to compute maximum optical depths instead of $\eps_{\mathrm{abs}}$ from Eq.~(\ref{eq:epsabs}) is simply to enable a straightforward comparison between all models, even those neglecting neutrino absorption.} attained by each model during its evolution, $\tau_{\mathrm{opt}}^{\mathrm{max}}$. We find higher values of $\tau_{\mathrm{opt}}^{\mathrm{max}}$ for models that lead to more compact torus configurations, namely for larger disk masses, smaller BH masses, higher BH spins, and lower values of the viscous $\alpha$-parameter. An enhanced role of absorption for faster spinning BHs has also been reported in \citet{Fernandez2015b}. It turns out that $\tau_{\mathrm{opt}}^{\mathrm{max}}$ is not only a useful measure for the importance of neutrino absorption in a given torus configuration, but it also correlates, though only approximately, with the average electron fraction of the torus and, therefore, of the ejecta, as can be seen in Fig.~\ref{fig:tabdat}. The reason for this correlation is simple: High optical depths tend to be found in tori with high densities and therefore more degenerate electron distributions and correspondingly low values of $\yeeqem$ (see, e.g., Fig.~\ref{fig:yeeq1}).

Lastly, to enable the comparison with popular neutrino leakage schemes we compute another measure for the importance of neutrino absorption, namely the quantity
\begin{align}\label{eq:absfac}
  \chi_{\mathrm{abs}} = \frac{\int_{r<r_1}  (\dot{n}_{\mathrm{em}} - \dot{n}_{\mathrm{abs}})\mathrm{d}V}
      {\int_{r<r_1} \dot{n}_{\mathrm{em}}\mathrm{d}V} 
\end{align}
that we call absorption factor here and which measures the fraction of neutrinos that are released by the system with respect to the total number of produced neutrinos (see panel (l) in Figs.~\ref{fig:globdat_mtorus}~and~\ref{fig:globdat_vis}). This quantity is an integral version of the quenching factor commonly used in neutrino leakage schemes (see, e.g., \citealp{Ruffert1996a} and the discussion in Sect.~\ref{sec:comp-with-prev}) to approximate the effective reduction of neutrino emission rates due to optical-depths effects. The value of $\chi_{\mathrm{abs}}$ vanishes if neutrino absorption exactly balances neutrino emission, while it approaches unity if neutrino absorption becomes negligible. For our disk models we find values of $\chi_{\mathrm{abs}}\sim 0.3-0.8$ during the neutrino-dominated phase, which are characteristic of neutrino transport in the semi-transparent regime.

\subsubsection{Absorption vs. emission}\label{sec:absorpt-vs.-emiss}

The relative importance of neutrino absorption with respect to neutrino emission can be assessed in a straightforward manner by adapting the quantities of Sect.~\ref{sec:neutr-emiss-weak} to the analysis of absorption, such as the average absorption timescale,
\begin{align}\label{eq:tauabstorus}
  \langle\tau_{\mathrm{abs}}\rangle = \frac{\int_{r<r_1}n_B\mathrm{d}V} {\int_{r<r_1}\dot{n}_{\mathrm{abs}}\mathrm{d}V} \, ,
\end{align}
and the fraction of the torus in which significant absorption rates are measured (i.e. where $\tauabs<1\,$s); see panels (d) and (e) in Figs.~\ref{fig:globdat_mtorus}~and~\ref{fig:globdat_vis}. The absorption timescale, $\langle\tau_{\mathrm{abs}}\rangle$, is an approximate measure of the time needed for absorption reactions to drive $\yeavg$ to its equilibrium value, $\yeeqavg$, recalling that $\yeeq$ is the equilibrium $Y_e$ for given thermodynamic state under the presence of neutrinos (cf. Eq.~(\ref{eq:yeeq})). Remarkably, even for the fiducial model, m01M3A8, with a rather low initial mass of 0.01\,$\Msol$ the absorption timescale is $\la 10\,$ms initially and shorter than the viscous timescales during the first $20-30\,$ms of evolution. This implies that even for this low-mass system absorption is at least at early times non-negligible in regulating $\yeeq$ and therefore $Y_e$. Absorption reactions become eventually unimportant around $t\sim t_{\mathrm{abs}}= 60\,$ms, where $t_{\mathrm{abs}}$ is defined as the time when only 10$\,\%$ of nucleons in the torus experience absorption rates larger than 1\,s$^{-1}$. When varying the global model parameters (cf. Table~\ref{table_weak}) $t_{\mathrm{abs}}$ scales about linearly with the emission freeze-out times, $t_{\mathrm{em}}$, such that roughly $t_{\mathrm{abs}}\sim 0.5\times t_{\mathrm{em}}$ and $t_{\mathrm{abs}}\la 100\,$ms in most models. The corresponding mass accretion rates measured at times $t_{\mathrm{abs}}$ are $\sim 2-6\times 10^{-2}\,\Msol$\,s$^{-1}$ (cf. Table~\ref{table_weak}).

\subsubsection{Impact of absorption on torus $Y_e$}\label{sec:impact-absorpt-torus}

We now discuss the impact of neutrino absorption on the evolution of $Y_e$ in the torus. We again take a look at Fig.~\ref{fig:torusye_time}, where torus-averaged versions are plotted of $Y_e$ (black lines), $\yeeq$ (green line), $\yeeqem$ (red line), $\yeeqabs$ (magenta line), and $\yeeqmu$ (blue line) and models with neutrino absorption (thick lines) are compared to models without (thin lines). In all models except the low-mass model, m001M3A8, neutrino absorption is responsible for a significant increment of $\yeeqavg$ (recalling that $\yeeqavg=\yeeqemavg$ for the no$\nu$ models), as a result of two effects.

The first effect is that captures of $\nu_e/\bar\nu_e$ on $n$/$p$ shift the kinetic $\beta$-equilibrium towards less neutron-rich conditions, i.e. $\yeeqavg>\yeeqemavg$ for models with neutrino absorption. The explanation is not far to seek and generic for neutrino-cooled disks: Since the emission timescales are sufficiently short in the neutrino-dominated phase, the torus emits roughly the same number of $\nu_e$ and $\bar\nu_e$ per unit of time. Neutrino irradiation by itself would thus saturate at $Y_e\rightarrow \yeeqabs \approx 0.5$ modulo quantitative corrections related to the neutrino spectra. Hence, $\yeeqabs>\yeeqem$ and therefore $\yeeq>\yeeqem$. At later times, well after neutrino absorption has become irrelevant, this situation changes, first because of the decline of $L_{N,\nu_e}/L_{N,\bar\nu_e}$ (which without recombination would lead to very low values of $\yeeqabs$), and second because of nuclear recombination, which ultimately drives both equilibrium quantities to the same value of $\yeeqem\approx\yeeqabs\approx Z_h/A_h\sim 0.5$, where $Z_h=25$ and $A_h=54$ are the charge number and mass number of the representative nucleus in our 4-species EOS.

The second effect on $Y_e$ related to the (non-)inclusion of neutrino absorption is of dynamical nature and is the reason why $\yeeqemavg$ is reduced in models neglecting neutrino absorptions compared to models including them. This reduction stems from the fact that cooling is enhanced in those models, because neutrinos can stream out freely from the torus without experiencing scatterings and re-absorption by the fluid. The boosted energy release rates imply higher densities (see panel (h) in Fig.~\ref{fig:globdat_mtorus}) and therefore more degenerate electron distributions and ultimately lower values of $\yeeqem$. This finding illustrates that a proper treatment of the energy transport can have just as important consequences for the evolution of $Y_e$ as the lepton transport in that an underestimation (overestimation) of net cooling rates leads to higher (lower) values of $Y_e$ in the torus.

Both aforementioned effects together lead to a reduction of the minimum value of the average torus $Y_e$, $\yemin$, which is given in Table~\ref{table_weak} and plotted in Fig.~\ref{fig:tabdat} for models not accounting for neutrino backreactions. The size of this reduction grows for models with overall higher optical depth, and it is most dramatic ($\Delta\yemin \approx 0.17$) for model m1M3A8, which has a torus mass of 0.1$\,\Msol$. Furthermore, $\yemin$ appears to decrease monotonically with the torus mass -- or probably with any other global parameter that leads to higher densities in the torus -- only if absorption reactions are absent. If taken into account, neutrino absorptions may limit or even reverse this trend for sufficiently high optical depths, as observed for the sequence of models with increasing torus mass (compare $\yemin$ of models m001M3A8, m01M3A8, and m1M3A8 with that of the corresponding no$\nu$ models).

On a final note, we point out that the blue lines in Fig.~\ref{fig:torusye_time} that represent $\yeeqmuavg$, i.e. $Y_e$ resulting if neutrinos were in thermodynamic equilibrium with vanishing chemical potential, bracket $\yeeqavg$ from above during almost the entire evolution.

\subsection{Outflow properties}\label{sec:outflow-properties}

\begin{table*}
  \centering
  \caption{Basic properties of the material ejected (i.e. reaching radii beyond $r=10^4\,$km) until the end of each simulation: ejecta mass, ejecta mass having $Y_e<0.25$, fraction of ejecta barely affected by weak interactions, average electron fraction and velocity of ejecta measured at $r=10^4\,$km, average entropy of ejecta measured for each trajectory at temperature $T=5\,$GK. Entries not in (in) parenthesis refer to models evolved with (without) neutrino absorption.}
  \label{table_ejecta}
  \begin{center}
  \begin{tabularx}{\textwidth}{lcccccc}
    %%% in Emacs use M-x align-current to align   
    \hline
    model                        & $m_{\mathrm{ej}}/m_{\mathrm{tor}}^0$ & $m_{\mathrm{ej}}^{Y_e<0.25}/m_{\mathrm{tor}}^0$ & $m_{\mathrm{ej}}^{\mathrm{inert}}/m_{\mathrm{ej}}$ & $\yeej (r=10^4\,\mathrm{km})$ & $v_{\mathrm{ej}}(r=10^4\,\mathrm{km})$ & $s_{\mathrm{ej}}(T=5\,\mathrm{GK})$ \\
    name                         & [$\%$]                               & [$\%$]                                          & [$\%$]                                             &                               & [$c$]                                    & [$k_b$/baryon]                      \\
    \hline
    m01M3A8(-no$\nu$)            & 22 (21)                              & 0.5 (10)                                        & 8.4  (9.2)                                         & 0.322   (0.276)               & 0.048   (0.043)                        & 26.8   (27.8)                       \\
    m1M3A8(-no$\nu$)             & 22 (17)                              & 0.1 (11)                                        & 5.3  (7.7)                                         & 0.333   (0.241)               & 0.045   (0.040)                        & 19.6   (22.1)                       \\
    m001M3A8(-no$\nu$)           & 28 (28)                              & 0   (0)                                         & 24   (24)                                          & 0.405   (0.403)               & 0.050   (0.048)                        & 37.3   (36.5)                       \\
    m01M5A8(-no$\nu$)            & 24 (23)                              & 0   (0)                                         & 14   (19)                                          & 0.358   (0.342)               & 0.045   (0.045)                        & 32.0   (30.2)                       \\
    m01M10A8(-no$\nu$)           & 27 (28)                              & 0   (0)                                         & 32   (37)                                          & 0.420   (0.421)               & 0.046   (0.044)                        & 39.1   (35.7)                       \\
    m01M3A4(-no$\nu$)            & 12 (11)                              & 1.5 (3.3)                                       & 21   (24)                                          & 0.325   (0.303)               & 0.037   (0.038)                        & 23.6   (24.9)                       \\
    m01M3A9(-no$\nu$)            & 28 (25)                              & 0   (9.2)                                       & 12   (9.1)                                         & 0.331   (0.280)               & 0.051   (0.051)                        & 27.4   (28.2)                       \\
    m01M3A8-$\alpha$03(-no$\nu$) & 19 (18)                              & 0   (7.2)                                       & 5.1  (5.8)                                         & 0.306   (0.276)               & 0.033   (0.032)                        & 27.2   (27.2)                       \\
    m01M3A8-$\alpha$1(-no$\nu$)  & 24 (22)                              & 3   (9.3)                                       & 13   (14)                                          & 0.338   (0.290)               & 0.057   (0.054)                        & 27.4   (28.5)                       \\
    m01M3A8-vis2(-no$\nu$)       & 31 (26)                              & 0   (0)                                         & 2.5  (2.9)                                         & 0.347   (0.337)               & 0.040   (0.048)                        & 30.5   (33.3)                       \\
    m01M3A8-mhd(-no$\nu$)        & 15 (14)                              & 1.7 (3.0)                                       & 2.5  (10.5)                                        & 0.304   (0.279)               & 0.125   (0.139)                        & 23.7   (23.6)                       \\
    m01M3A8-noQm(-no$\nu$)       & 21 (20)                              & 8.6 (17)                                        & 5.8  (11)                                          & 0.271   (0.241)               & 0.046   (0.042)                        & 28.4   (28.9)                       \\
    m01M3A8-wm                   & 23                                   & 0.6                                             & 8.2                                                & 0.318                         & 0.044                                  & 26.4                                \\
    m01M3A8-ye01(-no$\nu$)       & 22 (20)                              & 15  (16)                                        & 6.9  (14)                                          & 0.234   (0.203)               & 0.046   (0.047)                        & 26.7   (30.3)                       \\
    \hline
  \end{tabularx}
\end{center}
\end{table*}

\setlength{\tabcolsep}{6pt}
\begin{table*}
  \centering
  \caption{Quantities characterizing the nucleosynthesis yields in the ejecta and kilonova light curve: Mass fractions of lanthanides (with atomic charge number fulfilling $58\leq Z\leq71$), actinides (with atomic mass number fulfilling $232\leq A\leq 238$), 2nd-peak elements ($123\leq A\leq 137$), 3rd-peak elements ($A\geq 183$), ratio of 3rd-peak to 2nd-peak mass fractions normalized to the corresponding solar value, as well as time, luminosity, and temperature of the kilonova peak emission, where the peak is defined as the time when the luminosity becomes greater than the total heating rate powering the light curve. Kilonova properties have been obtained by assuming that all material at radii smaller than $10^4\,$km at the end of the simulation will become ejected. Entries not in (in) parenthesis refer to models evolved with (without) neutrino absorption.}
  \label{table_nuc}
  \begin{center}
  \begin{tabularx}{\textwidth}{lcccccccc}
    %%% in Emacs use M-x align-current to align   
    \hline
    model                        & $X_{\mathrm{lan}}$ & $X_{\mathrm{act}}$ & $X_{\mathrm{2nd}}$ & $X_{\mathrm{3rd}}$ & $[X_{\mathrm{3rd}}/X_{\mathrm{2nd}}]_\odot$ & $t_{\mathrm{peak}}$ & $L_{\mathrm{peak}}$ & $T_{\mathrm{peak}}$ \\
    name                         &                    &                    &                    &                    &                                             & [day]               & [$10^{39}$erg/s]    & [$10^3$K]           \\
    \hline                                                                                                                                
    m01M3A8(-no$\nu$)            & 0.032 (0.124)      & 2.9e-5 (2.1e-3)    & 0.280 (0.326)      & 8e-3  (0.118)      & 0.051 (0.628)                               & 2.00  (2.69)        & 5.30  (3.12)        & 2.70 (2.39)         \\
    m1M3A8(-no$\nu$)             & 3e-3  (0.093)      & 9.0e-8 (3.5e-3)    & 0.192 (0.287)      & 3e-4  (0.144)      & 0.003 (0.872)                               & 4.23  (7.32)        & 24.3  (9.68)        & 3.02 (2.15)         \\
    m001M3A8(-no$\nu$)           & 3e-3  (7e-4 )      & 2.4e-5 (1.8e-5)    & 0.025 (0.032)      & 8e-4  (4e-4 )      & 0.054 (0.021)                               & 0.54  (0.49)        & 9.02  (10.6)        & 5.81 (6.03)         \\
    m01M5A8(-no$\nu$)            & 8e-3  (0.021)      & 2.5e-5 (1.1e-5)    & 0.182 (0.237)      & 4e-3  (5e-3 )      & 0.039 (0.037)                               & 1.48  (1.72)        & 8.64  (6.57)        & 3.72 (3.20)         \\
    m01M10A8(-no$\nu$)           & 2e-4  (1e-4 )      & 4.8e-6 (4.8e-6)    & 0.011 (0.011)      & 2e-4  (0.000)      & 0.026 (8e-4 )                               & 0.70  (0.57)        & 35.5  (50.8)        & 6.67 (8.22)         \\
    m01M3A4(-no$\nu$)            & 0.055 (0.138)      & 4.5e-5 (2.3e-3)    & 0.310 (0.294)      & 0.017 (0.119)      & 0.096 (0.699)                               & 2.00  (2.44)        & 2.37  (1.72)        & 2.61 (2.33)         \\
    m01M3A9(-no$\nu$)            & 0.028 (0.098)      & 1.7e-5 (1.4e-3)    & 0.210 (0.270)      & 7e-3  (0.093)      & 0.056 (0.592)                               & 1.72  (2.56)        & 8.24  (4.16)        & 3.39 (2.46)         \\
    m01M3A8-$\alpha$03(-no$\nu$) & 0.030 (0.074)      & 4.5e-5 (7.4e-4)    & 0.248 (0.346)      & 0.011 (0.062)      & 0.075 (0.310)                               & 2.21  (2.98)        & 3.99  (2.69)        & 2.58 (2.55)         \\
    m01M3A8-$\alpha$1(-no$\nu$)  & 0.043 (0.111)      & 4.3e-5 (1.4e-3)    & 0.223 (0.267)      & 0.015 (0.114)      & 0.118 (0.736)                               & 2.00  (2.32)        & 5.33  (3.78)        & 2.66 (2.45)         \\
    m01M3A8-vis2(-no$\nu$)       & 8e-4  (3e-3 )      & 1.4e-5 (3.3e-5)    & 0.026 (0.078)      & 6e-4  (2e-3 )      & 0.038 (0.041)                               & 0.90  (1.15)        & 28.4  (17.6)        & 4.78 (3.93)         \\
    m01M3A8-mhd(-no$\nu$)        & 0.053 (0.080)      & 1.3e-3 (3.1e-3)    & 0.176 (0.146)      & 0.066 (0.125)      & 0.648 (1.485)                               & 1.15  (1.10)        & 15.6  (17.4)        & 3.46 (3.23)         \\
    m01M3A8-noQm(-no$\nu$)       & 0.157 (0.186)      & 1.7e-3 (7.9e-3)    & 0.337 (0.257)      & 0.144 (0.277)      & 0.739 (1.866)                               & 2.69  (2.98)        & 3.61  (2.89)        & 2.54 (2.29)         \\
    m01M3A8-wm                   & 0.043              & 3.9e-5             & 0.307              & 0.016              & 0.089                                       & 2.10                & 4.75                & 2.83                \\
    m01M3A8-ye01(-no$\nu$)       & 0.115 (0.199)      & 3.0e-3 (6.6e-3)    & 0.359              & 0.140 (0.244)      & 0.674 (1.343)                               & 2.44  (2.56)        & 3.88  (3.58)        & 2.51 (2.44)         \\
    \hline
  \end{tabularx}
\end{center}
\end{table*}

After having examined the evolution of $Y_e$ in the disks, we now investigate how neutrino absorptions influence the properties of the ejected material. Different types of outflows may be encountered in neutrino-cooled disks, e.g. neutrino-driven or viscous, or magnetohydrodynamically launched outflows \citep[e.g.][]{Fernandez2013b, Just2015a, Siegel2018c, Miller2020s}. In our set of rather low-mass models, viscous ejecta by far dominate neutrino-driven outflows \citep[e.g.][]{Just2015a}. For MHD models one might be able to distinguish material ejected through MHD-driven turbulent viscosity from ejecta expelled by means of other MHD related effects\footnote{We note that since our special relativistic MHD models cannot describe the general relativistic Blandford-Znajek process \citep{Blandford1977}, they do not exhibit electromagnetic jets along the polar axis.} \citep[see, e.g.,][]{Fernandez2019b}. However, finding suitable diagnostic criteria is not trivial and beyond the scope of our study, hence we do not discriminate between different ejecta components here.

\subsubsection{Ejecta masses}\label{sec:ejecta-masses}

Before considering the thermodynamic properties of the ejecta, we summarize our results concerning the ejecta masses: The total ejection efficiency, $m_{\mathrm{ej}}/m_{\mathrm{tor}}^0$, in the considered set of models (see Table~\ref{table_ejecta}) is always in the ballpark of $\sim 10-30\,\%$. These values are reduced by (at most) a few per cent in models where neutrino feedback is ignored, because more efficient neutrino cooling tends to counteract the ejection processes. Furthermore, our results are in broad agreement with previous studies \citep[e.g.][]{Fernandez2015a, Just2015a, Fernandez2020a, Fujibayashi2020a} concerning variations of $m_{\mathrm{tor}}^0, \MBH,$ and $\ABH$: $m_{\mathrm{ej}}/m_{\mathrm{tor}}^0$ decreases for more massive tori (mostly because those evolve longer in the neutrino-cooled phase), increases with BH spin (thanks to a reduction of the ISCO that favors mass ejection; see \citealp{Fernandez2015a}), and increases with BH mass (mainly because tori are less compact and therefore transition earlier into the adiabatic phase). We note that \citet{Fernandez2020a} report an opposite trend, namely decreasing ejecta masses with more massive BHs, but they also increase $\MBH/r_d$, i.e. the disk compactness (where $r_d$ is the radius of maximum density in the initial torus) together with $\MBH$, whereas we keep this quantity constant.

In the MHD models, the mass of material ejected until the final simulation times, $t_f$, is not finalized owing to computational limitations that have prevented us from continuing the simulation. The final ejecta mass can be estimated by assuming that all material at radii smaller than $r=10^4\,$km at $t=t_f$ will ultimately become ejected (cf. Tables~\ref{table_models} and~\ref{table_ejecta}), because BH accretion has basically ceased by $t=t_f$. This gives $m_{\mathrm{ej}}/m_{\mathrm{tor}}^0\approx 27.5\,\%$ and $29\,\%$ with and without neutrino absorption, respectively. The fact that our ejecta masses are lower than the $\approx 40\,\%$ reported by \citet{Fernandez2019b} may to some extent be explained by our weaker initial magnetic field \citep[such a tendency is supported by the results of][]{Christie2019a}, but could also be connected to our simplified treatment of GR in the form of the pseudo-Newtonian gravitational potential.

\subsubsection{Ejecta composition}\label{sec:aver-therm-prop}

\begin{figure*}
  \centering
  \includegraphics[width=0.99\textwidth]{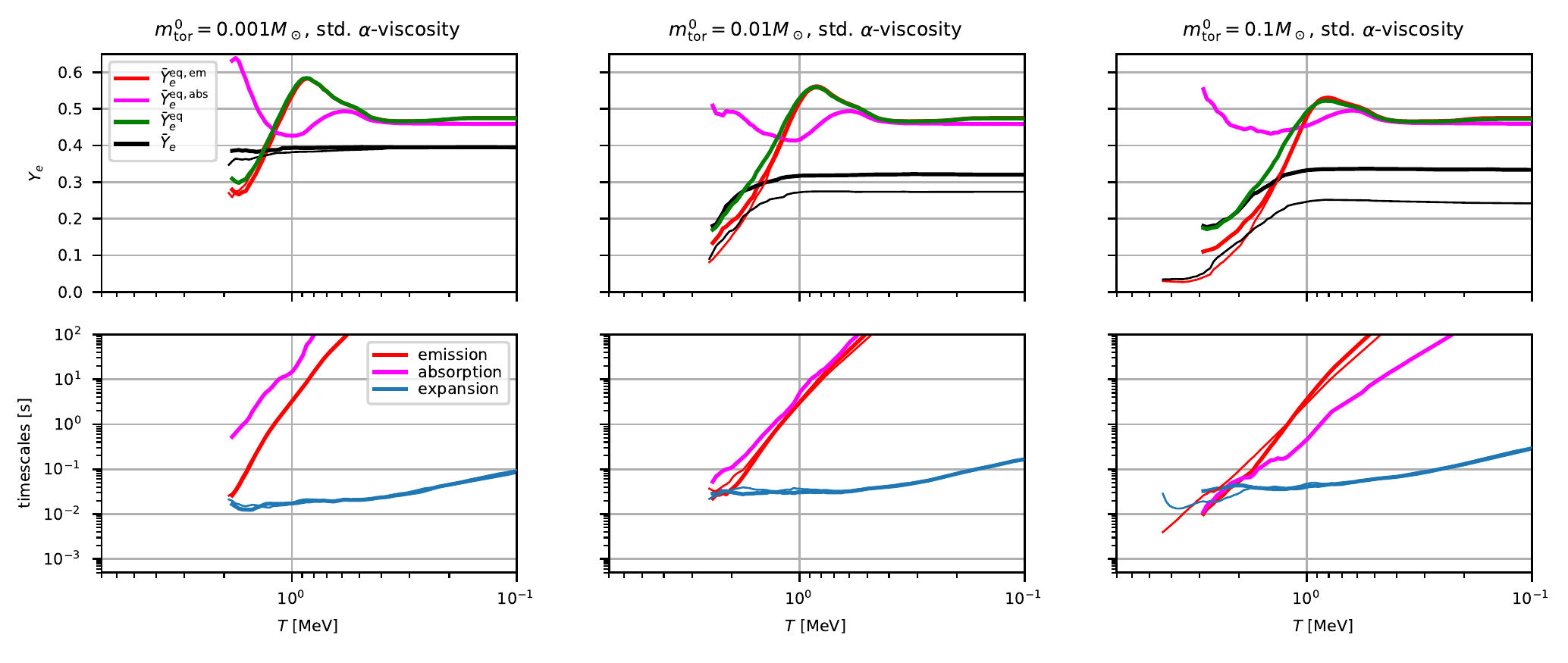}
  \includegraphics[width=0.66\textwidth]{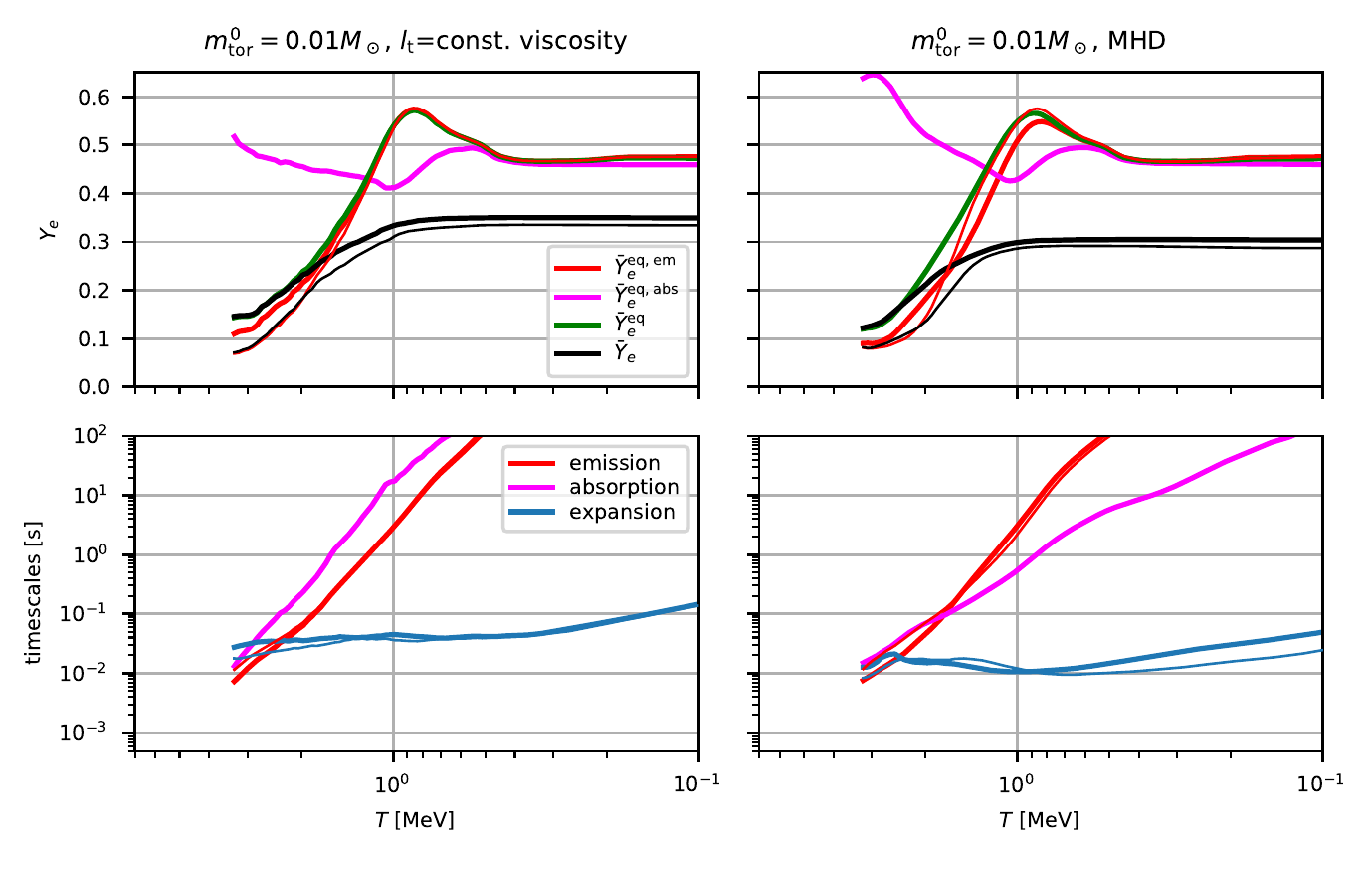}
  \caption{Averages of various quantities across the outflow trajectories at given temperatures for models m001M3A8, m01M3A8, and m1M3A8 (top row) as well as models m01M3A8-vis2 m01M3A8-mhd (bottom row). The top panels of each column depict averages of $Y_e$ (black lines) as well as its different limiting cases (see Sect.~\ref{sec:equil-cond-that} for the definitions). The bottom panels provide averages of the characteristic timescales of emission, $\bar\tau_{\mathrm{em}}$, absorption, $\bar\tau_{\mathrm{abs}}$, and expansion, $\bar{\tau}_{\mathrm{exp}}\equiv \bar{r}/\bar{v}$. On average, the effect of neutrino emission (absorption) starts to freeze out once $\bar\tau_{\mathrm{em}}$ ($\bar\tau_{\mathrm{abs}}$) becomes longer than $\bar{\tau}_{\mathrm{exp}}$. Averages are only computed for a given temperature if in terms of mass more than $50\,\%$ of the ejecta reach that temperature, hence $\bar{Y}_e$ does not start at the initial value of $Y_e^0=0.5$.}
  \label{fig:trajavg_mtorus}
\end{figure*}	

Next we take a look at the electron fraction of the ejecta. Although most features discussed in this section are generic for all models, we focus first on the model sequence with increasing torus mass, i.e. m001M3A8, m01M3A8, and m1M3A8, representing cases with negligible, modest, and significant impact of $\nu$-absorption, respectively. The dependence on other model parameters and modeling ingredients is addressed in the next section. Panels~(a) and~(m) in Fig.~\ref{fig:globdat_mtorus} show the mass-loss rate and the spherically averaged $Y_e$, respectively, measured at a fixed radius of $r=10^4\,$km. For the sequence of models with increasing torus mass and employing the standard $\alpha$-viscosity a reduction of the ejecta $Y_e$ in models without compared to models with absorption is found analogous to what was found for the torus $Y_e$ in Sect.~\ref{sec:impact-absorpt-torus}.

Since the extraction radius of $r=10^4\,$km is far away from the torus, properties measured at this radius cannot inform about the times when outflow material actually attained its final $Y_e$. This information is provided in the bottom panels of Fig.~\ref{fig:torusye_time}, where the fraction of ejected material with already finalized $Y_e$ values is plotted as function of time  ($m^{\mathrm{FO}}/m_{\mathrm{ej}}$) as well as the instantaneous $Y_e$ value of material freezing out at time $t$ ($Y_e^{\mathrm{FO}}$). In these plots we identify material for which $Y_e$ has frozen out by the criterion $\tau_\beta>10\,$s for the weak interaction timescale $\tau_\beta$. The figures reveal that material freezes out over an extended period of time, which is roughly centered around $t_{\mathrm{em}}$, and that freeze-out material is more neutron-rich for models without neutrino absorption even well beyond the time $t_{\mathrm{abs}}$ at which neutrino absorption becomes irrelevant. This last observation implies that ejecta trajectories can carry to a certain degree memory of the conditions prevalent when neutrino absorptions were still active, even if this material is expelled later and via viscous mechanisms that are completely unrelated to neutrino-absorption. We remark that \citet{Miller2020s} come to a similar conclusion in their investigation of neutrino-related effects in MHD disks.

More insight about the $Y_e$ evolution of outflow particles along their journey from deep inside the torus until free expansion can be gained by considering Fig.~\ref{fig:trajavg_mtorus}, where, as functions of temperature (with inverted $x$-axis to resemble the time evolution), the average $Y_e$ is plotted together with its equilibria (top row) as well as the characteristic timescales (bottom row) of neutrino absorption (magenta lines), emission (red lines), and expansion (blue lines). This analysis is inspired by a related one performed recently by \citet{Fujibayashi2020b} to investigate NS remnants of NS-NS mergers. In order to obtain an average $\bar{X}$ of a quantity $X$ at fixed temperature, we first map $X$ for each trajectory onto a temperature grid and then compute a mass-weighted average of $X$ if, and only if, for a given temperature more than $50\,\%$ of ejecta (in terms of mass) can be found. For the temperature mapping we only take into account the point connected to the latest evolution time if a certain temperature is reached multiple times (e.g. due to turnover motions). The characteristic timescales plotted in the lower panels of each row in Fig.~\ref{fig:trajavg_mtorus} are obtained by plugging previously computed quantities defined at fixed temperature into the defining equations, Eqs.~(\ref{eq:tauem}),~(\ref{eq:tauabs}), and $\tau_{\mathrm{expansion}}=\bar{r}/\bar{v}$ for emission, absorption, and expansion, respectively.

The results in Fig.~\ref{fig:trajavg_mtorus} corroborate our previous interpretation concerning the (equilibrium) conditions of the torus $Y_e$: At high temperatures, i.e. deep inside the torus (region B in Fig.~\ref{fig:evolphases}), the soon-to-be ejected outflow material is close to the respective equilibrium, $Y_e\rightarrow \yeeq>\yeeqem$  ($Y_e\rightarrow\yeeqem$) for models with (without) neutrino feedback. Soon after the ejection sets in and material leaves the densest layers of the torus, the expansion timescale becomes the shortest of the three timescales, which on average happens at temperatures $T\approx 2$\,MeV. At this point, $\bar{Y}_e$, starts to diverge from its equilibrium value, which continues to rise as a result of decreasing densities and temperatures. After a few more neutrino interactions and at slightly lower temperatures, $\bar{Y}_e$ then finally levels off to remain constant.

While the qualitative behavior of $\bar{Y}_e$ is similar for both types of models, with and without absorption, quantitative differences appear that are all the more pronounced for more massive tori. The reasons are in agreement with the findings of Sect.~\ref{sec:impact-absorpt-torus}: First, the starting values of $Y_e$ right before ejection are higher to begin with for more massive tori, namely about $\yeeqbar-\yeeqembar \sim 0.05-0.1$. Second, as can be deduced from the plotted timescales of emission and absorption, neutrino absorptions provide an additional boost in raising $\bar{Y}_e$ also after $\bar{Y}_e$ starts to decouple from $\bar{Y}_{e}^{\mathrm{eq}}$. In model m1M3A8, the absorption rates even dominate emission rates during this phase, indicating that a significant fraction of the ejecta have experienced conditions as in region A of Fig.~\ref{fig:evolphases} during their expansion. The third and last reason for elevated $Y_e$ values in models with absorption is that $\yeeqem$ itself is already higher as a result of finite diffusion timescales out of the torus and therefore less efficient neutrino cooling compared to the no-absorption case.

The top row of Fig.~\ref{fig:histnuc} shows the ejecta mass distribution in $Y_e$, measured at $r=10^4\,$km, as well as the abundance patterns of nucleosynthesis yields produced in the ejecta for the three models with increasing torus mass. In agreement with many existing results for viscous models of neutrino-cooled disks \citep[e.g.][]{Just2015a, Wu2017a, Fujibayashi2020a}, most models show a peak close to $Y_e\sim 0.2-0.3$, whereas the left boundary of the $Y_e$ distribution is systematically shifted to lower values for models without neutrino absorption, more so for higher torus masses. Since the production of lanthanides (and heavier elements) is typically activated for $Y_e$ near $0.25$ \citep[e.g.][]{Kasen2015}, the nucleosynthesis pattern of the heavier elements (with mass numbers $A\ga 130$) is accordingly quite sensitive to any modeling variation that induces changes in the vicinity of $Y_e=0.2-0.3$. It is therefore not surprising to observe a modest (substantial) boost in heavy-element and, in particular, lanthanide production when ignoring neutrino absorption for a torus mass of 0.01\,$\Msol$ (0.1\,$\Msol$). Remarkably, in both cases a nearly perfect solar-like abundance pattern (shown as circles in Fig.~\ref{fig:histnuc}) is obtained only when neglecting neutrino absorption. Another interesting observation is that when including neutrino absorption, and only when doing so, the efficiency of heavy-element production grows non-monotonically with torus mass: The 0.01\,$\Msol$ torus in the fiducial model m01M3A8 produces more heavy elements relative to its initial mass than both the lighter and heavier tori in models m001M3A8 and m1M3A8, respectively. We note that \citet{Just2015a, Fernandez2020a, Fujibayashi2020a} report a consistent trend for growing torus masses.

\subsubsection{Model dependence}\label{sec:model-dependence}

\begin{figure*}
  \newcommand\wid{0.45}
  \newcommand\hei{0.14}
  \centering
  \includegraphics[trim=5 8 5 5,clip,width=\wid\textwidth,height=\hei\textheight]{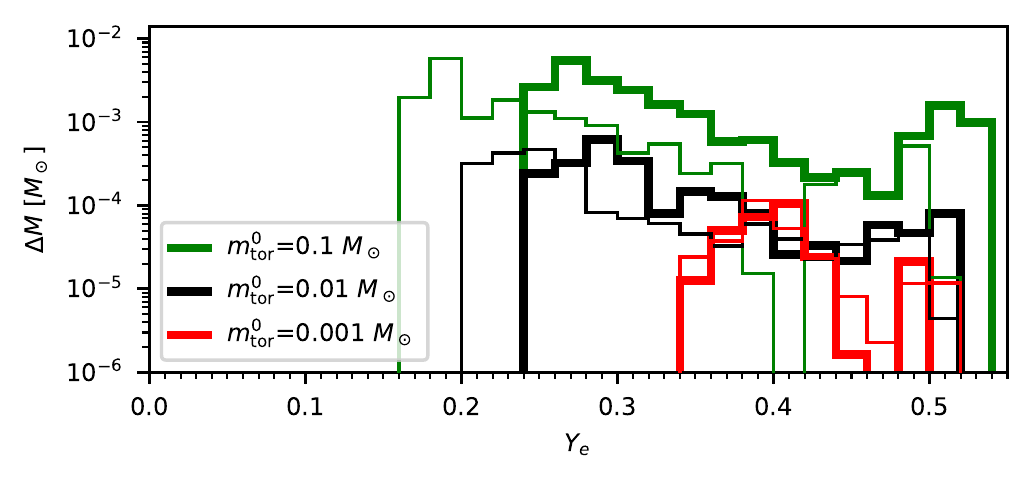}
  \qquad
  \includegraphics[trim=50 100 40 100,clip,width=\wid\textwidth,height=\hei\textheight]{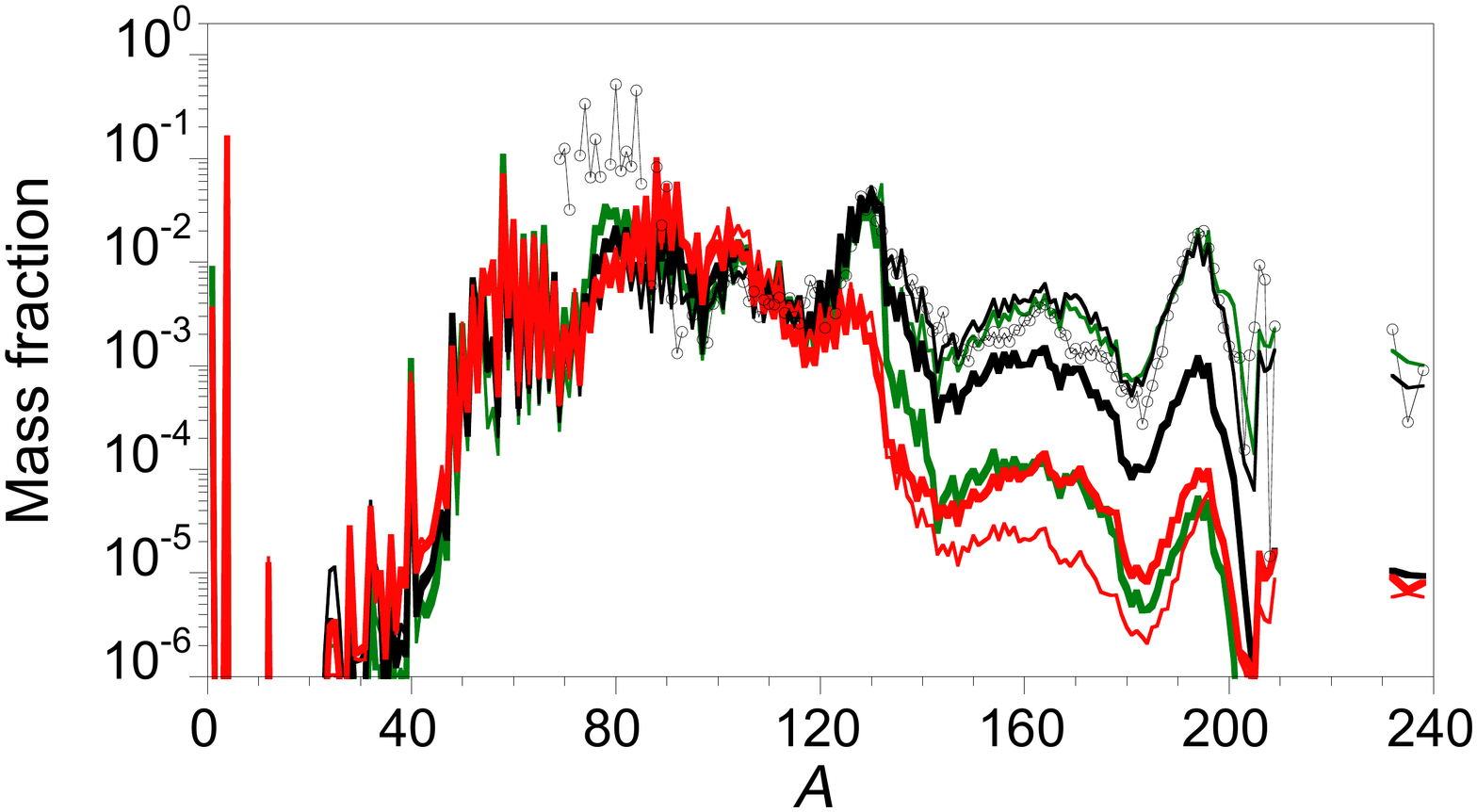}

  \includegraphics[trim=5 8 5 5,clip,width=\wid\textwidth,height=\hei\textheight]{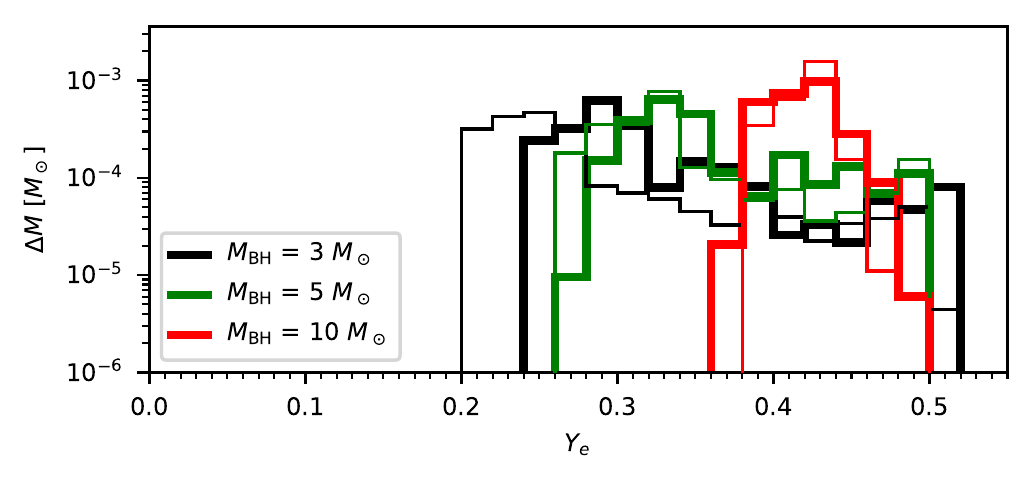}
  \qquad
  \includegraphics[trim=50 100 40 100,clip,width=\wid\textwidth,height=\hei\textheight]{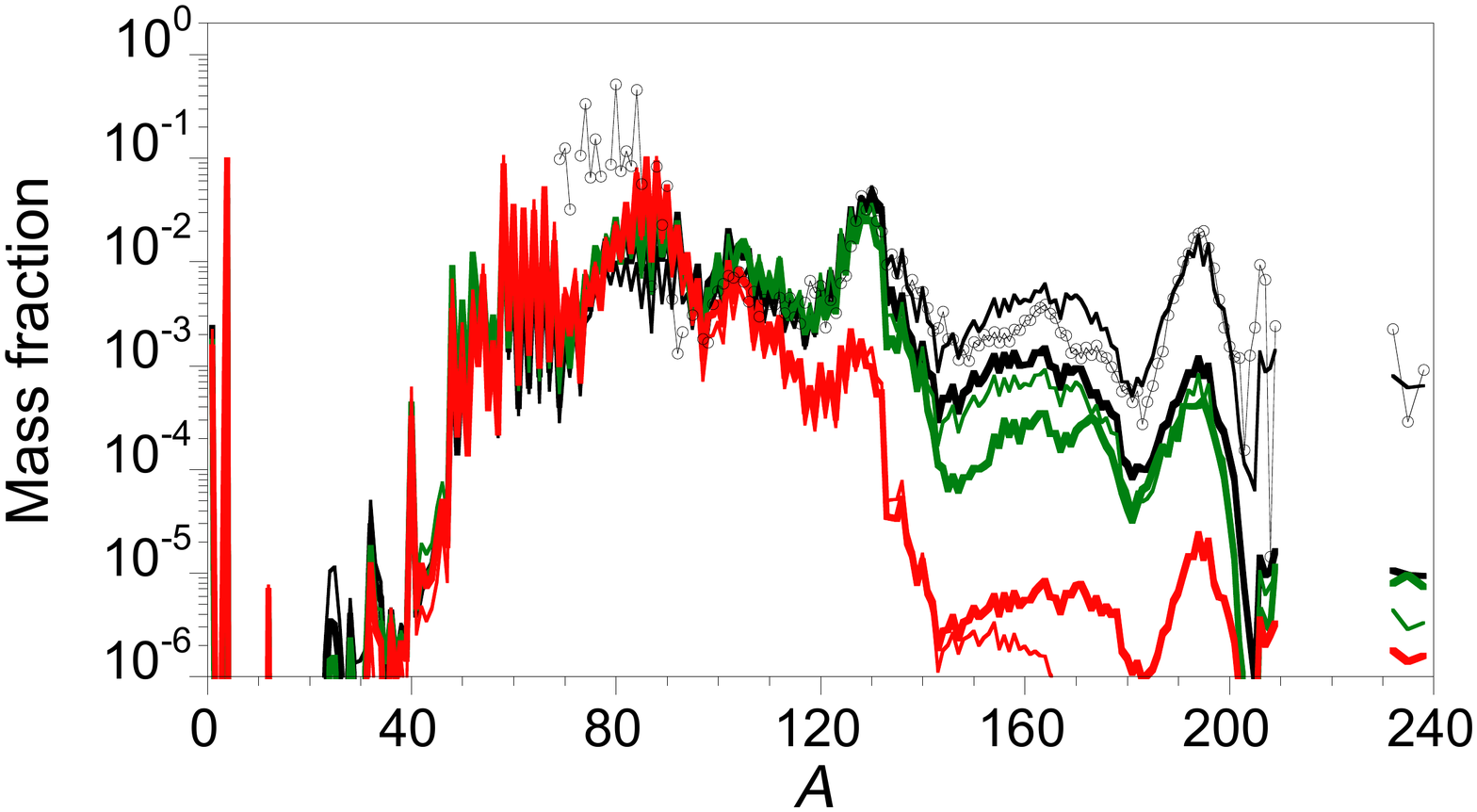}

  \includegraphics[trim=5 8 5 5,clip,width=\wid\textwidth,height=\hei\textheight]{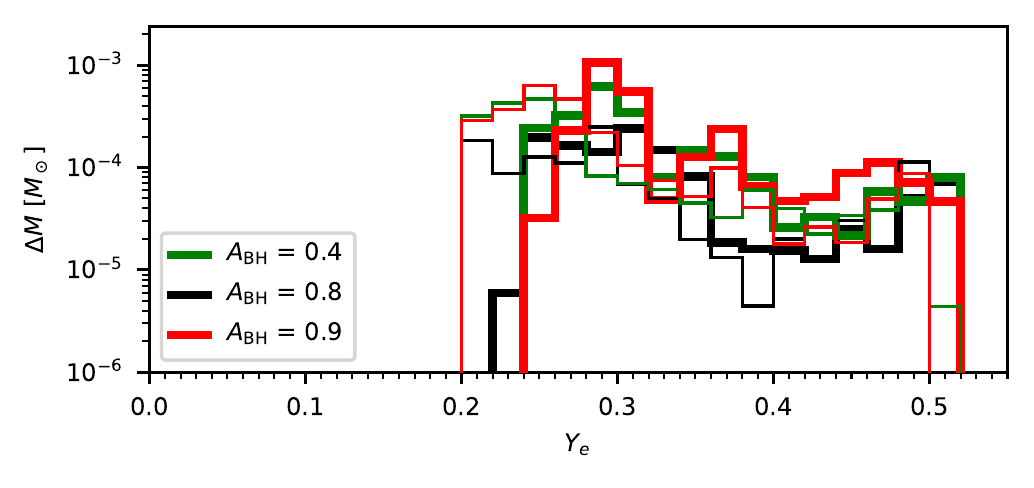}
  \qquad
  \includegraphics[trim=50 100 40 100,clip,width=\wid\textwidth,height=\hei\textheight]{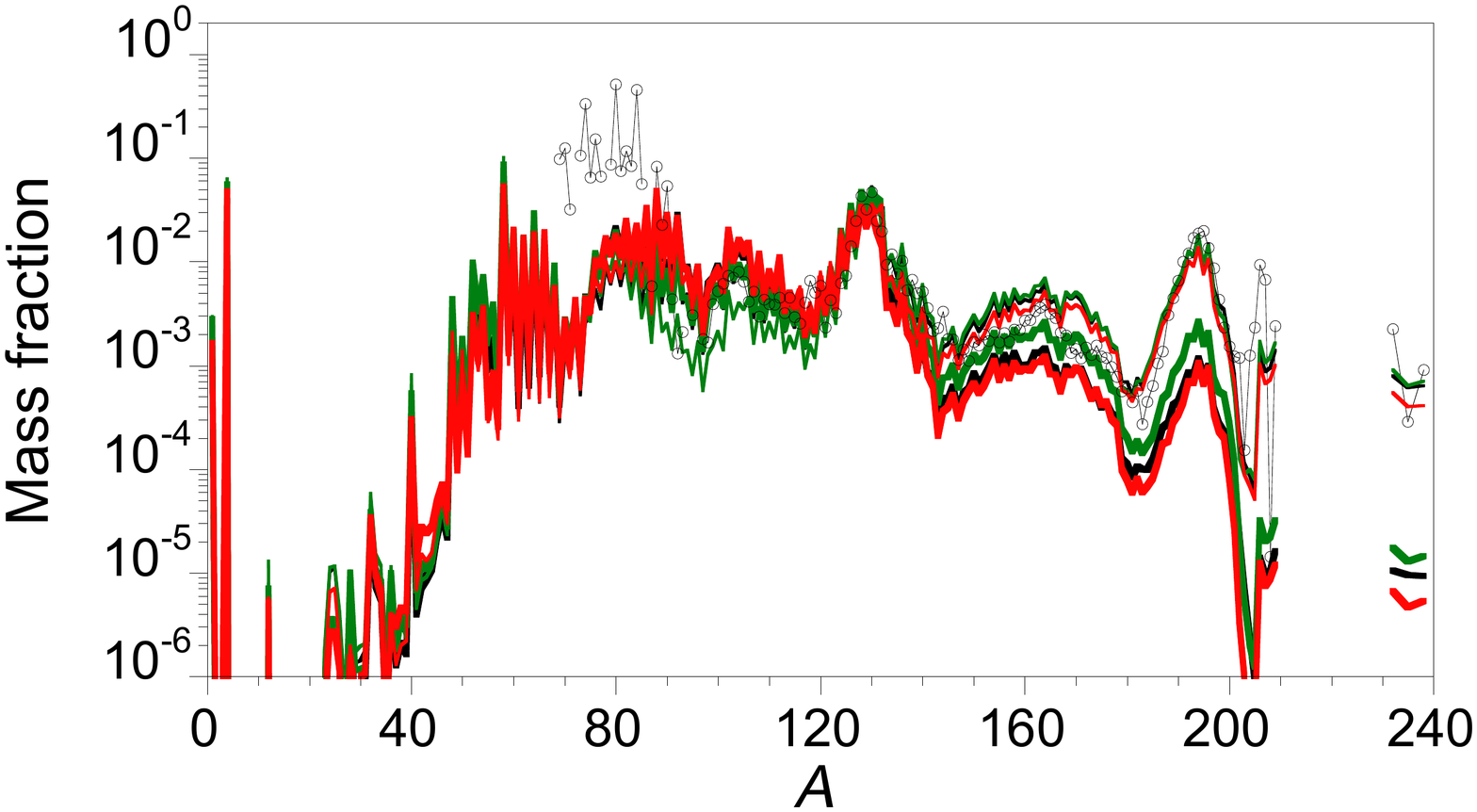}

  \includegraphics[trim=5 8 5 5,clip,width=\wid\textwidth,height=\hei\textheight]{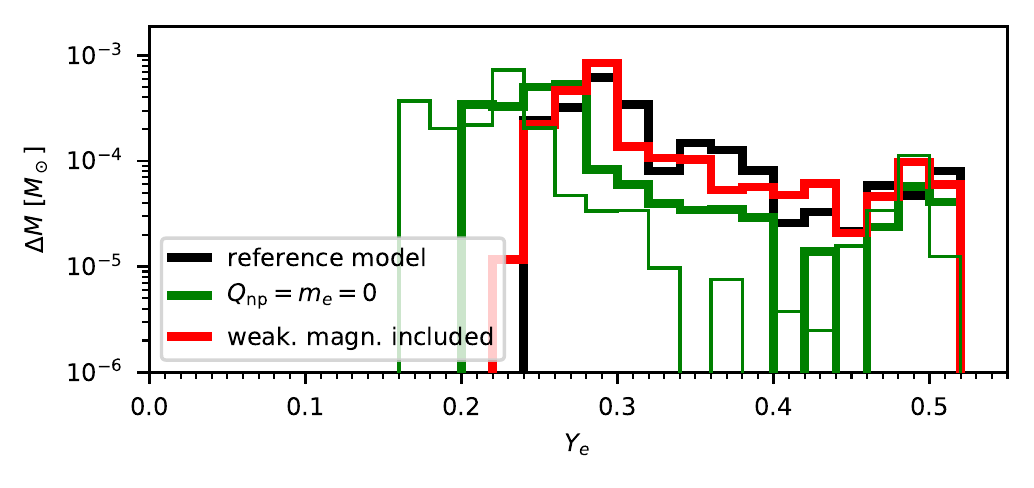}
  \qquad
  \includegraphics[trim=50 100 40 100,clip,width=\wid\textwidth,height=\hei\textheight]{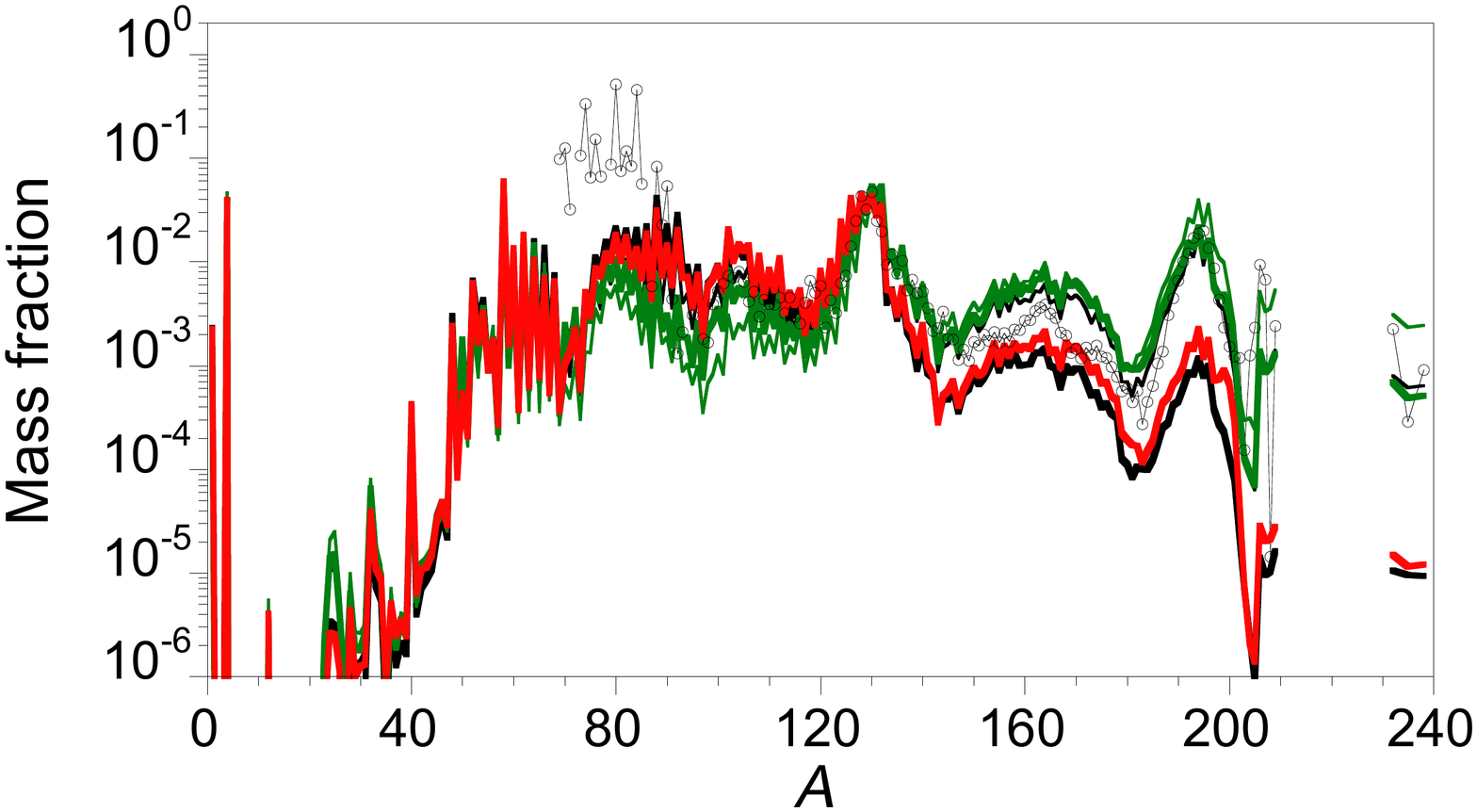}  

  \includegraphics[trim=5 8 5 5,clip,width=\wid\textwidth,height=\hei\textheight]{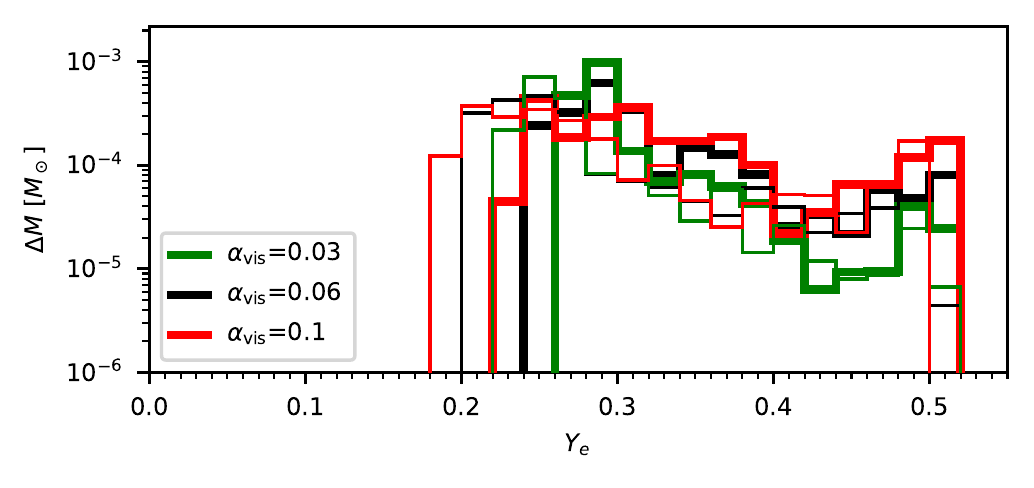}
  \qquad
  \includegraphics[trim=50 100 40 100,clip,width=\wid\textwidth,height=\hei\textheight]{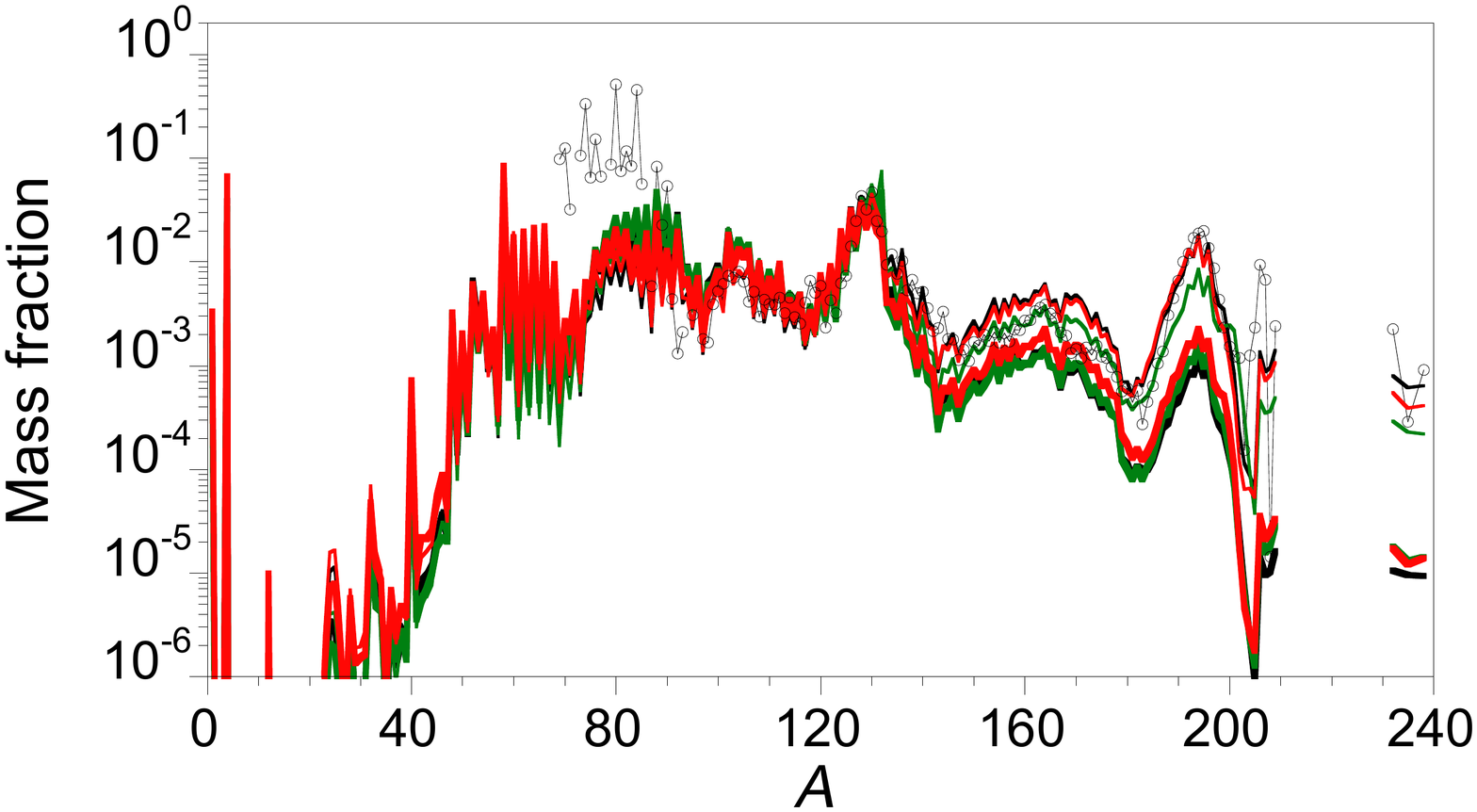}
  
  \includegraphics[trim=5 8 5 5,clip,width=\wid\textwidth,height=\hei\textheight]{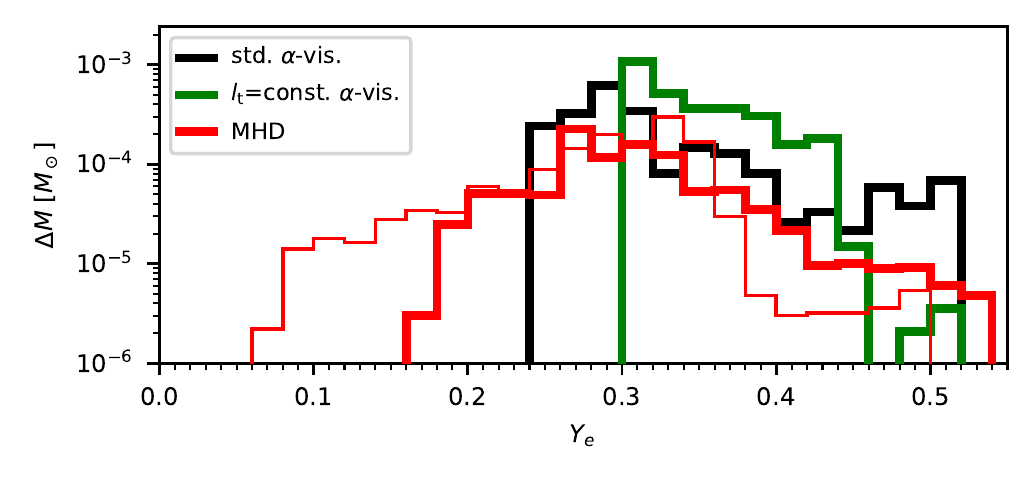}
  \qquad
  \includegraphics[trim=50 100 40 100,clip,width=\wid\textwidth,height=\hei\textheight]{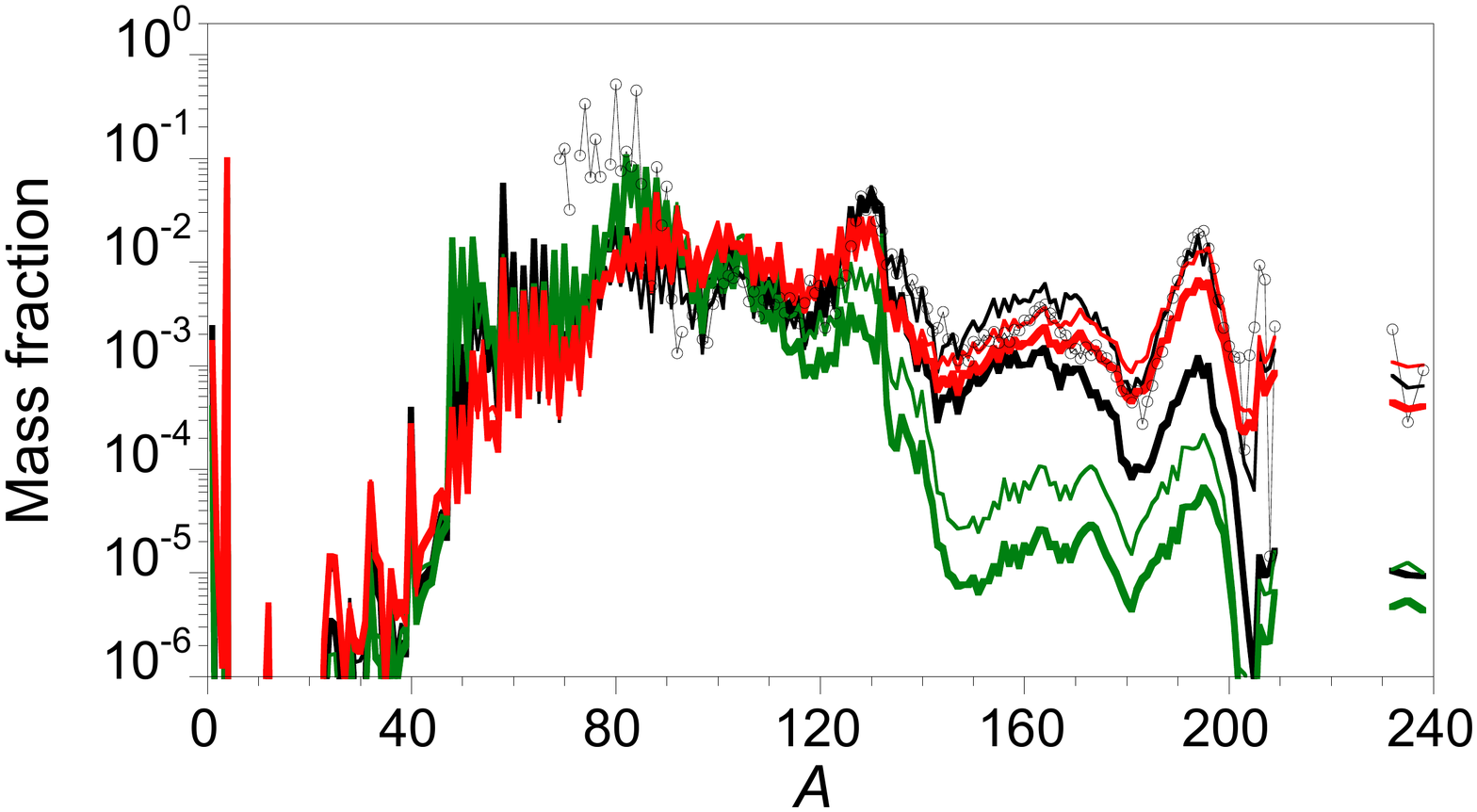}  

  \caption{\emph{Left column:} Histograms for the mass distribution versus electron fraction $Y_e$ as measured at radii of $10^4\,$km in the hydrodynamic simulations. \emph{Right column:} Corresponding abundance distributions of nuclei synthesized in the ejecta as function of mass number, $A$. The colors refer to the same models that are plotted on the left. Mass fractions corresponding to the models are normalized to sum up to unity, while the solar abundance pattern (depicted by open circles) is normalized to the $A=130$ mass fraction of model m01m3A8. In all panels the thick (thin) lines are used for models including (neglecting) neutrino absorption. The black lines always refer to the same, fiducial model, m01M3A8(-no$\nu$). From top to bottom (only) the following ingredients are varied with respect to those of the fiducial model: Initial torus mass, black hole mass, black hole spin, viscous $\alpha$ parameter, neutrino interaction physics ($Q_{np}$ and $m_e$ corrections (green lines) and weak magnetism correction (red lines)), treatment of turbulent viscosity ($\lturb$=const. viscosity (green lines), and MHD (red lines)).}
  \label{fig:histnuc}
\end{figure*}	

\begin{figure*}
  \centering
  \includegraphics[width=0.99\textwidth]{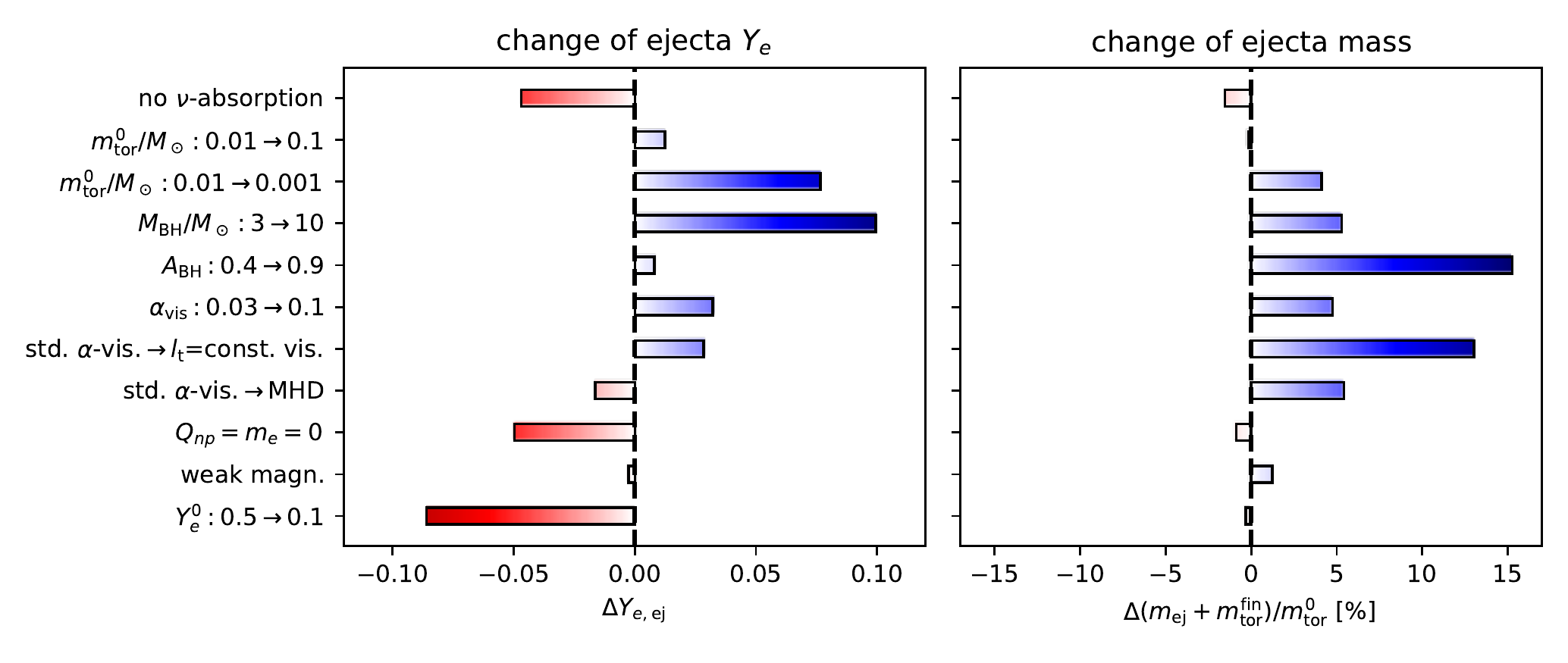}
  \caption{Bar diagram visualizing the absolute change of the average ejecta $Y_e$ (left panel) and the final ejecta mass (right panel) with variation of the indicated model parameters and modeling assumptions. Differences are measured with respect to the fiducial model, m01M3A8 (with initial properties of $m_{\rm tor}^0=0.01\,M_\odot$, $M_{\rm BH}=3\,M_\odot$, $A_{\rm BH}=0.8$, $Y_e(t=0)=0.5$, $\alpha_{\rm vis}=0.06$, and standard viscosity treatment, as well as ejecta properties of $\{Y_{e,\mathrm{ej}},(m_{\mathrm{ej}}+m_{\mathrm{tor}}^{\mathrm{fin}})/m_{\mathrm{tor}}^0\} = \{ 0.322, 22\,\%\}$), except for the variation of the BH spin and $\alpvis$, where the reference models are m01M3A4 (i.e. with $A_{\rm BH}=0.4$ instead of 0.8 and $\{Y_{e,\mathrm{ej}},(m_{\mathrm{ej}}+m_{\mathrm{tor}}^{\mathrm{fin}})/m_{\mathrm{tor}}^0\}=\{ 0.325, 12\,\%\}$) and m01M3A8-$\alpha$03 (i.e. $\alpha_{\rm vis}=0.03$ instead of $0.06$ and $\{Y_{e,\mathrm{ej}},(m_{\mathrm{ej}}+m_{\mathrm{tor}}^{\mathrm{fin}})/m_{\mathrm{tor}}^0\}=\{ 0.306, 19\,\%\}$), respectively.}
  \label{fig:reldif}
\end{figure*}

We now analyze the sensitivity of the ejecta properties and the impact of absorption to the variation of model parameters or modeling assumptions. Global properties of the ejecta can be found for all models in Table~\ref{table_ejecta}, namely the total mass, $m_{\mathrm{ej}}$, average electron fraction, $\yeej$, velocity, $v_{\mathrm{ej}}$, and entropy per baryon, $s_{\mathrm{ej}}$. Table~\ref{table_nuc} provides mass fractions of characteristic classes of elements, e.g. lanthanides and actinides. Some of the results of Tables~\ref{table_weak},~\ref{table_ejecta}, and~\ref{table_nuc} are visualized in Fig.~\ref{fig:tabdat} as function of torus optical depth and in Fig.~\ref{fig:reldif} as differences relative to the fiducial model.

\paragraph*{Disk mass.}

In the previous sections we already analyzed the effects related to neutrino absorption based on three models with initial disk masses of $\mtor^0=0.001$, $0.01$, and $0.1\,M_\odot$, which are representative of cases with small, medium, and significant impact of neutrino absorption. We will now briefly comment on the outcome for cases outside of this range of typical masses. In the rather trivial case of very low disk masses the temperatures are likely to be too low, and correspondingly the weak-interaction timescales too long, for the inner disk to reach weak equilibrium, $Y_e\rightarrow \yeeqem$. Hence, the values of $Y_e$ are effectively given by the initial condition, i.e. by $Y_e\ll 0.5$ and $Y_e\approx 0.5$ for disks in compact-object mergers and collapsars, respectively. In the oppposite case of very massive disks with correspondingly higher optical depths -- $\mtor^0\approx 0.2-0.4$ can be reached by NS-BH mergers with a combination of a high BH spin, low BH mass, and stiff nuclear EOS \citep[see, e.g.,][]{Foucart2012f} -- the extrapolation is not quite as straightforward. The trend of increasing values of $\yeeqavg$,  $\yeej$, and reduced yields of heavy r-process material seen for disk masses above $\sim 0.01\,M_\odot$ (see, e.g., Fig.~\ref{fig:tabdat}) is likely to continue. Additionally, neutrino-driven winds with typical electron fractions of $Y_e\sim 0.5$ will contribute more significantly to the ejecta and, hence, lead to a further increase of $\yeej$ \citep[e.g.][]{Just2015a}. Moreover, optically thick disks exhibit a more pronounced initial phase of inefficient neutrino cooling. Since inefficiently cooled accretion flows eject matter more readily than efficiently cooled ones (see, e.g., $m_{\rm ej}$ in Table~\ref{table_ejecta} for the m1M3A8 models with and without neutrino absorption for two models demonstrating this trend), very massive disks may exhibit systematically enhanced matter ejection efficiencies, $m_{\rm ej}/\mtor^0$, compared to disks that are only marginally optically thick.

\paragraph*{Black hole properties.}

When increasing the BH mass while keeping $r_d\propto \MBH$ (where $r_d$ is the radius at maximum density of the initial disk) the densities in the initial disk decrease. As a result, the temperatures are lower for more massive BHs than for less massive BHs, and therefore neutrino emission favors an equilibrium with higher electron fractions $\yeeqem$ (cf. Fig.~\ref{fig:yeeq1}). Moreover, due to the reduced optical depth the overall impact of neutrino absorption becomes less significant. Roughly speaking, increasing $\MBH$ has a similar effect on the $Y_e$ evolution as decreasing the torus mass.

On the other hand, varying the spin parameter of the BH between $0.4<\ABH<0.9$ (while keeping $r_d$=const.) appears to result in relatively small differences concerning the $Y_e$ evolution, which is likely related to the fact that the spin parameter does not lead to dramatically different thermodynamic conditions and optical depths in the torus. We note, however, that the sensitivity could be more significant for more massive tori, in which a larger fraction of the ejecta is launched as neutrino-driven winds \citep{Fernandez2015a}.

We caution the reader that since our approximate initial disk models are constructed by hand in a way to reproduce the most basic global parameters (i.e. $\mtor^0$, $\MBH$, and $\ABH$, as well as initial electron fraction and entropy), the local configuration does not perfectly agree with actual post-merger disks or disks in collapsar engines that exhibit the same values of $\mtor^0$, $\MBH$, and $\ABH$. For instance, as pointed out by \citet{Fernandez2020a}, the initial size of the disk, approximately measured by the radius of the density maximum $r_d$, represents an additional relevant degree of freedom characterizing the disk distribution\footnote{We are not aware of a study systematically discussing $r_d$ as a function of $\MBH$. For NS-BH mergers $r_d$ is likely correlated with the tidal disruption radius, $r_{\rm tide}$, for which $r_{\rm tide}/\MBH$ typically decreases with higher BH mass \citep[see, e.g.,][]{Lattimer1976, Foucart2018k}. We chose $r_d \propto \MBH$ here to keep our study simple and without specifically restricting to either case, NS-BH merger remnants or collapsars.}. Hence, some of the tendencies as function of $\MBH$ and $\ABH$ may depart from those found here when considering different initial disk distributions.

\paragraph*{Finite-mass corrections and weak magnetism.}

In models m01M3A8-noQm(-no$\nu$) we set the electron mass $m_e=0$ as well as the neutron-proton mass difference $Q_{np}=0$ in the $\beta$-reaction rates, cf. Eqs.~(\ref{eq:betarates}). In Sect.~\ref{sec:how-sens-y_em} we already found that this simplification can underestimate the electron fractions corresponding to emission equilibrium, $\yeeqem$, by up to $\sim 0.1$ particularly in the $\rho-T$ regime covered by the torus during the neutrino-dominated phase of evolution. It is therefore not surprising to find more neutron-rich ejecta for these models, in which $\yeej$ is reduced by about 0.05. The nucleosynthesis pattern matches very well the solar abundance pattern. Incidentally, for the fiducial model the noQm simplification has a similar net effect as ignoring neutrino absorption.

Considering now model m01M3A8-wm, in which weak magnetism is taken into account in contrast to the fiducial model m01M3A8, the differences appear to be minor and the reduction of the ejecta $Y_e$ is only 0.004. The small size of the impact of weak magnetism was anticipated already in Sect.~\ref{sec:how-sens-y_em} and can be ascribed to the relatively low neutrino energies involved in the disk evolution.

\paragraph*{$l_t$=const. viscosity treatment.}

\begin{figure}
  \centering
  \includegraphics[width=0.48\textwidth]{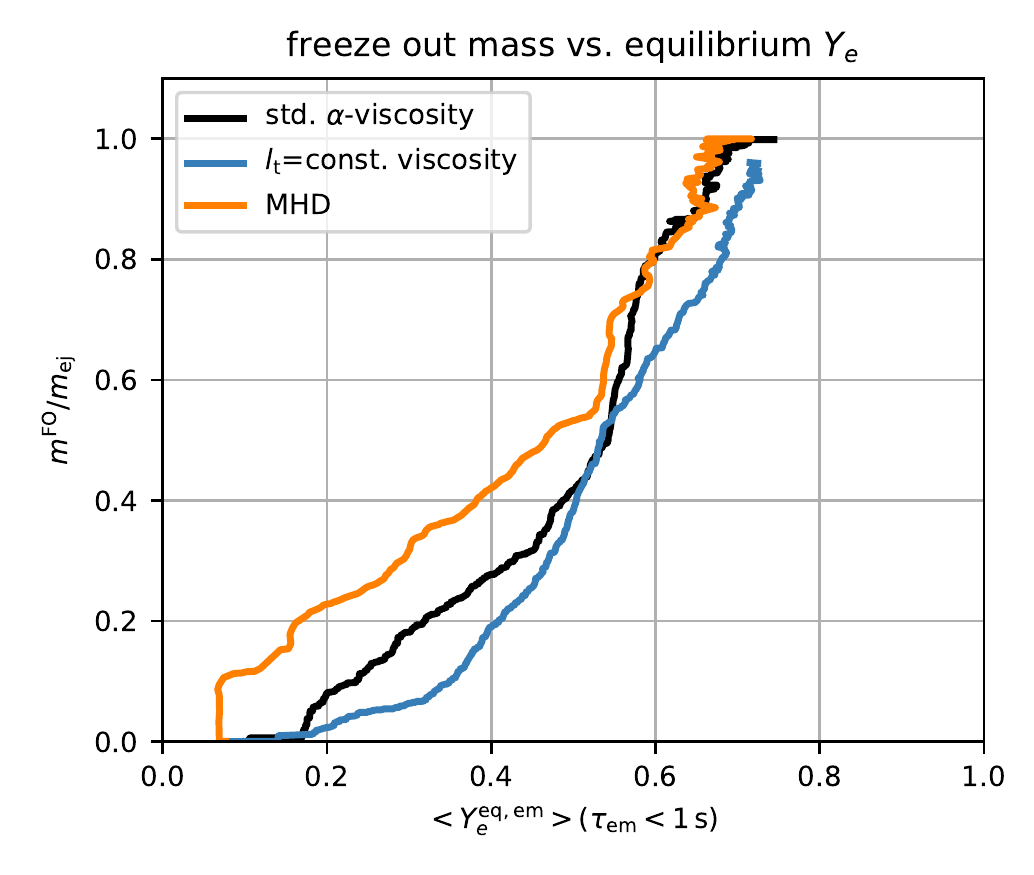}
  \caption{Total mass of all trajectories for which $Y_e$ is frozen out (i.e. with weak interaction  timescales $\tau_\beta>10\,$s) as function of the equilibrium $Y_e$ averaged over regions with efficient neutrino emission (i.e. where $\tauem<1\,$s) for three models differing only by the treatment of angular momentum transport, m01M3A8 (black), m01M3A8-vis2 (blue), and m01M3A8-mhd (orange). In the MHD model more material is able to freeze out while the bulk of the torus still has low $Y_e$, whereas in the model with $\lturb$=const. viscosity efficient freeze out only takes place at a higher torus $Y_e$.}
  \label{fig:myfreezeout}
\end{figure}	

Probably the most challenging modeling ingredient, apart from neutrino transport, is connected to the description of turbulent angular momentum transport. For the purpose of assessing how the $Y_e$ evolution and the impact of absorption depends on the treatment of angular momentum transport, we did not just vary the viscous $\alpha$-parameter, but also included models with two different viscosity prescriptions, Eqs.~(\ref{eq:viseta1})~and~(\ref{eq:viseta2}), as well as a 3D MHD model. The first and quite remarkable observation when comparing the $Y_e$ distributions and abundance patterns is that the ejecta properties appear to be less sensitive to varying the viscous $\alpha$-parameter than they are to the choice of the turbulent viscosity scheme (i.e. standard $\alpha$-viscosity vs. $\lturb$=const. viscosity vs. MHD). This result supports the notion that each viscosity scheme generates its own characteristic flow pattern with correspondingly different ejecta properties. In general, this notion implies that it might be impossible to reproduce all ejecta features of MHD models by just varying the $\alpha$-parameter in a certain viscosity scheme. \citet{Fernandez2019b} come to a similar conclusion when comparing viscous models with an MHD model.

We first discuss the alternative viscosity prescription expressed by Eq.~(\ref{eq:viseta2}). First of all, not only the average $Y_e$ is enhanced compared to the standard $\alpha$-viscosity model, but, probably even more importantly, also the bottom value in the $Y_e$--mass distribution, which lies at about $Y_e\approx 0.3$. As a consequence, for this model the abundances of elements with $A>130$ are dramatically reduced. Moreover, the overall difference between models with and without absorption is relatively small compared to the standard viscosity. Why is that? As already mentioned in Sect.~\ref{sec:diff-visc-prescr} the effective strength of the $\lturb$=const. viscosity employed in model m01M3A8-vis2 is reduced at later times compared to the case m01M3A8 using the standard $\alpha$-viscosity. This entails a slower and more gradual freeze out and emission of outflow material (cp. middle panels in Fig.~\ref{fig:torusye_time} with left panels in Fig.~\ref{fig:torusye_time_vis}). However, since neutrino absorptions become inefficient roughly around the same time in both models ($t_{\mathrm{abs}}= 75\,$ms and 60\,ms in the vis2 model and the reference model, respectively) the overall relevance of absorptions is naturally expected to be smaller in the vis2 model. This aspect, namely that more time is spent in a state where only neutrino emission is efficient, is one reason for the low sensitivity of neutrino absorption in this model. Another, related reason is connected by the circumstance that the equilibrium value for neutrino emission, $\yeeqem$, is time-dependent and grows from $\yeeqem\sim 0.1$ to $\yeeqem\sim \yeeqabs$ during the disk evolution, whereas neutrino absorption pushes $Y_e$ almost at all times towards $\yeeqabs\sim 0.5$. In other words, for fluid elements that freeze out at a very late stage of evolution it does not matter whether neutrino absorption is still efficient or not, because $Y_e$ in any case saturates at some high value $\ga 0.4-0.5$.

Comparing properties at the same evolution time when analyzing time-dependent features of the $Y_e$ evolution for models with different viscosity may be misleading because of the different viscous accretion timescales of each model. For that reason we plot in Fig.~\ref{fig:myfreezeout} the cumulative freeze-out masses, $m^{\mathrm{FO}}$, of material that becomes ejected as a function of the average equilibrium $Y_e$ of the torus, $\yeeqemavg$, i.e. $\yeeqem$ averaged over the region where $\tauem<1\,$s. Since $\yeeqemavg$ grows monotonically, except at very early and very late times, this quantity is a more suitable measure of the weak-interaction regime of the torus than the time coordinate. Looking at Fig.~\ref{fig:myfreezeout} we find additional support for the arguments of the previous paragraph: In the $\lturb$=const. model the freeze out takes place at systematically higher values of $\yeeqemavg$ compared to both alternative descriptions of angular momentum transport. 

\paragraph*{MHD treatment.}

Before looking at the results of the MHD models, we first comment on some technical difficulties and caveats. First, we point out that in some previous studies of neutrino-cooled MHD disks, outflows crossing a certain radius before a certain time were removed from the analysis \citep[e.g.][]{Siegel2018c, Miller2020s} in order to minimize spurious effects related to the early transient, during which the disk establishes a new dynamical equilibrium. While such a procedure is well motivated, it entails the risk of cutting out too much or too little material because of the difficulty to unambiguously demarcate spurious from proper ejecta. Unfortunately, this risk is particularly high for a low-mass torus that we consider herein, because the time during which $\yeeq\la 0.25$ is relatively short and largely overlaps with the initial transient phase (cf. top right panel in Fig.~\ref{fig:torusye_time_vis}), which probably lasts until about $t\la 10-20\,$ms. Moreover, it is a priori not clear whether such a cut-off truly captures all spurious ejecta, because it may happen that during the initial transient fluid elements with low $Y_e$ are artificially elevated to a higher, but still gravitationally bound orbit, on which weak interaction rates are low and $Y_e$ thus remains almost constant. Since those fluid elements are most likely ejected during the subsequent evolution, the early transient could thus have a non-negligible impact also on the late-time ejecta. Thus, we do not disregard ejecta based on the time of ejection in this study.
%% Finally, we stress that even if the early transient could somehow be filtered from the outflow analysis, this would not remove the dependence on the initial magnetic field configuration. The dependence on the initial $B$-field configuration is non-trivial and a delicate research topic on its own \citep[e.g.][]{Beckwith2008, Christie2019a}, of which the impact on $Y_e$ and on the relevance of absorption will have to be investigated more by future studies.
A second point to mention is, however, that we remove from the outflow analysis all ejecta crossing the sphere at $r=10^4\,$km at polar angles $\theta/\pi<0.1$ and $\theta/\pi>0.9$. Owing to our coarse $\theta$-resolution near the poles we noticed an artificial drag of outflow material towards the poles, and since the mass- and energy-density have to be reset near the poles whenever reaching low values, the hydrodynamic properties of axis-near material cannot be trusted. Therefore, all diagnostic quantities reported in this paper, except the ejecta masses ($m_{\mathrm{ej}}$ and $m_{\mathrm{ej}}^{Y_e<0.25}$, cf. Table~\ref{table_ejecta}), neglect material ejected in the aforementioned polar-angle intervals. The third issue arising in the MHD models is that, as already explained in Sect.~\ref{sec:mhd-models}, the no-absorption model shows a throttled MRI activity, most likely because of its systematically reduced disk thickness. Hence, the comparison to the case including absorption is somewhat distorted. Based on similar arguments as raised in the previous comparison between different viscosity prescriptions, it can be expected that a fully developed MRI and therefore more efficient angular momentum transport in the ``no$\nu$'' model would have facilitated earlier mass ejection and more neutron-rich conditions in the ejecta. If this assessment is correct, then the consequence would have been a more significant difference between the two cases with and without neutrino absorption than currently observed.

We now discuss the obtained results, keeping the aforementioned issues in mind. Compared to both types of viscous models, the ejecta $Y_e$--mass distribution for the MHD models (red lines in bottom panel of Fig.~\ref{fig:histnuc}) is broader and reaches down to significantly lower values of $Y_e$ even though the average $Y_e$ of the ejecta is not too different (cf. Table~\ref{table_ejecta}). Even when including neutrino absorptions, the $Y_e$ distribution seems to provide sufficiently neutron-rich conditions to produce heavy elements with a final abundance pattern that is almost resembling the solar one. The bottom right panels of Fig.~\ref{fig:trajavg_mtorus} as well as Fig.~\ref{fig:myfreezeout} suggest that the low end of the $Y_e$ distribution is connected to fluid elements that freeze out rather early, within a few tens of milliseconds when the equilibrium conditions in the torus are still more neutron rich. Matter ejection at early times thus appears to be more powerful in the MHD model than in the viscous models. This tendency, which was also reported by \citet{Fernandez2019b}, is a likely consequence of the fact that viscous models in the neutrino-dominated phase are, by design, barely turbulent at all. Most of the material ejected in viscous models is launched only once or after neutrino cooling has become inefficient at about $t\sim t_{\mathrm{em}}$ (e.g. lower panels of Figs.~\ref{fig:torusye_time}~and~\ref{fig:torusye_time_vis}). Since the neutrino-dominated phase is typically characterized by more neutron-rich equilibrium conditions, turbulent MHD tori thus seem to provide systematically more favorable conditions for the ejection of low-$Y_e$ material than viscous tori. Overall beneficial for the ejection of low-$Y_e$ material are also the shorter expansion timescales (cf. bottom right panel in Fig.~\ref{fig:trajavg_mtorus}), i.e. higher ejecta velocities compared to the viscous models. Apart from these differences, the global characteristics of our MHD models are remarkably similar to the viscous models, e.g. the behavior of $\yeeqavg$ and the freeze-out $Y_e$ (cf. Fig.~\ref{fig:torusye_time_vis}).

\paragraph*{Initial electron fraction.}

Most of our models start with a rather large electron fraction of $Y_e^0=0.5$, which was chosen intentionally in order to study precisely the r-process material that is produced self-consistently during and due to the disk evolution. While $Y_e=0.5$ is a suitable initial condition for collapsar BH-tori, remnant disks of compact-object mergers will in most cases have a lower electron fraction that is closer to $Y_e\sim 0.1$. For a direct comparison, we have also set up a model with $Y_e^0=0.1$ (while $\mtor^0=0.01\,M_\odot$, $\MBH=3\,\Msol$, and $\ABH=0.8$ are the same as in the fiducial model m01M3A8), which turns out to produce ejecta with average $Y_e$ of 0.23 (compared to 0.32 for $Y_e^0=0.5$). The sizable difference between both values indicates that there is a non-negligible fraction of material that does not have enough time for its electron fraction to adapt to the local equilibrium value, $\yeeq$. In order to estimate the fraction of the ejecta, $m^{\mathrm{inert}}_{\mathrm{ej}}/m_{\mathrm{ej}}$, that effectively remains unaffected by (or inert to) weak interactions we compute along each outflow trajectory the weak interaction timescale, $\tau_{\beta}=(\tauabs^{-1}+\tauem^{-1})^{-1}$, as well as the expansion timescale, $\tau_{\mathrm{exp}}=r/v_r$, and we count the outflow particle as unaffected if during its evolution from $t=0$ until ejection at $r=10^4\,$km $\tau_{\beta}$ never becomes shorter than the time $\Delta t$ that the particle spent in the torus before its ejection. We approximate $\Delta t$ by the total time during which $\tau_{\beta}<\tauexp$. We find values of $m^{\mathrm{inert}}_{\mathrm{ej}}/m_{\mathrm{ej}}$ (see Table~\ref{table_ejecta}) of about $\mathcal{O}(10\,\%)$ for most of our models. This implies that the nucleosynthesis yields for ten (or tens of) percent of the final ejecta are more determined by the initial conditions than by the conditions during the disk evolution. As expected, the inert fraction of the ejecta turns out to be larger for disks with lower optical depth (e.g. for lower disk mass, higher BH mass).

\subsection{Kilonova properties}\label{sec:kilonova-properties}

\begin{figure*}
  \centering
  \includegraphics[width=0.99\textwidth]{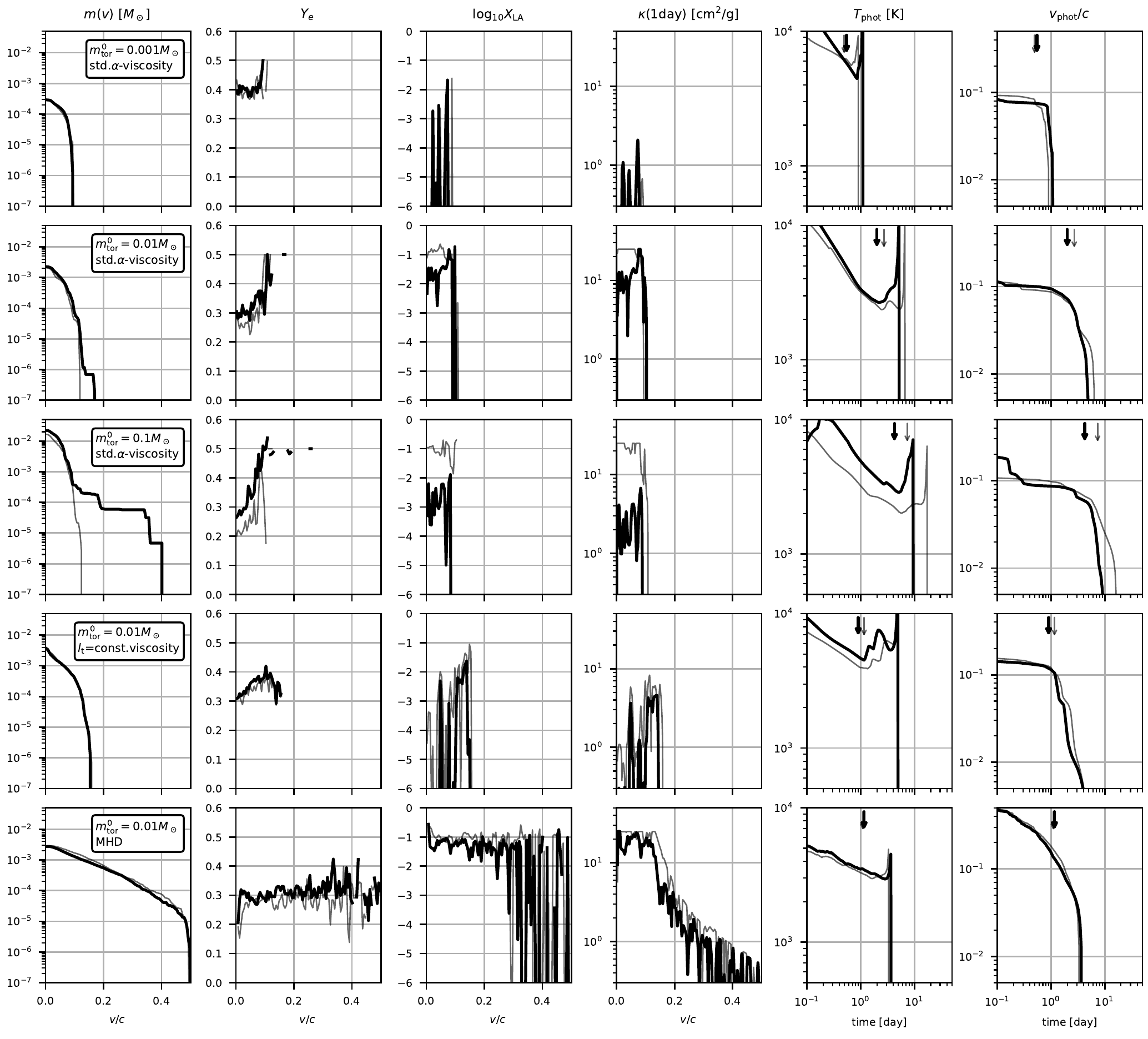}
  \caption{Spherically averaged ejecta properties (first four columns from left), namely mass distribution, electron fraction, lanthanide plus actinide fraction, and opacity measured at $t=1\,$day, as well as photospheric temperature (fifth column) and photospheric velocity (sixth column) for models (from top row to bottom row) m001M3A8, m01M3A8, m1M3A8, m01M3A8-vis2, and m01M3A8-mhd (thick lines) as well as the corresponding counterparts without neutrino feedback (thin lines). The functions of $v$ are obtained by assuming that $v$ remains constant for each trajectory after crossing $r=10^4\,$km. The high-velocity tail that is more pronounced for more massive, viscous tori (i.e. up to $v\approx 0.4c$ in model m1M3A8) represents a neutrino-driven wind, which owing to its brief appearance and relatively small mass is sampled rather poorly, explaining the step-like features in $m(v)$. The vertical arrows mark the times when the ejecta start to become optically thin, i.e. when the bolometric luminosity equals the instantaneous heating rate.}
  \label{fig:knprop}
\end{figure*}	

\begin{figure*}
  \centering
  \includegraphics[width=0.45\textwidth]{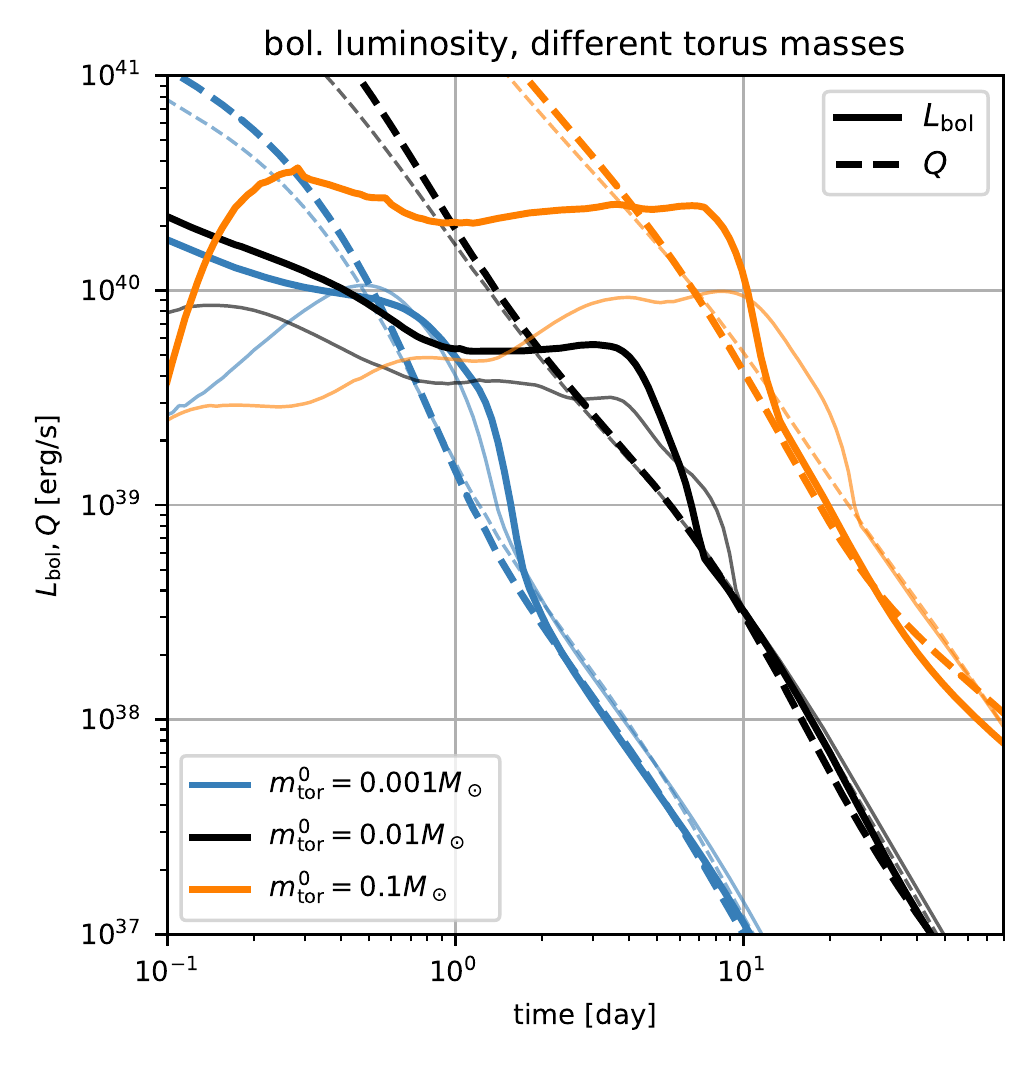}
  \includegraphics[width=0.45\textwidth]{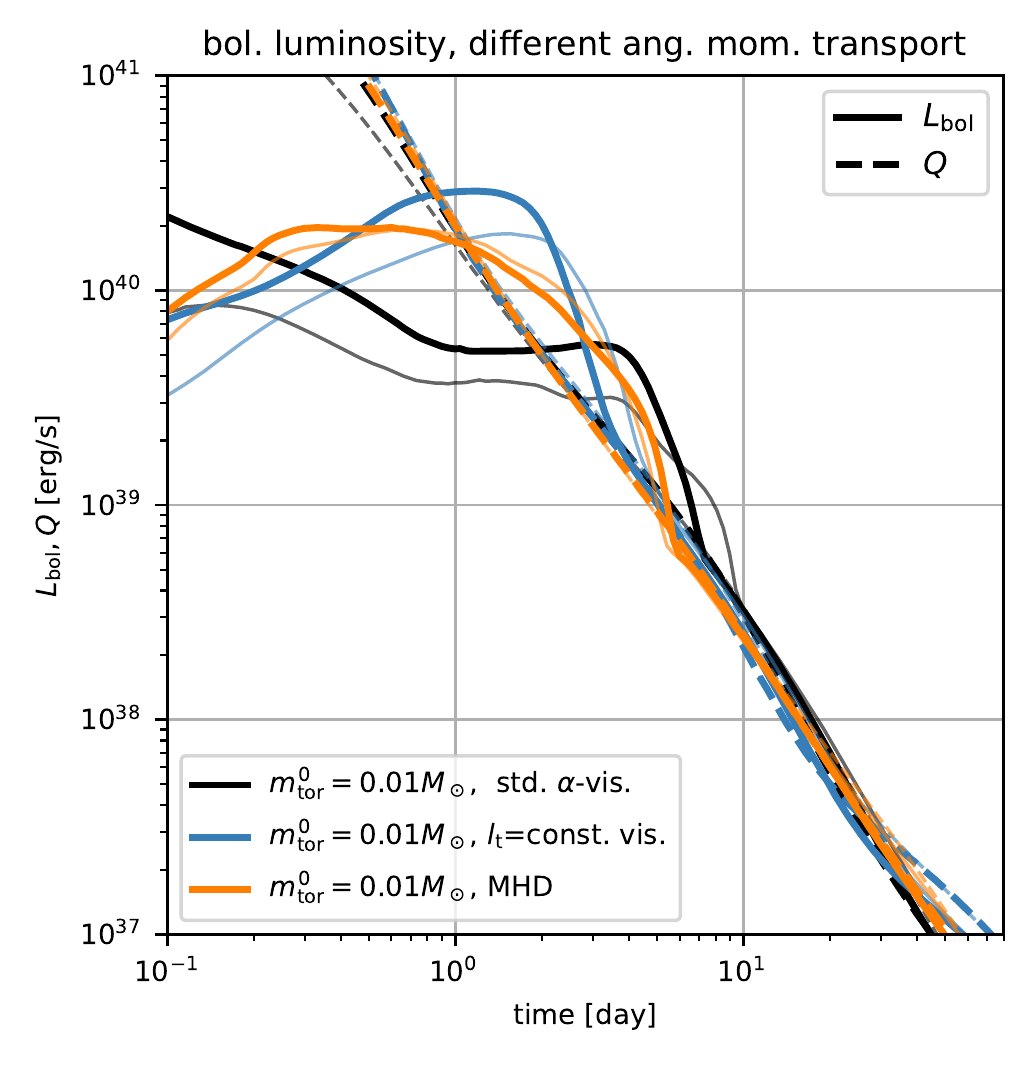}
  \caption{Bolometric kilonova luminosity, $L_{\mathrm{bol}}$ (solid lines), and total radioactive heating rate (including the effect of thermalization), $Q=\int \rho q\dd V$ (dashed lines), of ejecta computed using post-processed trajectory data of the hydrodynamic models. \emph{Left panel:} For models with increasing initial torus mass, namely m001M3A8(-no$\nu$) (blue lines), m01M3A8(-no$\nu$) (black lines), and m1M3A8(-no$\nu$) (orange lines). \emph{Right panel:} For models with different treatment of angular momentum transport, namely m01M3A8(-no$\nu$) (black lines), m01M3A8-vis2(-no$\nu$) (blue lines), and m01M3A8-mhd(-no$\nu$) (orange lines). Thick (thin) lines belong to models including (neglecting) neutrino absorption.}
  \label{fig:knlum_mtorus}
\end{figure*}	

\begin{figure}
  \centering
  \includegraphics[width=0.45\textwidth]{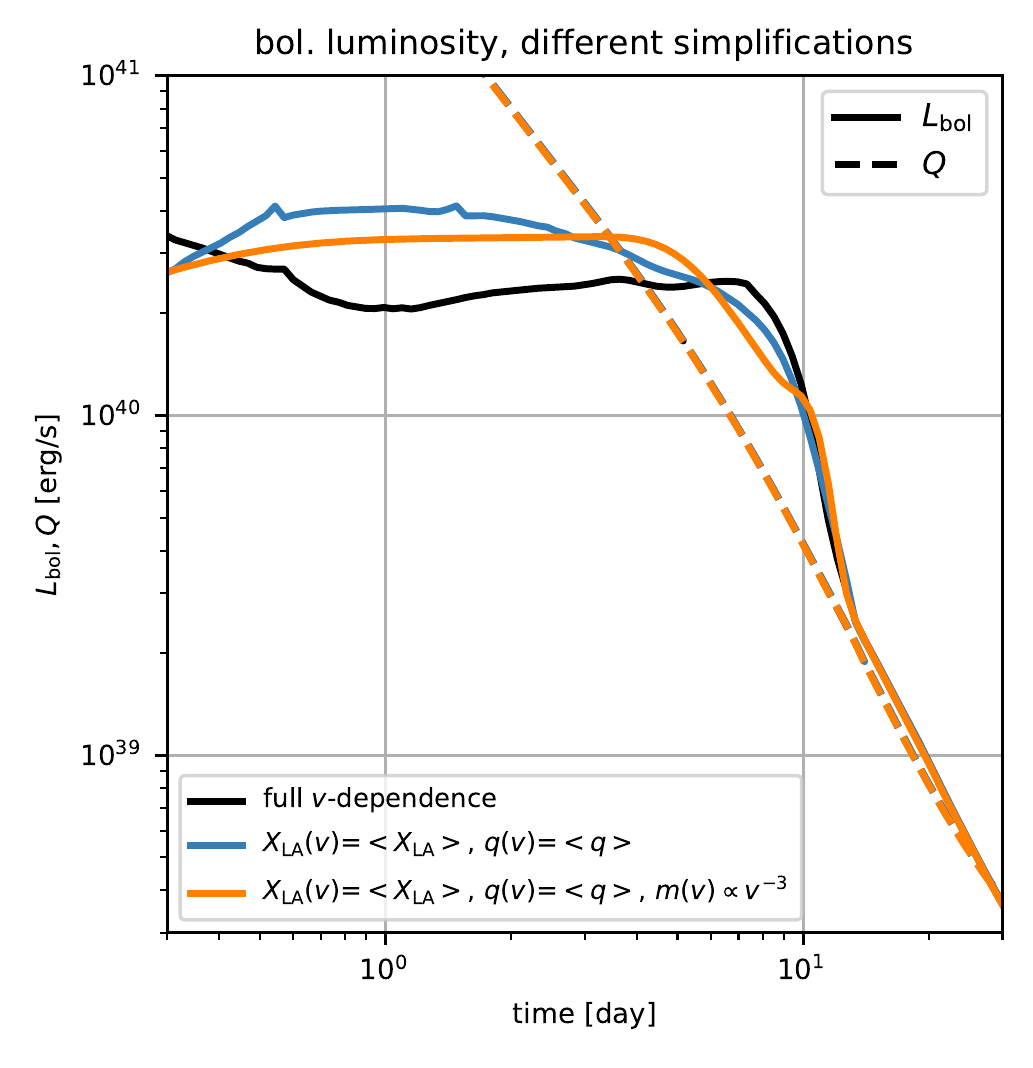}
  \caption{Same as Fig.~\ref{fig:knlum_mtorus} but only for model m1M3A8 and additionally showing the results for two simplifications made in the computation of the light curve: The blue line depicts the case where the composition and radioactive heating rate are taken to be homogeneous throughout the ejecta. The case represented by the orange line additionally replaces the original mass distribution by a power law such that $m(v)\propto v^{-3}$ while keeping the total mass unchanged.}
  \label{fig:knlum_approx}
\end{figure}	

In the last part of this study we briefly examine the bolometric kilonova light curves associated with the disk models. We employ the methods outlined in Sect.~\ref{sec:estim-kilon-lightc} and Appendix~\ref{sec:meth-comp-kilon}. The primary point of this exercise is not to provide detailed predictions for observers, but rather for modelers to get some basic idea about the yet poorly explored sensitivity of the three main kilonova properties (bolometric luminosity, photospheric temperature, photospheric velocity) with respect to modeling variations considered in this study. In order to remedy the fact that in some models the ejecta masses are not converged yet at the times when the simulation had to be stopped, $t_f$, and the heating rate would thus be underestimated, we rescale all trajectory masses by such that their sum equals the ejecta plus torus mass at $t=t_f$.

In Fig.~\ref{fig:knprop} we plot for five models the mass distribution, $m(v)$, electron fraction, $Y_e$, lanthanide plus actinide fraction, $X_{\mathrm{LA}}$, and photon specific opacity, $\kappa$, as function of the expansion velocity, $v/c$, as well as the photospheric temperatures, $T_{\mathrm{phot}}$ and the photospheric velocities, $v_{\mathrm{phot}}/c$, as functions of time. The quantities $T_{\mathrm{phot}}$ and $v_{\mathrm{phot}}$ are given by the temperatures and velocities at the location where the instantaneous optical depth of the ejecta equals unity. The bolometric luminosities, computed as $L_{\mathrm{bol}}(t)=4\pi(vt)^2F(t)$ with radiation flux density $F(t)$ and at constant velocity $v=0.5c$, as well as the integral heating rates, $Q\equiv \int \rho q\dd V$, are depicted for the same models in Fig.~\ref{fig:knlum_mtorus}. Thin lines in Figs.~\ref{fig:knprop} and~\ref{fig:knlum_mtorus} again denote the corresponding models that were evolved without neutrino absorption. The ejecta clearly exhibit a radial structure, with large variations in $\XLA$ and $\kappa$. As expected, the opacities for models ignoring absorption tend to be enhanced compared to models including absorption, all the more for models with increasing optical depth and therefore higher sensitivity to neutrino absorption.

Some of our light curves exhibit maximum values for the bolometric luminosities well before the ejecta become optically thin, namely at $t\la 0.1\,$d. Assuming that at such early times our opacities are inconsistent anyway (see, e.g., \citealp{Banerjee2020b} for a discussion of the early kilonova), we ignore the early light curve entirely in our discussion and only focus on the later phase when the bolometric luminosity ($L_{\mathrm{bol}}$, solid lines in Fig.~\ref{fig:knlum_mtorus}) becomes equal to the instantaneous heating rate ($Q$, dashed lines in Fig.~\ref{fig:knlum_mtorus}). At that time a large fraction of the ejecta becomes optically thin and radiation, which was produced earlier but remained trapped, becomes released at once, creating an excess of $L_{\mathrm{bol}}$ with respect to $Q$. This sudden release of radiation, i.e. the so-called diffusion wave \citet[e.g.][]{Waxman2018a, Kasen2019a, Hotokezaka2020a}, has been hypothesized by \citet{Hotokezaka2020a} to explain the light curve steepening observed after $t\sim 7\,$days for GW170817. Although not mathematically correct, we identify as the light curve peak the time when $L_{\mathrm{bol}}=Q$, and we provide the luminosity, temperature, and photospheric velocity at this time for all models in Table~\ref{table_nuc}. Using this definition of the peak, we measure typical peak times of a few days, peak luminosities of $10^{39}- 10^{40}\,$erg\,s$^{-1}$, and peak temperatures of $1-3\,\times 10^3\,$K. When comparing models with and without neutrino absorption, we observe relative differences for the peak luminosities and times that are scattered around $\sim 40-80\,\%$, where models including absorption throughout exhibit earlier and brighter peaks due to their systematically reduced lanthanide content. Moreover, models without absorption produce a peak at lower temperatures, i.e. in a red-shifted color regime. The relative differences are similarly large when comparing models with a different treatment of angular momentum transport (see right panel in Fig.~\ref{fig:knlum_mtorus}): Compared to the standard $\alpha$-viscosity treatment, the $\lturb$=const. viscosity produces an earlier and more luminous peak with photospheric temperatures enhanced by a factor of about two, which is explained by the systematically reduced fraction of lanthanide and actinide elements. The results for the MHD model end up somewhat in between those for the two different viscosities. A distinct feature of the MHD models is that the ejecta span a larger velocity range, up to $v\sim 0.5\,c$, whereas ejecta in the viscous models do not exceed $v\sim 0.2c$. This fast ejecta tail is probably the reason for the earlier and brighter peak in the MHD model compared to the fiducial model using the standard $\alpha$ viscosity despite the (for the most part) higher opacities.

A particular advantage of our kilonova model compared to others using manually constructed ejecta properties is the fact that the distribution of mass, heating rates and lanthanide fractions along $v$ comes directly from hydrodynamical simulations. For this reason it is worth exploring in Fig.~\ref{fig:knlum_approx} for model m1M3A8 the differences that would result when averaging $\XLA(v)$ and $q(v)$ over the ejecta (blue line) and when additionally replacing the original $m(v)$ by a power-law\footnote{Note that in this case, since $m(v)$ diverges at small velocities, a minimum velocity, $v_0$, must be imposed, below which $\dd m/\dd v=0$. $v_0$ is fixed by the condition that the average velocity of the power-law distribution is the same as for the original distribution.} as $m(v)\propto v^{-3}$ (orange line). Both approximations induce considerable changes on the $\sim 50\,\%$ level in the luminosity, at least at times $t\leq 10\,$days, i.e. before the optically thin phase of emission.

Even though the present examination of the kilonova is kept rather brief and simple and does not address spectroscopic properties, the results clearly advocate the importance of accurate neutrino-transport modeling as well as a careful treatment of turbulent angular momentum transport for reliable predictions of the kilonova light curve.

\section{Discussion}\label{sec:discussion}

\subsection{Comparison with the literature}\label{sec:comp-with-prev}

\setlength{\tabcolsep}{6pt}
\begin{table*}
  \centering
  \caption{Compilation of average electron fraction of the ejected material, $Y_{e,\mathrm{ej}}$, for models of neutrino-cooled disks available in the literature that consider the same astrophysical setup and that differ in modeling aspects investigated in this study, namely the initial electron fraction, $Y_e^0$, turbulent viscosity treatment, type of neutrino treatment, and inclusion or omission of $Q_{np}$ and $m_e$ terms. The names used for the neutrino treatment have the following meaning: ``leakage'' $\rightarrow$ classical leakage scheme as in \citet{Ruffert1996a}. ``leak.+abs.'' $\rightarrow$ same as before but augmented with a scheme to describe net $\nu$-absorption in optically thin regions. ``leak. ($\yeeq$=$\yeeqem$)'' $\rightarrow$ simplified leakage with $\chi_{\nu_e}=\chi_{\bar\nu_e}$ and therefore $\yeeq=\yeeqem$ (see Sect.~\ref{sec:neutr-absorpt-leak}). ``grey M1+leak.'' $\rightarrow$ combination of schemes in which leakage (M1) is employed in regions of high (low) optical depth. ``no abs. ($\yeeq$=$\yeeqem$)'' $\rightarrow$ neutrino absorption is completely ignored. See Sect.~\ref{sec:comp-with-prev} for more detailed explanations. The '*' indicates that the given value of $Y_{e,\mathrm{ej}}$ refers only to early ejecta produced within $t\la 100-200\,$ms. The '**' means the same as '*', but additionally the value of $Y_{e,\mathrm{ej}}$, as it was not provided, had to be estimated based on other information found in that reference.}
  \label{table_literature}
  \begin{center}
  \begin{tabularx}{\textwidth}{lcccccccc}
    %%% in Emacs use M-x align-current to align   
    \hline
    Reference                               & $\mtor^0$   & $\MBH$      & $\ABH$ & $Y_e^0$  & viscosity            & neutrino                                & $Q_{np}$ and $m_e$ & $Y_{e,\mathrm{ej}}$  \\
    {}                                      & [$M_\odot$] & [$M_\odot$] &        &          & treatment            & treatment                               & included?          & [1]                  \\
    \hline                                                                                                                                      
    \citet{Fernandez2020a}                  & 0.03        & 3           & 0.8    & 0.1      & std. $\alpha$-vis.   & leak.+abs.                              & yes                & 0.28                 \\
    \citet{Fernandez2019b}                  & 0.03        & 3           & 0.8    & 0.1      & std. $\alpha$-vis.   & leak. ($\yeeq$=$\yeeqem$)               & no                 & 0.20                 \\
                                                                                                                                                                                                         \\
    \citet{Just2015a}                       & 0.03        & 3           & 0.8    & 0.1      & std. $\alpha$-vis.   & spectral M1                             & yes                & 0.27                 \\
    m01M3A8      \emph{(this work)}         & 0.01        & 3           & 0.8    & 0.5      & std. $\alpha$-vis.   & spectral M1                             & yes                & 0.32                 \\
    m01M3A8-noQm-no$\nu$ \emph{(this work)} & 0.01        & 3           & 0.8    & 0.5      & std. $\alpha$-vis.   & no abs. ($\yeeq$=$\yeeqem$)             & no                 & 0.24                 \\
                                                                                                                                                                                                         \\
    \citet{Fujibayashi2020a}                & 0.1         & 3           & 0.8    & 0.07-0.5 & $\lturb$=const. vis. & grey M1+leak.                           & yes                & 0.31                 \\
                                            & 0.1         & 3           & 0.8    & 0.07-0.5 & $\lturb$=const. vis. & no abs. ($\yeeq$=$\yeeqem$)             & yes                & 0.30                 \\
                                                                                                                                                                                                         \\
    m01M3A8-vis2 \emph{(this work)}         & 0.01        & 3           & 0.8    & 0.5      & $\lturb$=const. vis. & spectral M1                             & yes                & 0.35                 \\
    m01M3A8-vis2-no$\nu$ \emph{(this work)} & 0.01        & 3           & 0.8    & 0.5      & $\lturb$=const. vis. & no abs. ($\yeeq$=$\yeeqem$)             & yes                & 0.34                 \\
                                                                                                                                                                                                         \\
    \citet{Siegel2018c}                     & 0.03        & 3           & 0.8    & 0.1      & MHD                  & leakage                                 & no                 & 0.18                 \\
    \citet{Siegel2019b}                     & 0.016       & 3           & 0.8    & 0.5      & MHD                  & leakage                                 & no                 & $\la 0.25^{**}$      \\
                                                                                                                                                                                                         \\
    \citet{Fernandez2019b}                  & 0.03        &             & 0.8    & 0.1      & MHD                  & leak. ($\yeeq$=$\yeeqem$)               & no                 & 0.16                 \\
                                                                                                                                                                                                         \\
    \citet{Miller2019a}                     & 0.12        & 2.58        & 0.69   & 0.1      & MHD                  & Boltzmann                               & yes                & $\sim 0.2-0.25^{**}$ \\
    \citet{Miller2020s}                     & 0.02        & 3           & 0.8    & 0.5      & MHD                  & Boltzmann                               & yes                & $0.36^{*}$           \\
                                                                                                                                                                                                         \\
    m01M3A8-mhd \emph{(this work)}          & 0.01        & 3           & 0.8    & 0.5      & MHD                  & spectral M1                             & yes                & 0.31                 \\
    \hline
  \end{tabularx}
\end{center}
\end{table*}

Given the relevance for heavy-element production, the available number and degree of sophistication of models for neutrino-cooled BH accretion disks is growing quickly. In the following we briefly put the results of our study into context by comparison with a variety of models available in the literature. The basic model setup used here (equilibrium torus, $\alpha$-viscosity or single-loop initial magnetic field configuration) is similar to what was used in most previous studies of viscous or MHD disks, and several features are in broad agreement with the literature. One difference to several existing models is that we initiated most of our simulations with $Y_e^0=0.5$, because we are only interested in the contribution that is produced self-consistently by the disk. For this reason our average ejecta $Y_e$, $\yeej$, is overall shifted compared to models starting with a $Y_e^0=0.1$ torus. Apart from this shift, however, the influence of other modeling variations on $Y_e$ should be similar in our simulations compared to simulations from the literature.

\subsubsection{Neutrino absorption in leakage schemes}\label{sec:neutr-absorpt-leak}

Before considering individual studies we first clarify some important points regarding neutrino leakage schemes, which otherwise may become sources of confusion. Classical leakage schemes, such as introduced by \citet{Ruffert1996a} and \citet{Rosswog2003}, are often mentioned to be unable to describe neutrino absorption. This is however only true for regions with positive \emph{net} absorption rates (i.e. absorption minus emission rates, such as in region A of the idealized model sketched in Fig.~\ref{fig:evolphases}). The impact of absorption in all other regions (such as region B in Fig.~\ref{fig:evolphases}) can in principle be captured by leakage schemes, which can be understood by the following considerations: Defining emission and absorption rates of $\nu_e$ and $\bar\nu_e$, respectively, as
\begin{subequations}\label{eq:leakemab}
  \begin{align}
    & R_{\nu_e}^{\mathrm{em}}  \equiv \lambda_{e^-}Y_p \, , \\
    & R_{\nu_e}^{\mathrm{abs}} \equiv \lambda_{\nu_e}Y_n \, , \\
    & R_{\bar\nu_e}^{\mathrm{em}}  \equiv \lambda_{e^+}Y_n \, , \\
    & R_{\bar\nu_e}^{\mathrm{abs}} \equiv \lambda_{\bar\nu_e}Y_p \, , 
\end{align}
\end{subequations}
the rate of change of $Y_e$ can be written as
\begin{align}\label{eq:dyeleak}
  \frac{\dd Y_e}{\dd t} & =
  - (R_{\nu_e}^{\mathrm{em}} - R_{\nu_e}^{\mathrm{abs}})
  + (R_{\bar\nu_e}^{\mathrm{em}} - R_{\bar\nu_e}^{\mathrm{abs}})  \nonumber\\
  & = -R_{\nu_e}^{\mathrm{eff}} + R_{\bar\nu_e}^{\mathrm{eff}}  \nonumber\\
  & = -R_{\nu_e}^{\mathrm{em}}\chi_{\nu_e} + R_{\bar\nu_e}^{\mathrm{em}}\chi_{\bar\nu_e} \, ,
\end{align}
where in the second line $R_{\nu}^\mathrm{eff}\equiv R_{\nu}^\mathrm{em}-R_{\nu}^\mathrm{abs}$ and in the third line $R_{\nu}^\mathrm{eff} = R_{\nu}^{\mathrm{em}}\chi_{\nu}$ are used for $\nu=\nu_e,\bar\nu_e$. Leakage schemes now compute $\dd Y_e/\dd t$ in the form provided by the third line of Eq.~(\ref{eq:dyeleak}) and approximate the local quenching factors $\chi_{\nu}$ based on (often problem-dependent) conditions related to the neutrino optical depth. Since all right-hand sides of Eq.~(\ref{eq:dyeleak}) are identical, the quenching factors are nothing but the difference between emission rate and absorption rate normalized to the emission rate, i.e.
\begin{align}\label{eq:chinu}
  \chi_\nu=(R_{\nu}^\mathrm{em}-R_{\nu}^\mathrm{abs})/R_{\nu}^\mathrm{em} \, ,
\end{align}
which is a local version of the factor $\chi_{\mathrm{abs}}$ of Eq.~(\ref{eq:absfac}) that is plotted in Figs.~\ref{fig:globdat_mtorus} and~\ref{fig:globdat_vis}. Therefore, classical leakage schemes are already capable of describing both the asymptotic $Y_e$ in neutrino-less optically thin conditions, $\yeeqem$ (since $\chi_\nu =1$ results in $R_{\nu_e}^{\mathrm{em}} = R_{\bar\nu_e}^{\mathrm{em}}$, which is equal to Eq.~(\ref{eq:yeeqem})), and its absorption-modified version, $\yeeq$ (since the third line of Eq.~(\ref{eq:dyeleak}) is equal to Eq.~(\ref{eq:yeeq}) up to errors entering the computation of $\chi_\nu$). From Eq.~(\ref{eq:dyeleak}) it becomes clear that $\yeeq$ depends sensitively on the assumptions entering the computation of the $\chi_\nu$ factors and, in particular, on the resulting ratio $\chi_{\nu_e}/\chi_{\bar\nu_e}$. If, for instance, the same quenching factors are chosen for both neutrino species, i.e. $\chi_{\nu_e}=\chi_{\bar\nu_e}$, then again $R_{\nu_e}^{\mathrm{em}} = R_{\bar\nu_e}^{\mathrm{em}}$ and therefore $\yeeq=\yeeqem$, which is equivalent to disregarding neutrino absorptions entirely concerning the equilibrium value of $Y_e$. To our knowledge $\yeeq$ has not been discussed so far in the context of leakage schemes, which is why the accuracy of $\yeeq$ in models of neutrino-cooled disks using leakage schemes is unknown at this point. The few comparisons of leakage schemes with transport methods that exist so far \citep{Richers2015a,Foucart2016a,Ardevol-Pulpillo2019a,Gizzi2019a} -- and performed in most cases only in the context of proto-neutron stars -- report rather modest agreement for the total luminosities and the ratio of $\nu_e$ to $\bar\nu_e$ luminosities when using the methodology of \citealp{Ruffert1996a} or \citealp{Rosswog2003} (in contrast to the good performance of more recent leakage variants by \citealp{Perego2016a, Ardevol-Pulpillo2019a}). However, in disks the agreement is unknown and might be both better or worse.

Various extensions to classical leakage schemes have been suggested in order to approximate net neutrino absorption in optically thin regions \citep{Fernandez2013b, Perego2016a, Siegel2018c, Ardevol-Pulpillo2019a}. The basic strategy often consists of irradiating (subdomains of) the fluid configuration with a neutrino field of constant luminosity and mean energy, similar to what is done in light-bulb schemes in the context of CCSNe \citep[e.g.][]{Bethe1985, Janka1996b}. While some of these extensions might be more accurate than others, rarely any such scheme has been benchmarked against a Boltzmann solution, which at this point makes it basically impossible to assess whether available simulations using such schemes typically under- or overestimate $Y_e$ in region A of neutrino-cooled disks (cf. Fig.~\ref{fig:evolphases}). Frankly, in such regions the accuracy of the M1 scheme (employed in our study) is not well known either. The results by \citet{Miller2020s} (cf. next section for a discussion) suggest that $Y_e$, and therefore the quantitative impact of absorption, may be underrated in leakage and M1 schemes.

Finally, although this aspect is probably less relevant for semi-transparent disks, for the sake of completeness we mention that classical neutrino leakage schemes cannot properly describe dynamic conditions of very high optical depth, in which neutrinos are trapped and become advected by the fluid. This shortcoming is tackled in implementations of \cite{Sekiguchi2012a, Perego2016a, Ardevol-Pulpillo2019a} by additionally solving an advection equation for trapped neutrinos and assuming these trapped neutrinos to equilibrate with the fluid under conditions of high optical depth.

\subsubsection{Comparison with selected studies}

Using a leakage scheme in combination with neutrino irradiation from a manually constructed neutrino field, \citet{Fernandez2020a} investigated the impact of absorption based on post-processing methods, namely by comparing the net contributions of each of the four $\beta$-reactions (cf. Eq.~(\ref{eq:betarates})) along outflow trajectories. Lacking at this point detailed information about their equilibrium values $\yeeqem$ or $\yeeq$, we cannot directly compare the impact of absorption in region B (cf. Fig.~\ref{fig:evolphases}) of their simulations to that found for our models. Nevertheless, they obtain remarkably similar ejecta $Y_e$ as we do and draw the same conclusion regarding the impact of neutrino absorption, namely that neutrino absorption becomes relevant for disk masses above $\sim 0.01\,\Msol$. In \citet{Fernandez2019b} similar viscous models are presented, which in contrast to those of \citet{Fernandez2020a} assume $\chi_{\nu_e}=\chi_{\bar\nu_e}$ (and therefore $\yeeq=\yeeqem$) and ignore terms containing $m_e$ and $Q_{np}$ in the $\beta$-rates. In agreement with our findings that both of these measures reduce $Y_e$ in the ejecta, they obtain an average electron fraction in the ejecta, $\yeej$, that is reduced by 0.08 (cf. Table~\ref{table_literature}).

\citet{Fujibayashi2020a}, employing the $\lturb$=const. scheme to describe viscous angular momentum transport, also switch off neutrino irradiation for one of their models using a combination of grey leakage and M1 transport (cf. \citealp{Sekiguchi2012a} for details of this neutrino scheme). They find a very small impact of absorption, which is fully compatible to the minor difference that we find between our models m01M3A8-vis2 and m01M3A8-vis2-no$\nu$. This agreement suggests that the small impact of absorption observed in \citet{Fujibayashi2020a} is to a lesser extent a consequence of GR effects (which in that study were newly included in contrast to previous simulations of viscous disks) but mainly a ramification of the $\lturb$=const. scheme, which compared to the standard $\alpha$-viscosity scheme underrates the impact of neutrino absorption by decelerating the evolution during the neutrino-dominated phase as discussed in Sects.~\ref{sec:diff-visc-prescr} and~\ref{sec:model-dependence}. As a result of the extended neutrino-dominated phase, the torus $Y_e$ follows $\yeeq$ until higher values are reached, which shifts the final $Y_e$ pattern of the ejecta and correspondingly reduces the abundances of elements with $A>130$. The fact that the impact of neutrino absorption turns out to vary significantly when switching to another viscosity scheme represents an example for non-linear coupling of modeling ingredients.

The first studies discussing outflows from three-dimensional MHD models of neutrino-cooled disks \citep{Siegel2018c, Siegel2019b, Fernandez2019b, Christie2019a} report very neutron-rich ejecta with $Y_e$ distributions peaking between $Y_e=0.1-0.2$ and therefore enabling heavy-element production in a solar-like fashion all the way up to the 3rd r-process peak (at least for tori more massive than $\sim 0.01\,\Msol$). Our MHD models confirm the basic tendency that more neutron-rich material can be ejected compared to viscous models, owing mainly to the more vigorous turbulence during the neutrino-dominated phase and the shorter expansion timescales. However, the average $Y_e$ in our ejecta (cf. $\yeej$ in Table~\ref{table_ejecta}) is still considerably higher than theirs. A part of this discrepancy can be accounted to the initial $Y_e$, which is $Y_e^0=0.5$ in our models compared to 0.1 in all aforementioned studies except \citet{Siegel2019b}. However, an additional reduction of $Y_e$ in those studies can be attributed to the fact that they employ the simplification of $Q_{np}=m_e=0$ (assuming that our understanding of their implementation is correct), which effectively decreases $\yeeqem$ in the torus by about 0.05-0.1. Moreover, the enhancement of $\yeeq$ with respect to $\yeeqem$ due to neutrino absorptions is not captured in \citet{Fernandez2019b} and \citet{Christie2019a}, because, like for the viscous models mentioned above, they use the same optical depth for $\nu_e$ and $\bar\nu_e$.

At the time of this writing the most complete simulations of neutrino-cooled disks -- including GRMHD and a Monte-Carlo Boltzmann-solver for neutrino transfer -- have been presented by \citet{Miller2019a, Miller2020s}, while \citet{Miller2020s} particularly focused on the role of neutrino absorption in raising $Y_e$ in the outflow. Using a similar initial torus as \citet{Siegel2019b} with a mass of $0.02\,\Msol$, \citet{Miller2020s} find ejecta with significantly reduced neutron densities, which is in line with the notion suggested by our results, namely that neutrino absorption as well as a correct treatment of the terms including $Q_{np}$ and $m_e$ in the $\beta$-rates drive $Y_e$ to considerably larger values. The absence of any lanthanides or heavier elements in the ejecta reported in \citet{Miller2020s} points to an even more significant impact of neutrino absorption than we find in the present study. It cannot be ruled out that our M1 approximation for the neutrino transport, or neglecting GR metric effects, or something else could attenuate the impact of absorption in our models. Nevertheless, the differences between \citet{Miller2020s} and our models might not be so dramatic after all, because \citet{Miller2020s} only follow ejected material for evolution times between $t\sim 70$ and $150\,$ms. The fact that the average torus $Y_e$ is still as low as $Y_e\sim 0.25$ at the end of their simulation (cf. their Fig.~1) leaves open the possibility that neutron-rich material with $Y_e\la 0.25$ may still be expelled at later times. Moreover, in contrast to our study the impact of absorption is not tested by conducting two separate simulations with and without neutrino absorption. Instead,  \citet{Miller2020s} estimate the impact of absorption by comparing the $Y_e$ pattern of the full transport simulation with a second pattern that is obtained by integrating $\dd Y_e/\dd t$ along the same fluid trajectories using the same neutrino emission rates but neglecting the absorption rates (assuming that we correctly understood their procedure). By doing so, however, the emission rates -- which themselves depend on $Y_e$ -- become inconsistent with $Y_e$, most likely leading to an overestimated difference between the two cases with and without neutrino absorption.

We note that, apart from the aforementioned ones, more studies of neutrino-cooled disks exist, to which however a detailed comparison of results for the ejected material is more difficult and therefore omitted here (as well as in Table~\ref{table_literature}). Neutrino-cooled disks have been the subject of a significant number of one-dimensional models mainly in the context of gamma-ray bursts \citep[e.g.][]{Popham1999, DiMatteo2002, Kohri2002, Chen2007, Kawanaka2007, Metzger2009b, Liu2017n}. The models by \citet{Hossein-Nouri2018a}, which were one of the first three-dimensional GRMHD models and employed a neutrino leakage scheme, cover only the first $\sim 70$\,ms and do not discuss properties of the outflow. Moreover, \citet{Janiuk2019a} simulated neutrino-cooled disks in 2D GRMHD, but instead of evolving $Y_e$ using a conservation equation for the electron-number, they assume an instantaneous emission--absorption equilibrium to hold. Hence, $Y_e$ in their outflows does not freeze out but keeps changing during the ejection, rendering a measurement of the final $Y_e$ of ejected material ambiguous and a comparison with our results therefore difficult.

\subsection{Are neutrino-cooled disks major sites of 3rd-peak r-process elements?}\label{sec:can-neutrino-cooled}

\begin{figure*}
  \centering
  \includegraphics[width=0.95\textwidth]{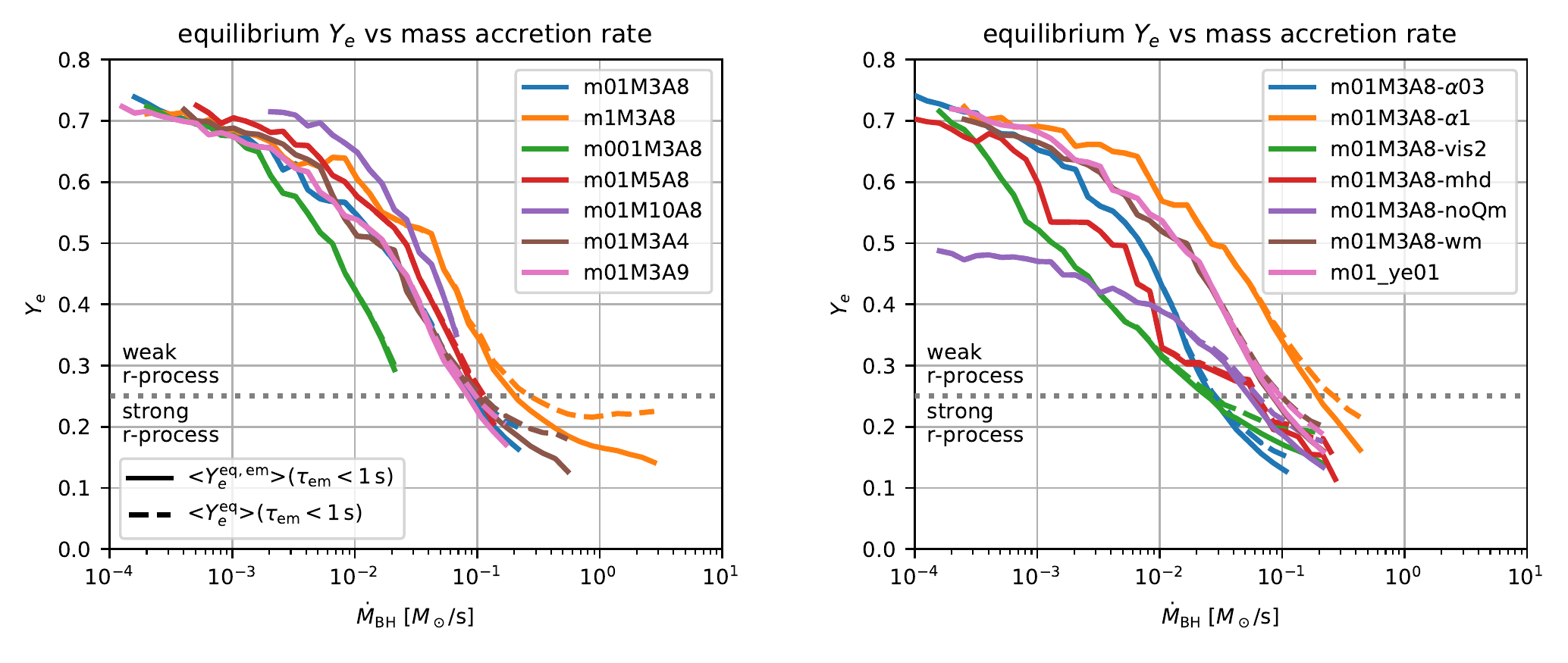}
  \caption{Equilibrium electron fractions, averaged over emission-efficient regions where $\tauem<1\,$s, as function of the mass accretion rate for all evolved models with (dashed lines) and without (solid lines) neutrino absorption. The arrangement into two panels was done only to keep each line distinguishable from others. Time runs from right to left, i.e. the tori protonize with time. Assuming that our finite-mass disks would be representative for collapsar disks, one can infer that the mass accretion rate needs to be greater than $\sim 2\times 10^{-2}\,\Msol\,$s$^{-1}$ in order for the torus to have $\langle\yeeqem\rangle <0.25$ and thus the possibility to enable a strong r-process in the ejecta.}
  \label{fig:mdotye}
\end{figure*}	

An important question is whether neutrino-cooled disks, either as remnants of NS-NS or NS-BH mergers or as collapsar engines, could represent major sources of the heaviest r-process elements, namely lanthanides and 3rd-peak nuclei. A first, interesting observation in this regard is that for otherwise similar global model parameters (i.e. $\MBH, \ABH, \mtor^0$) a disk formed after a compact-object merger ($Y_e^0 \sim 0.1$) would produce heavy r-process elements more efficiently than a collapsar disk ($Y_e^0 \sim 0.5$), just because of the initially lower $Y_e$. While this difference, attributed entirely to the initial $Y_e$, needs to be kept in mind when comparing the entire element production in disk ejecta, in the following we are only interested in the ejecta fraction with conditions shaped self-consistently during the disk evolution.

The abundance distributions plotted in Fig.~\ref{fig:histnuc} suggest that neutrino-cooled disks can be prolific sources of heavy (i.e. $A>130$) elements for a wide range of global model parameters. However, in order for these systems to be main contributors of 3rd-peak elements, the yields of lighter elements needs to be equal to or less than those in the solar reference distribution. A necessary -- though not sufficient -- condition is that the abundance ratio of 3rd-peak to 2nd-peak elements needs to lie close to or above the corresponding ratio of the solar distribution (i.e. $[X_{\mathrm{3rd}}/X_{\mathrm{2nd}}]_\odot \geq 1$; see Table~\ref{table_nuc} and Fig.~\ref{fig:tabdat}). Basically all of the viscous models with neutrino absorption, $Y_e^0=0.5$, and without the $Q_{np}=m_e=0$ simplification are deficient in this ratio by factors of a few. Hence, ignoring other (e.g. nuclear) uncertainties, one may be inclined to believe that viscous disks are not the perfect candidates for main 3rd-peak sources in contrast, for instance, to tidal ejecta in NS-NS and NS-BH mergers \citep[e.g.][]{Hotokezaka2013b, Bauswein2014, Goriely2015, Foucart2015, Radice2018b}, which generically have very low $Y_e$.

The situation seems to look more favorable in the case of MHD disks, because lower values of $Y_e$ can be reached by a larger fraction of outflow material than in viscous disks. Indeed, we find higher values of $[X_{\mathrm{3rd}}/X_{\mathrm{2nd}}]_\odot$ than for the corresponding viscous models and an abundance pattern that is only mildly sub-solar in the $A>130$ regime, mild enough for the gap to be explained by other, for instance nuclear uncertainties. Hence, the results indicate that MHD disks could indeed produce 3rd-peak material prolifically enough to represent a major source, may it be in collapsars or NS mergers. However, we refrain from drawing strong conclusions at this point, on the one hand because of the poorly constrained uncertainties connected to the initial conditions and the initial transient. On the other hand, at this point we have not investigated the dependence of the thermodynamic conditions in the disk -- and therefore of $\yeeq$ and of the ejecta $Y_e$ -- on the numerical resolution, simply because of computational limitations. Another reason to be careful not to overinterpret our results is the apparent tension between our rather optimistic results with the pessimistic results by \citet{Miller2020s}, which may have its origin in our approximate M1 neutrino treatment or the crude approximation of GR effects.

For the case of a collapsar disk an additional major uncertainty, apart from those related to neutrino transport and MHD turbulence, is connected to the formation and resulting structure of the collapsar disk. In contrast to the isolated, finite-mass disks studied in this work (as well as in \citealp{Siegel2019b} and \citealp{Miller2020s}), the mass reservoir in collapsar disks is continuously fed from infalling and circularizing stellar material, such that the evolution close to the BH is quasi-stationary and mostly a function of the infalling mass flow rate, $\dot{M}_{\mathrm{infall}}$, which is presumably comparable to the BH accretion rate, i.e. $\dot{M}_{\mathrm{infall}}\approx \dot{M}_{\mathrm{BH}}$. In the most optimistic case neutrino absorption could, for whatever reason, be barely relevant and the mass ejection processes so swift that also neutrino emission is unable to raise the electron fraction while ejecta travel from deep  inside the torus to large radii. Assuming such an idealized situation, and supposing that our tori can be interpreted as collapsar disks, one can estimate the lower limit of the ejecta $Y_e$ by considering the values $\yeeqem$ corresponding to emission equilibrium (cf. Eq.~(\ref{eq:yeeqem})) attained in the torus for given mass accretion rates onto the central BH. To this end we plot in Fig.~\ref{fig:mdotye} for all models of this study $\yeeqem$ (as well as $\yeeq$) averaged over regions where weak interactions are efficient (i.e. where $\tauem<1\,$s) as function of the mass accretion rate measured at corresponding times.

We notice a considerable variation of $\yeeqem(\dot{M}_{\mathrm{BH}})$ between different models. This variation provides an idea about the large uncertainty that is connected to the unknown structure of collapsar disks: Disks with similar values of $\dot{M}_{\mathrm{BH}}$ can carry quite different bulk electron fractions depending on their evolution history. This result illustrates the limitations of our finite-mass disk models and emphasizes the importance of a global model following self-consistently the disk circularization process.

Nevertheless, if we assume that our set of models is somewhat representative (which is by no means guaranteed) to bracket the spectrum of possible $\yeeqem(\dot{M}_{\mathrm{infall}})$ curves realized in collapsars, then we might infer that the infall rate needs to be at least as high as $\dot{M}_{\mathrm{infall}}\ga 2\times 10^{-2}\,\Msol\,$s$^{-1}$ in order for $Y_e$ to attain a sufficiently low value that can enable strong r-processing, $Y_e\approx 0.25$. For lower mass-accretion rates, the r-process could operate with sufficient efficiency only up to the 2nd peak or, for $\dot{M}_{\mathrm{infall}}\la 10^{-3}\,\Msol\,$s$^{-1}$, not at all. We stress again that this condition on $\yeeq$ in the torus is a rather conservative limit concerning $Y_e$ in the ejecta. The latter will most likely be driven to higher values along the outflow trajectory due to the effects of neutrino emission and absorption.

If we assume that our set of models is somewhat representative (which is by no means guaranteed) to bracket the spectrum of possible $\yeeqem(\dot{M}_{\mathrm{infall}})$ curves realized in collapsars, then we might infer that the infall rate needs to be at least as high as $\dot{M}_{\mathrm{infall}}\ga 2\times 10^{-2}\,\Msol\,$s$^{-1}$ in order for $Y_e$ to attain a sufficiently low value that can enable strong r-processing, $Y_e\approx 0.25$. For lower mass-accretion rates, the r-process could operate with sufficient efficiency only up to the 2nd peak or, for $\dot{M}_{\mathrm{infall}}\la 10^{-3}\,\Msol\,$s$^{-1}$, not at all. We stress again that this condition on $\yeeq$ in the torus is a rather conservative limit concerning $Y_e$ in the ejecta. The latter will most likely be driven to higher values along the outflow trajectory due to the effects of neutrino emission and absorption.

By means of Fig.~\ref{fig:mdotye}, our models thus provide an additional, indirect constraint on the progenitor structure, which needs to be supplemented to the two (obvious) conditions that a BH must be formed and that material carries sufficient angular momentum to stay on circular orbits and form a disk. While the model E15 of \citet{Heger2000a} was estimated in Fig.~1 of \citet{Siegel2019b} to fulfill these constraints, \citet{Aloy2020a} expect rather high BH masses of $\MBH\approx 7.5$ and $17.3$, and therefore probably too low mass accretion rates, for their models 35OC and 350B (which were taken from \citealp{Woosley2006a}), respectively. Moreover, the sensitivity of $\yeeq$ with respect to the mass-accretion rate could imply a natural dependency of the produced r-process pattern on the progenitor structure. For instance, one might speculate that progenitors with sufficient angular momentum to create disks early during the collapse might tend to generate a stronger r-process than low angular momentum progenitors, for which disks only form late and with low accretion rates.

\section{Summary}\label{sec:summary-conclusions}

Neutrino-cooled disks are likely to be important production sites of heavy elements, but our understanding of the detailed composition of the ejecta and its sensitivity with respect to the physics of the neutrino treatment is still incomplete. The aim of this study was to obtain a better quantitative understanding of the neutrino emission and absorption effects during the long-term evolution of a disk. To this end, we investigated, for the first time systematically over a wide range of global parameters, the impact of including or not including neutrino absorption in regulating the composition in the disk and the ejecta. We varied the disk mass, BH mass and spin, details of the neutrino interaction rates, and the initial electron fraction of the disk. We furthermore tested the sensitivity to varying the viscous $\alpha$-parameter in the standard $\alpha$-viscosity formalism, to using an entirely different viscosity prescription, and to replacing the axisymmetric viscous evolution by a 3D MHD evolution. In a post-processing step, we computed the nucleosynthesis yields of the ejecta and estimated basic properties of the expected kilonova. In contrast to many existing models of neutrino-cooled disks in the context of NS mergers that typically assume an initial electron fraction of $Y_e^0\sim 0.1-0.2$, our fiducial model is set up with $Y_e^0=0.5$. By doing so, we specifically probe only the amount and properties of r-process viable material that is self-consistently produced by the disk evolution, because we effectively remove low-$Y_e$ material from the outflow that is ejected before being able to reach (any type of) weak equilibrium.

The main results of our study are:
\begin{enumerate}
\item We identify four characteristic regions for weak interactions in typical neutrino-cooled disks (sketched in Fig.~\ref{fig:evolphases}): At high mass accretion rates, $\dot{M}_{\mathrm{BH}}\ga 10^{-2}\,\Msol\,$s$^{-1}$, neutrino absorption competes with neutrino emission in the bulk of the torus and it effectively raises the $Y_e$ equilibrium value from the pure-emission case, $\yeeqem$ (cf. Eq.~(\ref{eq:yeeqem}) and Fig.~\ref{fig:yeeq1}) to $\yeeq =  \yeeqem+\Delta Y_e$ (cf. Eq.~(\ref{eq:yeeq})), where typically $\Delta Y_e\approx 0.05-0.1$. In the surface layers close to the symmetry axis neutrino absorption dominates and drives $Y_e$ towards $\yeeqabs\sim 0.5$ (cf. Eq.~(\ref{eq:yeeqabs})). At lower accretion rates with $\dot{M}_{\mathrm{BH}}\ga 10^{-3}\,\Msol\,$s$^{-1}$ neutrino absorption becomes irrelevant compared to emission and $\yeeq\approx \yeeqem$. Finally, below $\dot{M}_{\mathrm{BH}}\sim 10^{-3}\,\Msol\,$s$^{-1}$ also neutrino emission ceases and $Y_e$ freezes out.

\item Neutrino absorption also has an indirect leverage on the values of $\yeeqem$, because absorption attenuates the rates of neutrino cooling and therefore keeps the electron degeneracy lower than without absorption (compare thick and thin red lines in Fig.~\ref{fig:torusye_time}). Moreover, $\yeeqem$ is artificially reduced by $\sim 0.05-0.1$ when neglecting the neutron-proton mass difference, $Q_{np}$, and electron mass in the neutrino emission rates, $\lambda_{e^\pm}$ (cf. Eq.~(\ref{eq:betarates})), e.g. in models that employ neutrino emission rates as formulated in \citet{Ruffert1996a} or \citet{Rosswog2003} (see Table~\ref{table_literature}). In contrast, $\yeeqem$ is almost insensitive to the inclusion of weak-magnetism effects (cf. Fig.~\ref{fig:yeeq1} and Table~\ref{table_weak}).

\item The electron fraction of the torus (approximately given by $\yemin$ in Table~\ref{table_weak}) roughly anti-correlates with the torus optical depth to neutrinos, $\tau_{\mathrm{opt}}^{\mathrm{max}}$ (cf. Fig.~\ref{fig:tabdat}), for $\tau_{\mathrm{opt}}^{\mathrm{max}}\la 1-10$, because a higher optical depth tends to be reached by more compact tori exhibiting a higher level of electron degeneracy. Increasing the BH mass, $\MBH$, while keeping the initial disk size proportional to $\MBH$, lowers the optical depth and neutron density of the torus and of the ejecta, whereas varying the BH spin only has a weak impact. For larger optical depths,  $\tau_{\mathrm{opt}}^{\mathrm{max}}\ga 1-10$, the anti-correlation between optical depth and $Y_e$ saturates and tends to be reversed due to the counteracting effects of neutrino absorption. As a result of this non-monotonic behavior, we observe the highest efficiency of heavy element production for the model with intermediate torus mass of $ 0.01\,M_\odot$ within the sequence of models varying the torus mass (see, e.g., Fig.~\ref{fig:reldif}).
 
\item Assuming a constant length $\lturb$ (parametrizing the scale of turnover motions) instead of the conventional prescription for the $\alpha$-viscosity effectively slows down angular momentum transport at late times when the torus has expanded to radii $r\gg \lturb$. This late-time deceleration extends the phase of efficient neutrino emission and allows $Y_e$ to trace its secularly increasing equilibrium value until it has reached higher values. As a consequence, the ejecta are less neutron rich and their $Y_e$ is less sensitive to the effects of neutrino absorption (as in \citealp{Fujibayashi2020a}). The fact that the two viscosity prescriptions lead to qualitatively different torus dynamics and nucleosynthesis results highlights the importance of a self-consistent MHD description of angular momentum transport.

\item\label{mhdconclusions} Replacing the viscous 2D treatment by a 3D MHD description, we observe a broader $Y_e$--mass distribution reaching down to lower $Y_e$ values. This trend, which is in agreement with \citet{Siegel2018c, Fernandez2019b} and \citet{Miller2019a}, is connected to the circumstance that the flow pattern of the MHD models is violently turbulent already during the neutrino-dominated phase, whereas the flow pattern of viscous models remains rather laminar during that phase. Our results are overall less optimistic for producing low $Y_e$ material than those of \citet{Siegel2018c} and \citet{Fernandez2019b} (who employ a more approximate treatment of neutrinos as well as $Q_{np}=m_e=0$) but more optimistic than those of \citet{Miller2020s} (who apply a more sophisticated neutrino solver than we do but only evolve until about $150\,$ms). Compared to the models using the standard $\alpha$-viscosity we observe a relatively weak impact of neutrino absorption, but as a reason we suspect insufficient grid resolution to capture the MRI in the model ignoring neutrino absorption owing to a geometrically thinner disk compared to the case including neutrino absorption.

\item For our set of models with disk masses up to $0.1\,M_\odot$, neutrino absorption results in a rise of the average ejecta $Y_e$ by about $0.02-0.05$ and in a reduction of the lanthanide and 3rd-peak mass fractions by factors of about $2-30$. However, despite this reduction the abundance pattern in the $A>130$ mass range is still rather close to the solar pattern for sufficiently compact and degenerate disks (e.g. as in our fiducial model with $\mtor^0=0.01\,\Msol$, $\MBH=3\,M_\odot$, and $\ABH=0.8$), particularly when using MHD instead of the $\alpha$-viscosity. Thus, our results support, if only marginally, the possibility of neutrino-cooled disks in NS mergers or collapsars being major sources of $A>130$ elements.

\item The mass fraction of outflow material that is unable to reach any kind of weak equilibrium before being ejected -- and basically retaining its initial values of $Y_e\approx Y_e^0$ -- is significant, namely $m^{\rm inert}_{\rm ej}/m_{\rm ej}\sim\mathcal{O}(10\,\%)$ (cf. Table~\ref{table_ejecta}). As a consequence, the average electron fraction of ejected material, $\yeej$, is systematically higher (by $\sim 0.08$ for our choice of parameters) for high initial values of $Y_e^0=0.5$ compared to the model with $Y_e^0=0.1$.
  
\item Since $\yeeqem<\yeeq\sim Y_e$ one can obtain an estimate of the lowest possible $Y_e$ that can be reached in ejecta of collapsar disks for given mass accretion rates by mapping the equilibrium electron fractions of all our models to the corresponding BH accretion rates (cf. Fig.~\ref{fig:mdotye}). If our models were representative for conditions in collapsars, the resulting $\yeeqem(\dot{M}_{\mathrm{infall}})$ relations would imply that $Y_e<0.25$ can only be reached for infall mass fluxes of $\dot{M}_{\mathrm{infall}}>2\times 10^{-2}\,M_\odot\,$s$^{-1}$ and, correspondingly, in progenitors providing sufficient angular momentum for a disk to be formed at such high mass infall rates. Since neutrino absorption becomes relevant right around this threshold, $\yeeqem$ systematically underestimates the $Y_e$ values expected in disks with higher mass-accretion rates.

\item Using an approximate kilonova model, we find variations of $40-80\,\%$ in the peak times and luminosities when comparing models with neutrino transport to models without, and variations of similar order of magnitude when changing the prescription for angular momentum transport. We observe a considerable sensitivity also to the detailed ejecta structure, $m(v)$, as well as when averaging the composition and heating rates over the entire ejecta in velocity space. These modeling uncertainties must be taken into account when decyphering future kilonova observations.
 
\end{enumerate}

The results of our study demonstrate that the nucleosynthesis in outflows from neutrino-cooled disks is by far not universal and may delicately depend both on the astrophysical parameters as well as on the modeling assumptions. We stress again that the impact of absorption is even more significant for heavier disks that typically occur in NS mergers than found here for the $0.01\,M_\odot$ disk models. Considering that in most cases the ejected lanthanide fractions differ by a factor of a few and more between models with and without neutrino absorption, we conclude that reliable nucleosynthesis and kilonova predictions require continued efforts to improve the quantitative understanding of weak interactions in neutrino-cooled disks. Having said that, a profound knowledge of the MHD processes that determine the thermodynamic state of the fluid, and by that regulate the equilibrium values $\yeeqem$ and $\yeeq$ in the disk, is every bit as important in order to faithfully predict $Y_e$ in the ejecta.

Our study, although comprehensive in a great variety of modeling aspects, still contains a number of limitations. Due to its approximate nature, the M1 method used here might over- or underestimate the impact of absorption, and the omission of GR effects (apart from the ones captured by the Artemova-potential) represents an additional source of uncertainty. Moreover, future investigations will have to overcome the shortcomings connected to the manually constructed initial configuration of the gas and, even more importantly, of the magnetic field \citep[see][for a first such study not accounting for neutrino absorption]{Christie2019a}. Last but not least, resolution studies will have to elaborate the conditions on resolution and/or grid configuration required in order to obtain converged equilibrium values $\yeeqem$ and $\yeeq$ in neutrino-MHD disks.

\section*{Acknowledgments}
We are grateful to the anonymous referee for constructive comments that improved the manuscript. We thank Miguel Aloy, Hirotaka Ito, Martin Obergaulinger, and Gabriel Mart\'inez-Pinedo for stimulating discussions. OJ was supported by the Special Postdoctoral Researchers (SPDR) program RIKEN. OJ and AB acknowledge support by the European Research Council (ERC) under the European Union's Horizon 2020 research and innovation programme under grant agreement No. 759253. SG is an FRS-FNRS research associate and his work was supported by the Fonds de la Recherche Scientifique (FNRS) and the Fonds Wetenschappelijk Onderzoek-Vlaanderen (FWO) under the EOS Project No O022818F. HTJ acknowledges funding by the Deutsche Forschungsgemeinschaft (DFG, German Research Foundation) through grants SFB-1258 ``Neutrinos and Dark Matter in Astro- and Particle Physics (NDM)'' and under Germany’s Excellence Strategy through Excellence Cluster ORIGINS (EXC 2094)—390783311. SN is partially supported by JSPS Grants-in-Aid for Scientific Research KAKENHI (A) 19H00693, Pioneering Program of RIKEN for Evolution of Matter in the Universe (r-EMU), and Interdisciplinary Theoretical and Mathematical Sciences Program (iTHEMS) of RIKEN. AB acknowleges support by Deutsche Forschungsgemeinschaft (DFG, German Research Foundation) - Project-ID 279384907 - SFB 1245 and - Project-ID 138713538 - SFB 881 (``The Milky Way System'', subproject A10). We acknowledge computational resources by the HOKUSAI supercomputer at RIKEN and by the Max Planck Computing and Data Facility (MPCDF). Nucleosynthesis calculations benefited from computational resources made available on the Tier-1 supercomputer of the F\'ed\'eration Wallonie-Bruxelles infrastructure funded by the Walloon Region under the grant agreement no. 1117545.

\emph{Data availability:} The data underlying this article will be shared on reasonable request to the corresponding author.

\appendix

\section{Equations for the hydrodynamic models}\label{sec:evolv-equat-hydr}

In this section we briefly summarize the equations that are solved in our hydrodynamic simulations. For the viscous models we solve the Newtonian Navier-Stokes equations in axisymmetry:
\begin{subequations}\label{eq:visevo}
\begin{align}
  & \partial_t \rho + \nabla_j(\rho v^j)  = 0  \, , \\
  & \partial_t (\rho Y_e) + \nabla_j(\rho Y_e v^j) = Q_{\mathrm{N}}\, , \\
  & \partial_t (\rho v^i) + \nabla_j(\rho v^i v^j + P_{\mathrm{g}} - T_{\mathrm{vis}}^{ij}) 
    = - \rho\nabla^i\Phi +Q_{\mathrm{M}}^i  \, , \\   
  & \partial_t e_{\mathrm{t}} + \nabla_j(v^j e_{\mathrm{t}} + v^j P_{\mathrm{g}} - v_i T_{\mathrm{vis}}^{ij})
    \nonumber \\
   & \hspace{3cm} = -\rho v_j\nabla^j\Phi + Q_{\mathrm{E}} + v_j Q_{\mathrm{M}}^j \, ,
\end{align}
\end{subequations}
where $\rho,v^i,Y_e,P_{\mathrm{g}}, e_{\mathrm{t}}, \Phi, T_{\mathrm{vis}}^{ij}$ are the baryonic mass density, velocity, electron fraction, gas pressure, total (i.e. internal plus kinetic) energy density, gravitational potential, and viscous stress tensor, respectively, while $Q_{\mathrm{N}},Q_{\mathrm{M}}^i, Q_{\mathrm{E}}$ stand for the source terms related to the exchange of lepton number, momentum, and energy between the gas and neutrinos, respectively (cf. Eqs.~(\ref{eq:qterms})). The viscous stress tensor is generally given by
\begin{align}
T_{\mathrm{vis}}^{ij} = \eta_{\mathrm{vis}}(\nabla^iv^j + \nabla^j v^i - \frac{2}{3}\delta^{ij}\nabla_k v^k) \, ,
\end{align}
where we take into account only the $T_{\mathrm{vis}}^{r\phi}$ and $T_{\mathrm{vis}}^{\theta\phi}$ components.

For the MHD models we solve in all three spherical polar coordinates the special relativistic MHD counterpart of Eqs.~(\ref{eq:visevo}), namely
\begin{subequations}\label{eq:mhdevo}
\begin{align}
   &\partial_t D + \nabla_j(D v^j)  = 0  \, , \\
   &\partial_t (D Y_e) + \nabla_j(D Y_e v^j) = Q_{\mathrm{N}}\, , \\
   &\partial_t (S^i) + \nabla_j(S^i v^j + P_{\mathrm{g}}^* - b^i B^j/W) \nonumber \\
    & \hspace{2.5cm} = - D\nabla^i\Phi +Q_{\mathrm{M}}^i  \, , \\
   & \partial_t \tau + \nabla_j( v^j \tau + v^j P_{\mathrm{g}}^* - b^0B^j /W) \nonumber \\
   & \hspace{2.5cm} = -D v_j\nabla^j\Phi +  Q_{\mathrm{E}} +  v_j Q_{\mathrm{M}}^j \, , \\
   & \partial_t B^i + \nabla_j(v^j B^i - v^i B^j) = 0 \, , \label{eq:indct} \\
   & \nabla_i B^i = 0       \,  , \label{eq:divb} 
\end{align}
\end{subequations}
in which $W=(1-v_jv^j/c^2)^{-1/2}$ is the Lorentz factor, $P_{\mathrm{g}}^*= P_{\mathrm{g}}+ b^2/2$ the total pressure, and $B^i$ ($b^i$) the three-vector (four-vector) of the magnetic field in the laboratory (comoving) frame. The relationship between conserved and primitive variables is given by:
\begin{subequations}\label{eq:conprim}
\begin{align}
   & D   = \rho W  \, , \\
   & S^i  = \rho h^* W^2 v^i - b^0b^i/c  \, , \\
   & \tau  = \rho h^* W^2c^2 - P_{\mathrm{g}}^* - b^0b^0 - \rho W c^2  \, , \\
   & b^0   = W(v_j B^j)/c  \, , \\
   & b^i   = Wv^i(v_j B^j)/c^2 + B^i/W   \, ,
\end{align}
\end{subequations}
with $h^*=(\rho c^2 + e_{\mathrm{i}}+P_{\mathrm{g}}+b^2)/(\rho c^2)$ being the total specific enthalpy and $e_{\mathrm{i}}$ the gas internal energy density.

The above equations are solved along with the two-moment equations of neutrino transport, which read:
\begin{subequations}\label{eq:momeq}
\begin{align}
  &\partial_tE + \nabla_j(F^j + v^j E) \nonumber \\
  & \hspace{1.5cm} + P^{ij}\nabla_iv_j  - \partial_\eps(\eps P^{ij} \nabla_iv_j ) =  S^{(0)} \, , \\
  &\partial_tF^i + \nabla_j( c^2 P^{ij} + v^j F^i) 
   \nonumber\\
  &\hspace{1.5cm} + F^j\nabla_jv^i - \partial_\eps(\eps Q^{ijk}\nabla_j v_k) =  S^{(1),i} \, ,
\end{align}
\end{subequations}
where $E, F^i, P^{ij}, Q^{ijk}$ are the angular moments of rank $0,1,2,3$, respectively, of the energy distribution of neutrinos and depend on the neutrino species and neutrino energy, $\eps$. The higher moments, $P^{ij}, Q^{ijk}$, are estimated by expressing them as local functions of $E, F^i$, in the form suggested by \citet{Minerbo1978} with the corresponding 3rd-order moment given in \citet{Just2015b}. Since Eqs.~(\ref{eq:momeq}) are Newtonian and we want to avoid unphysical effects related to the frame-dependent terms (i.e. terms containing the fluid velocities $v^j$), we limit the fluid velocities entering Eqs.~(\ref{eq:momeq}) to not exceed $0.1c$ by absolute value and neglect the azimuthal components. By doing so, we capture neutrino trapping and advection into the BH (acknowledging that advection terms can become dominant in the dynamic diffusion regime; see, e.g., \citealp{Mihalas1984}), but we neglect advection in the azimuthal direction and Doppler- and aberration effects in optically thin regions, mainly because we want to avoid spurious effects that might appear in high-velocity regions that are not well described by our $\mathcal{O}(v/c)$ scheme with approximate M1 closure. We believe that the uncertainties introduced by these restrictions are not more severe than the uncertainties related to other modeling ingredients, such as the $\alpha$-viscosity or the initial magnetic field configuration. However, future comparisons with reference solutions based on relativistic Boltzmann solvers will have to scrutinize this assumption. The coupling between hydrodynamics and neutrino transport is accomplished by means of the source terms, which are related by:
\begin{subequations}\label{eq:qterms}
\begin{align}
 Q_{\mathrm{N}} &= -  m_{\mathrm{B}}\int_0^\infty(S^{(0)}_{\nu_e}-S^{(0)}_{\bar\nu_e})\frac{\mathrm{d}\eps}{\eps} \, , \\
 Q_{\mathrm{M}}^i& =  - \frac{1}{c^2}\sum_{\nu}\int_0^\infty S^{(1),i}_{\nu}\,\mathrm{d}\eps   \, , \\
 Q_{\mathrm{E}}& =  - \sum_{\nu}\int_0^\infty S^{(0)}_{\nu}\,\mathrm{d}\eps   \, ,
\end{align}
\end{subequations}
where $m_{\mathrm{B}} = 1.66\times 10^{-24}\,$g.

Both the (magneto-/viscous-)hydrodynamics equations as well as the transport equations are evolved using PPM reconstruction \citep{Colella1984} in the formulation of \citet{Mignone2014a}, the HLLE Riemann solver, and 2nd-order Runge-Kutta time integration. In order to satisfy the divergence constraint, Eq.~(\ref{eq:divb}), at all times, the magnetic fields are defined on a staggered grid and the induction equation, Eq.~(\ref{eq:indct}), is evolved using the UCT scheme as in \citet{DelZanna2007}. For the recovery of the primitive variables from the conserved variables in the MHD models we employ a hybrid scheme, which for the first few iterations tries to find a solution with the 3D Newton-Raphson method as in \citet{Cerda-Duran2008a} and, if not successful, reverts to the more robust 1D bracketing scheme recently presented by \citet{Kastaun2021a} in combination with the 1D root-finding procedure of \citet{Chandrupatla1997a}.

\section{Method of light curve computation}\label{sec:meth-comp-kilon}

\begin{figure}
  \centering
  \includegraphics[width=0.45\textwidth]{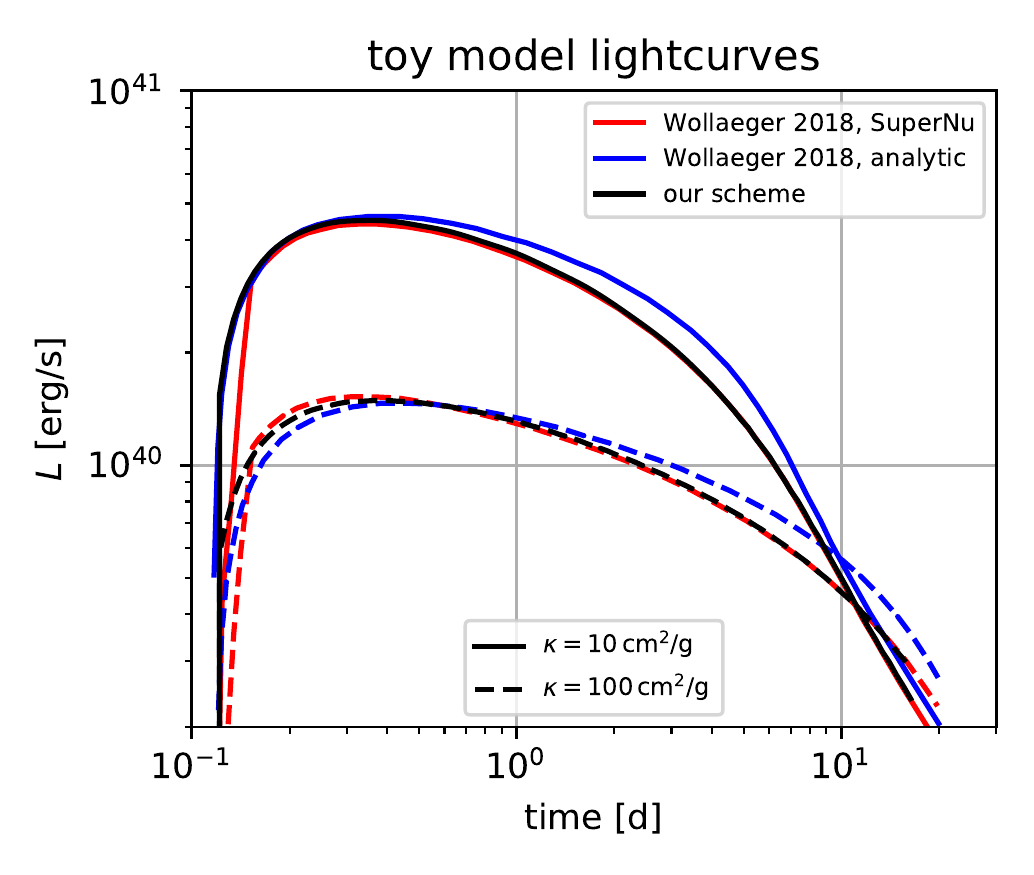}
  \caption{Comparison of light curves obtained using our two-moment scheme, Eqs.~(\ref{eq:knmomeq2}), with reference solutions from \citet{Wollaeger2018a} (data taken from their Fig.~5) for a toy model of expanding ejecta with a constant opacity of $\kappa=10\,$cm$^{2}\,$g$^{-1}$ and $100\,$cm$^{2}\,$g$^{-1}$. Black lines refer to our scheme, while red lines and blue lines refer to the Monte-Carlo and the analytic scheme of \citet{Wollaeger2018a}, respectively.}
  \label{fig:kn_toymodel}
\end{figure}	

\begin{figure*}
  \centering
  \includegraphics[width=0.95\textwidth]{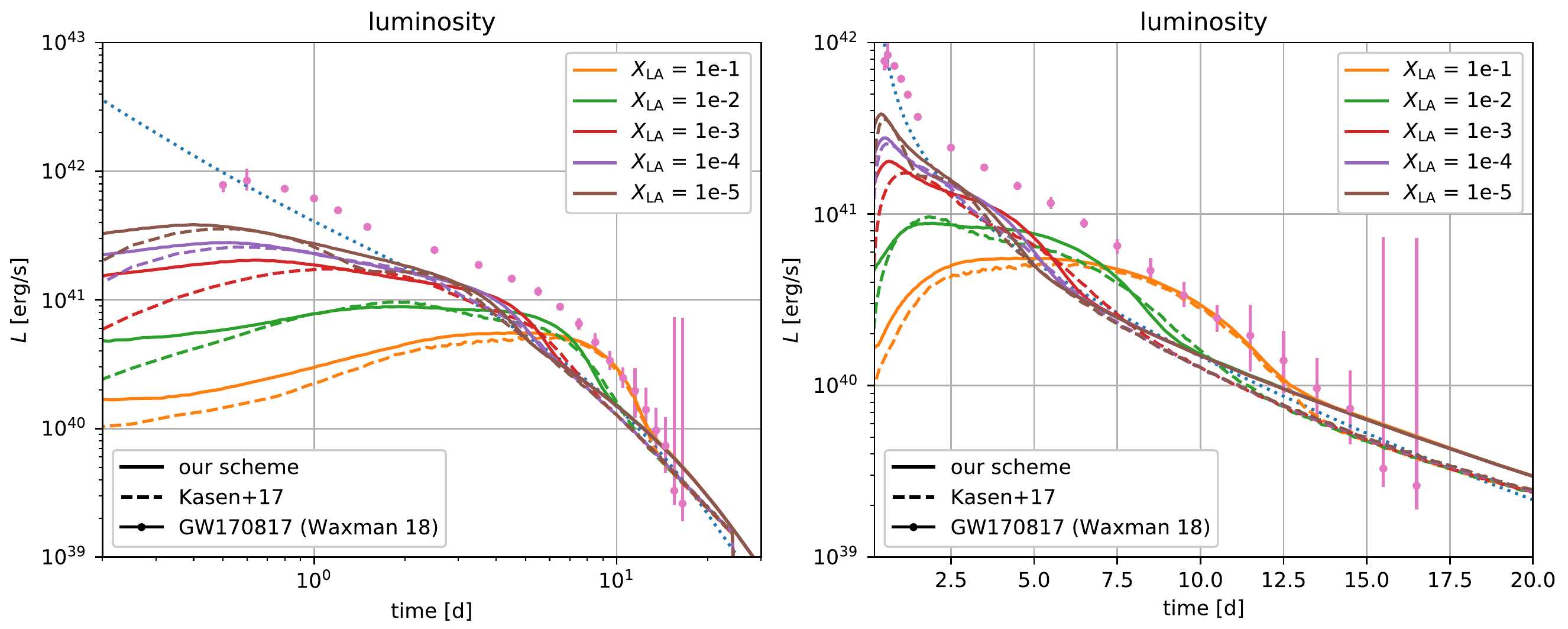}
  \caption{Comparison of bolometric luminosity obtained using our two-moment scheme, Eqs.~(\ref{eq:knmomeq2}), with the simplified opacity prescription of Eq.~(\ref{eq:kappa}) with radiative transfer solutions including detailed atomic opacities \citep{Kasen2017a} for a power-law ejecta distribution bearing different lanthanide mass fractions, $\XLA$. Left and right panels show the same data but use different scaling of the time axis in order to facilitate the comparison between the data sets. The blue dotted line shows the heating rate, $q$, powering all light curves. Consistent with the reference solutions, our light curves exhibit earlier and brighter peaks for decreasing values of $\XLA$. Dots with error bars show for comparison the bolometric luminosity observed for GW170817 \citep{Waxman2018a}.}
  \label{fig:kn_kasen}
\end{figure*}	

In Sect.~\ref{sec:estim-kilon-lightc} we already outlined the basic assumptions and the procedure for obtaining the distribution of mass, $m(v)$, specific nuclear heating rate, $q(v)$, and lanthanide plus actinide mass fraction, $\XLA(v)$, along the homologous velocity coordinate $v$. In this section we explain the procedure to compute the kilonova light curve. We start with the same assumptions as were invoked in \citet{Pinto2000a, Wollaeger2018a, Rosswog2018a} for analytic schemes to construct the evolution equation for the radiation energy density, $E$. However, we relax the Eddington approximation, $P\rightarrow E/3$, and instead express the Eddington factor $P/E$ as function of $E$ and energy-flux density, $F$, using the closure relation of \citet{Minerbo1978}. Hence, we end up with a set of two-moment equations for $E,F$ that is very similar to the neutrino-transport system in Eqs.~(\ref{eq:momeq}). In contrast to the latter, however, we evolve the energy-integrated equations with a velocity explicitly given by $v/c=r/(ct)\equiv x$ as a function of a single coordinate, $x$. This leads to:
\begin{subequations}\label{eq:knmomeq1}
\begin{align}
  &\frac{\dd E}{\dd t} + \frac{1}{c t x^2}\frac{\partial}{\partial x}( x^2 F) + 4 \frac{E}{t} = q \rho \, , \\
  &\frac{1}{c}\frac{\dd F}{\dd t} + \frac{1}{t x^2}\frac{\partial}{\partial x} (x^2 P) + 4 \frac{F}{ct} = - \kappa \rho F \, ,
\end{align}
\end{subequations}
where $\dd/\dd t=\partial/\partial t + v\partial/\partial r$ is the Lagrangian time derivative and all quantities only depend on $x$ and $t$. After multiplying by $t$ and substituting $t$ by the dimensionless coordinate
\begin{align}
  \tilde{t} = \ln \frac{t}{t_0} \, ,
\end{align}
where $t_0$ is a fiducial timescale, we end up with:
\begin{subequations}\label{eq:knmomeq2}
\begin{align}
  &\frac{\dd E}{\dd \tilde{t}} + \frac{1}{c x^2}\frac{\partial}{\partial x} (x^2 F)
  = q \rho t_0 e^{\tilde{t}} - 4 E  \, , \\
  &\frac{1}{c}\frac{\dd F}{\dd \tilde{t}} + \frac{1}{x^2}\frac{\partial}{\partial x} (x^2 P)
  = - \kappa \rho F t_0 e^{\tilde{t}} - 4 \frac{F}{c} \, .
\end{align}
\end{subequations}
The motivation for the above manipulation was to bring Eqs.~(\ref{eq:knmomeq2}) into a form where the left-hand side is formally equivalent to the two-moment equations in the laboratory frame but with the radial coordinate, $r$, replaced by the velocity coordinate, $x$. This allows to employ the same numerical methods as used for the neutrino transport and described in \citet{Just2015b}. We verify the method and corresponding numerical scheme by comparing the result for a simple toy model of expanding ejecta with a constant opacity ($\kappa=10\,$cm$^{2}$\,g$^{-1}$ or $\kappa=100\,$cm$^{2}$\,g$^{-1}$) with reference solutions by \citet{Wollaeger2018a}; cf. Fig.~\ref{fig:kn_toymodel} and see \citet{Wollaeger2018a} for the detailed model specifications. The agreement with the transfer scheme SuperNu of \citet{Wollaeger2018a} is excellent and slightly better than for the analytic method used in \citet{Wollaeger2018a}.

The final ingredient needed to compute light curves is the translation between ejecta properties and photon opacities, $\kappa$. In reality, $\kappa$ is a function of the detailed composition as well as the thermodynamic properties and ionization state \citep[e.g.][]{Kasen2013, Tanaka2020a}. In this study we employ a simplified parametrization of $\kappa$ in terms of only $\XLA$ considering that lanthanides, if present in the ejecta, tend to dominate the opacity. Additionally, we approximately take into account the reduction of possible line-transitions, and therefore of $\kappa$, due to electron recombination below temperatures $T\la 5000\,K$. We employ the following functional dependence:
\begin{align}\label{eq:kappa}  
  \kappa(\XLA,T) = &\max\{\kappa_0,\min\{\kappa_1,1.3\times (\XLA/10^{-3})^{\alpha(\XLA)}\} \nonumber \\
  &  \times\min\{(T/5000\,\mathrm{K})^3,1\}
\end{align}
where $\kappa_0=0.2$\,cm$^2$\,g$^{-1}$, $\kappa_1=0.65$\,cm$^2$\,g$^{-1}$, and $\alpha=0.2$ ($=0.65$) for $\XLA<10^{-3}$ ($>10^{-3}$). The temperature is computed assuming that radiation is in thermal equilibrium with the gas in regions of significant optical depth. The form of $\kappa$ as in Eq.~(\ref{eq:kappa}) was chosen with the motivation to reproduce a set of light curves by \citet{Kasen2017a}, in which the lanthanide content of the radiating ejecta was varied from $\XLA=0.1$ down to $\XLA=10^{-5}$. The ejecta mass and bulk velocity for this reference case are $0.02\,\Msol$ and $0.1\,c$, respectively; see \citet{Kasen2017a} for the precise form of the analytic mass distribution. With the aim to reproduce the light curves by \citet{Kasen2017a}, we use for this test (and only for this test) a radioactive heating rate that is presumably similar, though possibly not identical, to the one used in \citet{Kasen2017a}. It is taken from \citet{Lippuner2015a} and given by $q_{\mathrm{rad}}$\,[erg\,g$^{-1}$\,s$^{-1}$]$=(1.0763\times 10^{10} t_{\mathrm{day}}^{-1.518}+9.5483\times 10^9 e^{-t_{\mathrm{day}}/4.947})$ with $t_{\mathrm{day}}$ being the time in units of days. Thermalization is included as described in Sect.~\ref{sec:estim-kilon-lightc}. As can be seen in Figure~\ref{fig:kn_kasen}, our radiation solver, Eqs.~(\ref{eq:knmomeq2}), combined with the opacity prescription of Eq.~(\ref{eq:kappa}), can reasonably well reproduce the original light curves by \citet{Kasen2017a} and their dependence on $\XLA$. For comparison, Fig.~\ref{fig:kn_kasen} also shows the bolometric light curve of GW170817 \citep{Waxman2018a}.

%%%%%%%%%%%%%%%%%%%%%%%%%%%%%%%%%%%%%%%%%%%%%%%%%%%%%%%%%%%%%
%% Bibliography
%%%%%%%%%%%%%%%%%%%%%%%%%%%%%%%%%%%%%%%%%%%%%%%%%%%%%%%%%%%%%

%% \bibliographystyle{mnras}
%% \bibliography{articles_bibdesk}

\end{document}